\documentclass[runningheads]{lncse}
\usepackage{multicol}
\usepackage{makeidx}

\makeindex

\def\today{19~February~2000}
\def \ie {i.e.~} 
\def \etal{\textit{et al.}}
\def \rhs {rhs} 
\def \BT{B\"acklund transformation}
\def \KdV{Korteweg-de Vries}
\def\CRAS{C.~R.~Acad.~Sc.~Paris}
\def\PTP{Prog.~Theor.~Phys.~}
\def\SAM{Stud.~Appl.~Math.~}
\def \D {\hbox{d}}
\def \Ai  {\mathop{\rm Ai}\nolimits}
\def \Log {\mathop{\rm Log}\nolimits}
\def \ns  {\mathop{\rm ns}\nolimits}
\def \cs  {\mathop{\rm cs}\nolimits}
\def \sinh{\mathop{\rm sinh}\nolimits}
\def \sech{\mathop{\rm sech}\nolimits}
\def \grad{\mathop{\rm grad}\nolimits}
\def \diag{\mathop{\rm diag}\nolimits}
\def \Re  {\mathop{\rm Re}\nolimits}
\def \Im  {\mathop{\rm Im}\nolimits}
\def \mod#1{\vert #1 \vert}
\def \bfE {{\bf E}}
\def \bfP {{\bf P}}
\def \bfQ {{\bf Q}}
\def \bfu {{\bf u}}
\def \bfx {{\bf x}}
\def \bfY {{\bf Y}}
\def     \Metric{\sigma}
\def\GL{Ginz\-burg-Lan\-dau}
\def\GLA{u}

\def\pc{\overline{p}}
\def\qc{\overline{q}}

\def \expon {\alpha} % \Theta_{01}
\begin{document}

\frontmatter

%\tableofcontents

\mainmatter

\title
 {Exact solutions of nonlinear partial differential equations
  by singularity analysis}

\titlerunning{Exact solutions by singularity analysis \today}

\author{R.~Conte}

\institute{
Service de physique de l'\'etat condens\'e,
CEA Saclay,\\
F--91191 Gif-sur-Yvette Cedex,
France
}
\authorrunning{R.~Conte}

\maketitle
\index{Conte@Robert Conte}
\begin{abstract}
Whether integrable, partially integrable or nonintegrable,
nonlinear partial differential equations (PDEs)
can be handled from scratch 
with essentially the same toolbox,
when one looks for analytic solutions in closed form.
The basic tool is the appropriate use of the singularities of the solutions,
and this can be done without knowing these solutions in advance.
Since the elaboration of the \textit{singular manifold method}
by Weiss \textit{et al.},
many improvements have been made.
After some basic recalls,
we give an interpretation of the method
allowing us to understand why and how it works.
Next, we present the state of the art of this powerful technique,
trying as much as possible to make it a (computerizable) algorithm.
Finally,
we apply it to various PDEs in $1+1$ dimensions,
mostly taken from physics, some of them chaotic~:
sine-Gordon,
Boussinesq,
Sawada-Kotera,
Kaup-Kupershmidt,
complex Ginzburg-Landau,
Kuramoto-Sivashinsky,
etc.
\end{abstract}


\noindent \textit{Keywords}
\par Exact solutions
\par Nonperturbative methods
\par Painlev\'e analysis
 
\vfill
\textit{Direct and inverse methods in nonlinear evolution equations},
CIME school, Cetraro, 5--12 September 1999,
ed.~A.~Greco, Springer Verlag (2000).
\hfill \break
\today \hfill S99/072, nlin-SI/0009024

% ==========================================================================
\section{Introduction}
\label{sectionIntroduction}
\indent

Our interest is to find explicitly the ``macroscopic'' quantities
which materialize the integrability of a given \textit{nonlinear}
differential equation,
such as particular solutions or first integrals.
We mainly handle partial differential equations (PDEs),
although some examples are taken from ordinary differential equations (ODEs).
Indeed, the methods described in these lectures apply equally to both cases.

These methods are based on the \textit{a priori} study of the singularities
of the solutions.
The reader is assumed to possess
a basic knowledge of the singularities of nonlinear \textit{ordinary}
differential equations,
the Painlev\'e property for ODEs
and the Painlev\'e test.
All this prerequisite material is well presented in a
book by Hille \cite{Hille}
while Carg\`ese lecture notes \cite{Cargese96Conte}
contain a detailed exposition of the methods, 
including the Painlev\'e test for ODEs.
Many applications are given in a review \cite{RGB1989}.

As a general bibliography on the subject of these lectures,
we recommend Carg\`ese lecture notes \cite{Cargese96Musette}
and a shorter subset of these with emphasis on the various so-called
truncations \cite{ConteBari1997}.

Throughout the text, we exclude linear equations, unless explicitly stated.

% ==========================================================================
\section{Various levels of integrability for PDEs, definitions}
\label{sectionVariousLevels}
\indent

In this section,
we review the required definitions
(exact solution,
B\"acklund transformation,
Lax pair,
Darboux transformation,
etc).

The most important point is the global nature of the information which is 
looked for.
The existence theorem of Cauchy (for ODEs) or Cauchy-Kowalevski (for PDEs)
is of no help for this purpose.
Indeed, it only states a local property and says nothing on what happens
outside the disk of definition of the Taylor series.
Therefore it cannot distinguish between chaotic equations and integrable ones.

Still from this point of view,
Laurent series are not better than Taylor series.
For instance, the Bianchi IX cosmological model is a six-dimensional
dynamical system
\begin{eqnarray}
& &
\Metric^2 (\Log A)'' = A^2 - (B-C)^2,\
\hbox{ and cyclically},\
\Metric^4=1,
\label{eqBianchiIX}
\end{eqnarray}
which is undoubtedly chaotic \cite{ScheenDemaret}.
Despite the existence of the Laurent series \cite{CGR1993}
\begin{eqnarray}
& &
A/\Metric= \chi^{-1} + a_2 \chi + O(\chi^3),\ \chi=\tau-\tau_1,
\nonumber
\\
& &
B/\Metric= b_0 \chi + b_1 \chi^2 + O(\chi^3),\
\label{eqLaurentFamilyOneOrder0}
\\
& &
C/\Metric= c_0 \chi + c_1 \chi^2 + O(\chi^3),\
\nonumber
\end{eqnarray}
which depends on six independent arbitrary coefficients,
$(\tau_1,b_0,c_0,b_1,c_1,a_2)$,
a wrong statement would be to conclude to the absence of chaos.

This leads us to the definition of the first one of several needed global 
mathematical objects.

\begin{definition}
One calls
\textbf{exact solution} \index{solution!exact} of a nonlinear PDE
any solution defined in the whole domain of definition of the PDE
and which is given in 
\textit{closed form},
 \index{closed form}
i.e.~as a finite expression.
\end{definition}

The opposite of an exact solution is of course not a wrong solution,
but what Painlev\'e calls ``une solution illusoire'',
such as the above mentioned series.

Note that a multivalued expression is not excluded.
{}From this definition, an exact solution is 
\textit{global}, \index{solution!global}
as opposed to
\textit{local}. \index{solution!local}
This generically excludes series or infinite products,
unless their domain of validity can be made the full domain of definition,
like for linear ODEs.

\begin{example}
The Kuramoto-Sivashinsky (KS) equation
\index{Kuramoto-Sivashinsky equation}
\begin{eqnarray}
& &
u_t + u u_x + \mu u_{xx} + \nu u_{xxxx}=0,\ \nu \not=0,
\label{eqKS}
\end{eqnarray}
describes, 
for instance, the fluctuation of the position of a flame front, 
or the motion of a fluid going down a vertical wall,
or a spatially uniform oscillating chemical reaction in a homogeneous medium
(see Ref.~\cite{Manneville1988} for a review),
and it is well known for its chaotic behaviour.
An exact solution is 
the solitary wave of Kuramoto and Tsuzuki \cite{KuramotoTsuzuki}
in which the wavevector $k$ is fixed
\begin{eqnarray}
u & = &
  120 \nu \left({k \over 2} \tanh {k \over 2} \xi\right)^3
 +\left({60 \over 19} \mu - 30 \nu k^2\right){k \over 2} \tanh {k \over 2} \xi
 +c,\
\nonumber
\\
& &
\xi=x-ct-x_0,\
k^2={11 \mu \over 19 \nu} \hbox{ or } -{\mu \over 19 \nu},
\label{eqKuramotoTsuzuki}
\end{eqnarray}
which depends on two arbitrary constants $(c,x_0)$.
On the contrary, the Laurent series
\begin{eqnarray}
u & = & 120 \nu \xi^{-3} + {60 \over 19} \mu \xi^{-1} + c 
- {120.11 \over 19^2} \mu^2 \xi 
 + u_6 \xi^3 +  O(\xi^4),
\label{eqKSLaurent}
\end{eqnarray}
which depends on three arbitrary constants $(c,x_0,u_6)$,
is not an exact solution,
since no closed form expression is yet known for the sum of this series.
\end{example}

There exists a powerful tool to build exact solutions,
this is the \BT.
For simplicity, but this is not a restriction,
we give the basic definitions for a PDE defined as a single scalar equation
for one dependent variable $u$ and two independent variables $(x,t)$.

\begin{definition}
(Refs.~\cite{DarbouxSurfaces} vol.~III chap.~XII, \cite{MatveevSalle})
%\cite{Backlund1883} 
A \textbf{B\"acklund transformation} 
(BT)
between two given PDEs 
\index{B\"acklund transformation} 
\begin{equation}
 E_1(u,x,t)=0,\ E_2(U,X,T)=0
\end{equation}
is 
a pair of relations
\begin{equation}
 F_j(u,x,t,U,X,T)=0,\ j=1,2                                % no \lambda here
\label{eqBT}
\end{equation}
with some transformation between $(x,t)$ and $(X,T)$,
in which $F_j$ depends on the derivatives of $u(x,t)$ and $U(X,T)$,
such that the elimination of $u$ (resp.~$U$) between $(F_1,F_2)$
implies
$E_2(U,X,T)=0$ (resp.~$E_1(u,x,t)=0$).
The BT is called the \textbf{auto-BT} or the \textbf{hetero-BT}
according as the two PDEs are the same or not.
\index{B\"acklund transformation!auto--}
\index{B\"acklund transformation!hetero--}
\end{definition}

\begin{example}
The sine-Gordon equation
\index{sine-Gordon equation}
(we identify sine-Gordon and sinh-Gordon since an affine transformation on $u$
does not change the integrability nor the singularity structure)
\begin{eqnarray}
& & 
\hbox{sine-Gordon }: E(u) \equiv u_{xt} + 2 a \sinh u=0
\label{eqSG}
\label{eqsG}
\end{eqnarray}
admits the auto-BT
\begin{eqnarray}
& & 
(u+U)_x +  4         \lambda  \sinh{u-U \over 2}=0,
\label{eqSGBTx}
\\
& &
(u-U)_t - {2 a \over \lambda} \sinh{u+U \over 2}=0,
\label{eqSGBTt}
\end{eqnarray}
in which $\lambda$ is an arbitrary complex constant,
called the \textit{B\"acklund parameter}. \index{B\"acklund parameter}
\end{example}

Given the obvious solution $U=0$ 
(called \textit{vacuum}), 
\index{vacuum}
the two equations (\ref{eqSGBTx})--(\ref{eqSGBTt}) are Riccati ODEs 
with constant coefficients for the unknown $e^{u/2}$,
\begin{eqnarray}
& & 
(e^{u/2})_x = \lambda (1 - (e^{u/2})^2),
\\
& &
(e^{u/2})_t = - a (1 - (e^{u/2})^2)/(2 \lambda),
\end{eqnarray}
therefore their general solution is known in closed form 
\begin{eqnarray}
& & 
e^{u/2}=\tanh \theta,\
\theta=\left(\lambda x - {a \over 2 \lambda} t - z_0\right),\
\label{eqSGOneSoliton}
\end{eqnarray}
with $(\lambda,z_0)$ arbitrary.
This solution is called the 
\textit{one-soliton solution}, \index{one-soliton solution}
it is also written as 
\begin{eqnarray}
& & 
\tanh (u/4) = - e^{\displaystyle{-2 \theta}},\
u_x = 4   \lambda      \sech 2 \theta,\
u_t =-2 a \lambda^{-1} \sech 2 \theta.
%\label{eqSGOneSolitonBis}
\end{eqnarray}
By iteration, this procedure gives rise to the $N$-soliton solution
                                        \index{$N$-soliton solution}
\cite{Lamb1967,AblowitzClarkson},
an exact solution depending on $2N$ arbitrary complex constants
($N$ values of the B\"acklund parameter $\lambda$,
 $N$ values of the shift $z_0$),
with $N$ an arbitrary positive integer.
A remarkable feature of the SG-equation,
due to the fact that at least one of the two ODEs 
(\ref{eqSGBTx})--(\ref{eqSGBTt})
is of order one,
is that this $N$-soliton can be obtained from $N$ different copies of the
one-soliton by a simple algebraic operation,
i.e.~without integration (see Musette's lecture \cite{CetraroMusette}).

\begin{example}
The Liouville equation
\index{Liouville equation}
\begin{eqnarray}
& & 
\hbox{Liouville: } E(u) \equiv u_{xt} + \alpha e^u=0
\label{eqLiouville}
\end{eqnarray}
admits two BTs.
The first one 
\index{B\"acklund transformation!hetero--}
\begin{eqnarray}
& &
(u-v)_x = \alpha \lambda e^{(u+v)/2},\
\label{eqLiouHeteroBTx}
\\
& &
(u+v)_t = - 2 \lambda^{-1} e^{(u-v)/2},\
\label{eqLiouHeteroBTt}
\end{eqnarray}
is a BT to a linearizable equation called the d'Alembert equation
\index{d'Alembert equation}
\begin{eqnarray}
& & 
\hbox{d'Alembert: } E(v) \equiv v_{xt}=0.
\label{eqdAlembert}
\end{eqnarray}
The second one is an auto-BT
\begin{eqnarray}
& &
(u+U)_x = - 4 \lambda \sinh {u - U \over 2},\
\label{eqLiouAutoBTx}
\\
& &
(u-U)_t = \lambda^{-1} \alpha e^{(u + U)/2}.
\label{eqLiouAutoBTt}
\end{eqnarray}
\end{example}

The first of these two BTs allows one to obtain the general solution of the
nonlinear Liouville equation, see Section \ref{sectionTruncationOneFamily}.

This ideal situation (generation of the general solution) is exceptional
and the generic case is the generation of particular solutions only,
as in the sine-Gordon example.
\index{sine-Gordon equation}

The importance of the BT is such that it is often taken as a definition of
\textit{integrability}.

\begin{definition}
A PDE in $N$ independent variables is \textbf{integrable} 
if at least one of the following properties holds.
\index{integrability}
\begin{enumerate}
\item
It is linearizable.

\item
For $N>1$, it possesses an auto-BT which, if $N=2$,
depends on an arbitrary complex constant, the B\"acklund parameter.
\index{B\"acklund parameter}

\item
It possesses a hetero-BT to another integrable PDE.
\index{B\"acklund transformation!hetero--}

\end{enumerate}
\label{definitionIntegrable}
\end{definition}

Although partially integrable and nonintegrable equations,
i.e.~the majority of physical equations,
admit no BT, they retain part of the properties of (fully) integrable PDEs,
and this is why the methods presented in these lectures apply to both cases
as well.
For instance, the KS equation 
\index{Kuramoto-Sivashinsky equation}
admits the \textit{vacuum} solution $u=0$
and, in Section \ref{sectionVariousLevels}, an iteration will be built
leading from $u=0$ to the solitary wave
(\ref{eqKuramotoTsuzuki});
the nonintegrability manifests itself in the finite number of times
this iteration provides a new result ($N=1$ for the KS equation,
and one cannot go beyond (\ref{eqKuramotoTsuzuki}) \cite{CM1989}).

For various applications of the BT, see Ref.~\cite{RogersShadwick}.

When a PDE has some good reasons to possess such features,
such as the reasons developed in Section \ref{sectionPDEPtest},
we want to find the BT if it exists,
since this is a generator of exact solutions,
or a degenerate form of the BT if the BT does not exist,
and we want to do it by singularity analysis \textit{only}.

Before proceeding, we need to define some other elements of integrability.

\begin{definition}
\label{defLaxPair}
Given a PDE,
a \textbf{Lax pair} \index{Lax pair}
is a system of two linear differential operators
\begin{equation}
\hbox{Lax pair}:\ 
L_1(U,\lambda),\
L_2(U,\lambda),
\label{eqLax}
\end{equation}
depending on a solution $U$ of the PDE and,
in the $1+1$-dimensional case, on an arbitrary constant $\lambda$,
called the \textbf{spectral parameter},
            \index{spectral parameter}
with the property that the vanishing of the commutator $[L_1,L_2]$ is
equivalent to the vanishing of the PDE $E(U)=0$.
\end{definition}

A Lax pair can be represented in several, equivalent ways.

% Ablowitz and Segur (1991) page 9.

The \textit{Lax representation} \cite{Lax}
\index{Lax pair!Lax representation}
is a pair of linear operators $(L,P)$ (scalar or matrix) defined by
\begin{eqnarray}
& &
L_1 = L - \lambda,\
L_2 = \partial_t - P,\
L_1 \vec \psi=0,\
L_2 \vec \psi=0,\
\lambda_t=0,
\label{eqLaxForm}
\end{eqnarray}
in which the elimination of the scalar $\lambda$ yields
\begin{eqnarray}
& &
L_t=\lbrack P,L \rbrack,
\label{eqLaxRep}
\end{eqnarray}
\ie,
thanks to the \textit{isospectral} condition $\lambda_t=0$,
a time evolution analogous to the one in Hamiltonian dynamics.
               \index{isospectral}

The \textit{zero-curvature representation}
is a pair $(L,M)$ of linear operators independent of $(\partial_x,\partial_t)$
     \index{Lax pair!zero-curvature representation}
\begin{eqnarray}
& &
L_1 = \partial_x - L,\
L_2 = \partial_t - M,\
L_1 \vec \psi=0,\
L_2 \vec \psi=0,\
\nonumber
\\
& &
\lbrack \partial_x - L,\partial_t - M \rbrack = L_t - M_x + L M - M L =0.
\label{eqLaxZeroCurvRep}
\end{eqnarray}
The common order $N$ of the matrices is called
the \textit{order} of the Lax pair. \index{Lax pair!order}

The \textit{projective Riccati representation}
is a first order system of $2 N - 2$ Riccati equations in the unknowns
$\psi_j / \psi_1, j=2,\dots,N$,
equivalent to the zero-curvature representation (\ref{eqLaxZeroCurvRep}).
\index{Lax pair!Riccati representation}

The \textit{scalar representation} 
is a pair of scalar linear PDEs, one of them of order higher than one,
\index{Lax pair!scalar representation}
\begin{eqnarray}
& &
L_1 \psi = 0,\
L_2 \psi = 0,\
\nonumber
\\
& &
X \equiv \lbrack L_1, L_2 \rbrack = 0.
\label{eqLaxScalarRep}
\end{eqnarray}
In $1+1$-dimensions, one of the PDEs can be made an ODE
(i.e.~involving only $x$- or $t$-derivatives),
in which case the order of this ODE is called the order of the Lax pair.

The \textit{string representation} or \textit{Sato representation}
\cite{JM1983}
\index{Lax pair!string representation}
\begin{eqnarray}
& &
\lbrack P,Q \rbrack = 1.
\end{eqnarray}
This very elegant representation,
reminiscent of Hamiltonian dynamics,
uses the Sato definition of a
\textit{microdifferential operator}
(a differential operator with positive and negative powers of the
differential operator $\partial$)
and of its \textit{differential part} denoted $()_{+}$
(the subset of its nonnegative powers),
e.g.
\index{differential part}
\index{microdifferential operator}
\begin{eqnarray}
& &
Q=\partial_x^2-u,\
\\
& &
L=Q^{1/2},
\\
& &
\left(L^3 \right)_{+}=\partial_x^3-(3/4) \lbrace u,\partial_x \rbrace,
\\
& &
\left(L^5 \right)_{+}=\partial_x^5-(5/4) \lbrace u,\partial_x^3 \rbrace
+(5/16) \lbrace 3 u^2 + u_{xx},\partial_x \rbrace,
\end{eqnarray}
in which $\lbrace a,b \rbrace$ denotes the anticommutator $a b + ba$.
See Ref.~\cite{Cargese96DiFrancesco} for a tutorial presentation.

\begin{example}
The sine-Gordon equation (\ref{eqsG}) admits the zero-curvature representation
\index{sine-Gordon equation}
\begin{eqnarray}
& &
(\partial_x - L) \pmatrix{\psi_1 \cr \psi_2 \cr}=0,\
L=
\pmatrix{U_x/2 & \lambda \cr \lambda & - U_x/2 \cr},\
\label{eqSGLaxx}
\\
& &
(\partial_t - M) \pmatrix{\psi_1 \cr \psi_2 \cr}=0,\
M= - (a/2) \lambda^{-1} \pmatrix{0 & e^{U} \cr e^{- U} & 0 \cr},
\label{eqSGLaxt}
\end{eqnarray}
equivalent to the Riccati representation, with $y=\psi_1 / \psi_2$,
\begin{eqnarray}
& &
y_x = \lambda +  U_x y - \lambda y^{2},
\label{eqSGRiccatiZx}
\\
& &
y_t = -{a \over 2} \lambda^{-1} e^{U}
      +{a \over 2} \lambda^{-1} e^{-U} y^2.
\label{eqSGRiccatiZt}
\end{eqnarray}
\end{example}

\begin{example}
The matrix nonlinear Schr\"odinger equation
\index{nonlinear Schr\"odinger equation}
\begin{eqnarray}
& &
 i Q_t +(b/a) Q_{xx} - 2 a b Q R Q =0,\
-i R_t +(b/a) R_{xx} - 2 a b R Q R =0,
%
%i            \pmatrix{0 & Q_t \cr -R_t & 0 \cr}
% +(b/a)  \pmatrix{0 & Q_{xx} \cr R_{xx} & 0 \cr}
% - 2 a b \pmatrix{0 & Q R Q \cr R Q R & 0 \cr}.
\end{eqnarray}
in which 
$(Q,R)$ are rectangular matrices of respective orders $(m,n)$ and $(n,m)$,
and $(i,a,b)$ constants,
admits the zero-curvature representation
(\cite{MP1982} Eq.~(5))
%\cite{MP1982} Eq.~(5)      $(L,M)$ representation
%\cite{Manakov1973} Eq.~(6) $(L,A)$ representation
% the pair is not explicit in \cite{ZS1982}, and absent from \cite{ZS1974}
\begin{eqnarray}
& &
(\partial_x -L)\psi=0,\
(\partial_t -M)\psi=0,\
\\
& &
L=a P + \lambda G,\
M=(-a G P^2 + G P_x + 2 \lambda P +(2/a) \lambda^2 G) b/i,
\end{eqnarray}
in which $\lambda$ is the spectral parameter,
$P$ and $G$ matrices of order $m+n$ defined as 
\begin{eqnarray}
& &
P=\pmatrix{0 & Q \cr -R & 0 \cr},\
G=\pmatrix{1_m & 0 \cr 0 & - 1_n \cr}.
\end{eqnarray}
The matrix $G$ characterizes the internal symmetry group
${\rm GL}(m,{\mathcal C}) \otimes {\rm GL}(n,{\mathcal C})$.
The lowest values
\begin{eqnarray}
& &
m=1,n=1,\ Q=\pmatrix{u \cr},\ R=\pmatrix{U \cr},\
\label{eqNLS11}
\end{eqnarray}
define the AKNS system (Section \ref{sectionNLS}),
whose reduction $U=\bar u$ is the usual scalar nonlinear 
Schr\"odinger equation.
\end{example}

\begin{example}
The $2+1$-dimensional Ito equation \cite{Ito}
               \index{Ito equation}
\begin{eqnarray}
{\hskip -7.0 truemm}
E(u) & \equiv &
\left(
u_{xxxt} + 6 \alpha^{-1} u_{xt} u_{xx} + a_1 u_{tt} + a_2 u_{xt} + a_3 u_{xx}
 + a_4 u_{ty}\right)_x=0
\end{eqnarray}
has a Lax pair whose scalar representation is
\begin{eqnarray}
& &
L_1 \equiv
\partial_x^3 + a_1 \partial_t + (a_2 + 6 \alpha^{-1} U_{xx}) \partial_x
+ a_4 \partial_y - \lambda
\\
& &
L_2 \equiv
\partial_x \partial_t - \mu \partial_x + ({a_3 \over 3} +2 \alpha^{-1} U_{xt})
\\
& &
\alpha \lbrack L_1,L_2 \rbrack =
2 E(U) + 6 U_{xxx} L_2.
\end{eqnarray}
%[Weak Lax pair = bad example?] \hfill \break
In the $2+1$-dimensional case $a_4 \not=0$,
the parameter $\lambda$ can be set to $0$ by the change
$\psi \mapsto \psi e^{\lambda y}$.
This is the reason of the precision at the end of item 2 in definition 
\ref{defLaxPair}.
This pair has the order four in the generic case $a_1 \not=0$,
although neither $L_1$ nor $L_2$ has such an order.
\end{example}

\begin{example}
The string representation of the Lax pair of the derivative of the first
Painlev\'e equation is
\index{Painlev\'e!equation (P1)}
\begin{eqnarray}
& &
\lbrack P,Q \rbrack 
=\lbrack \left( (\partial_x^2-u)^3\right)_{+},\partial_x^2-u \rbrack
= -(1/4) u_{xxx} +(3/4) u u_x
= 1.
\end{eqnarray}
\end{example}

\begin{example}
The Sato representation of the Lax pair 
for the whole Korteweg-de Vries hierarchy is
\index{Korteweg-de Vries!hierarchy}
\begin{eqnarray}
& &
\partial_{t_m} L = \lbrack \left(L^{2 m -1} \right)_{+},L \rbrack,\
L=Q^{1/2},\
Q=\partial_x^2-u,\
m=1,3,5,\dots
\end{eqnarray}
\end{example}

{}From the singularity point of view,
the Riccati representation is the most suitable, as will be seen.

The last main definition we need is the Darboux transformation (DT).
The working definition given below is very simplified (this is an involution)
as compared to the one of Darboux \cite{Darboux1882},
but it is sufficient for our purpose.
The full definition is given in Musette's lecture \cite{CetraroMusette}.

\begin{definition}
Given a PDE,
a \textbf{Darboux transformation} \index{Darboux transformation} 
is a transformation between two solutions $(u,U)$ of the PDE
\begin{equation}
\hbox{DT}:\ 
u = \sum_f {\mathcal D}_f \Log \tau_f + U
% NDLR It seems to never be a sum, but just one term (1 Riccati=2 linear)
\label{eqDTGeneral}
\end{equation}
linking their difference 
to a finite number of linear differential operators ${\mathcal D}_f$ 
($f$ like family)
acting on the logarithm of functions $\tau_f$.
% The linear operators ${\mathcal D}_f$ are easy to derive from the test
\end{definition}

In the definition (\ref{eqDTGeneral}), it is important to note that, 
despite the notation,
each function $\tau_f$ is in fact the ratio of the ``tau-function'' of $u$
by that of $U$.
\index{tau-function}

Lax pairs, B\"acklund and Darboux transformations are not independent.
In order to exhibit their interrelation,
one needs an additional information,
namely the link
\begin{eqnarray}
& & \forall f\ : \ {\mathcal D}_f \Log \tau_f = F_f(\psi),
\label{eqLinkTauPsi}
\end{eqnarray}
which most often is the identity $\tau=\psi$,
between the functions $\tau_f$ and the function $\psi$
in the definition of a scalar Lax pair.

\begin{example}
The (integrable) sine-Gordon equation (\ref{eqsG}) 
admits the Darboux transformation
\index{sine-Gordon equation}
\begin{eqnarray}
& &
u=U - 2 (\Log \tau_1 - \Log \tau_2),
\label{eqSGDTu}
\end{eqnarray}
in which $(\tau_1,\tau_2)$ is a solution $(\psi_1,\psi_2)$
of the system (\ref{eqSGLaxx})--(\ref{eqSGLaxt}).
\end{example}

Then its BT (\ref{eqSGBTx})--(\ref{eqSGBTt}) is the result of the elimination
\cite{Chen1974}
of $\tau_1 / \tau_2$
between the DT (\ref{eqSGDTu}) and the Riccati form of the Lax pair
(\ref{eqSGRiccatiZx})--(\ref{eqSGRiccatiZt}),
with the correspondence $\tau_f=\psi_f,f=1,2$.
This elimination reduces to the substitution $y=e^{-(u-U)/2}$ in the
Riccati system (\ref{eqSGRiccatiZx})--(\ref{eqSGRiccatiZt}),
and this is one of the advantages of the Riccati representation.
Therefore the B\"acklund parameter and the spectral parameter are
identical notions.

\begin{example}
The (nonintegrable) Kuramoto-Sivashinsky equation 
\index{Kuramoto-Sivashinsky equation}
admits the degenerate Darboux transformation
\begin{eqnarray}
& &
u=U + (60 \nu \partial_x^3 + (60/19) \mu \partial_x) \Log \tau,
\label{eqKSSolitaryWithD}
\end{eqnarray}
in which $U=c$ (vacuum) and $\tau$ is
the general solution $\psi$ of the linear system
(a degenerate second order scalar Lax pair)
\begin{eqnarray}
& &
L_1 \psi \equiv (\partial_x^2 - k^2/4) \psi=0,
\\
& &
L_2 \psi \equiv (\partial_t + c \partial_x) \psi=0,
\\
& &
\lbrack L_1,L_2 \rbrack \equiv 0.
\end{eqnarray}
\end{example}

The solution $u$ defined by (\ref{eqKSSolitaryWithD}) is then
the solitary wave (\ref{eqKuramotoTsuzuki}),
and this is a much simpler way to write it,
because the logarithmic derivatives in (\ref{eqKSSolitaryWithD}) 
take account of the whole nonlinearity.

Since, roughly speaking,
the BT is equivalent to the couple (DT, Lax pair),
one can rephrase as follows the iteration to generate new solutions.
Let us symbolically denote 
\begin{description}
\item
$E(u)=0$ the PDE,

\item
$ \hbox{Lax}(\psi, \lambda, U) = 0 $ a scalar Lax pair,

\item
$F$ the link (\ref{eqLinkTauPsi}) ${\mathcal D} \Log \tau = F(\psi)$ 
from $\psi$ to $\tau$,

\item
$u=\hbox{Darboux}(U,\tau)$ the Darboux transformation.

\end{description}
The iteration is the following, see e.g.~\cite{GRTW}.

\begin{enumerate}
\item
(initialization)
Choose $u_0=$ a particular solution of $E(u)=0$;
set $n=1$;
perform the following loop until some maximal value of $n$;

\item
(start of loop)
Choose $ \lambda_n=$ a particular complex constant;

\item
Compute, by integration,
a particular solution $\psi_n$ of the linear system
\hfill \break \noindent
$ \hbox{Lax}(\psi, \lambda_{n}, u_{n-1}) = 0$;

\item
Compute, without integration, ${\mathcal D} \Log \tau_n = F(\psi_n)$;

\item
Compute, without integration, $u_n=\hbox{Darboux}(u_{n-1},\tau_n)$;

\item
(end of loop) Set $n=n+1$.

\end{enumerate}

Depending on the choice of $\lambda_n$ at step 2, and of $\psi_n$ at step 3,
one can generate 
either the $N$-soliton solution,
or     solutions rational in $(x,t)$,
or     a mixture of such solutions. % \cite{Matveev} positons

%\vfill \eject

% ==========================================================================
\section{Importance of the singularities~: 
a brief survey of the theory of Painlev\'e}
\label{sectionImportance}
\indent

A classical theorem states that a function of one complex variable without any
singularity in the analytic plane (\ie\ the complex plane compactified by
addition of the unique point at infinity)
is a constant.
Therefore a function with singularities is characterized,
as shown by Mittag-Leffler,
by the knowledge of its singularities in the analytic plane.
Similarly,
if $u$ satisfies an ODE or a PDE,
the structure of singularities of the general solution 
characterizes the level of integrability of the equation.
This is the basis of the theory of the (explicit) integration of
nonlinear ODEs built by Painlev\'e, which we only briefly introduce here
[for a detailed introduction, see Carg\`ese lecture notes~:
Ref.~\cite{Cargese96Conte} for ODEs,
Ref.~\cite{Cargese96Musette} for PDEs].

To integrate an ODE is to acquire a global knowledge of its general solution,
not only the local knowledge ensured by the existence theorem of Cauchy.
So, the most demanding possible definition for the ``integrability'' of an ODE
is the single valuedness of its general solution,
so as to adapt this solution to any kind of initial conditions.
Since even linear equations may fail to have this property,
e.g.~$2 x u' + u=0,\ u=c x^{-1/2}$,
a more reasonable definition is the following one.

\begin{definition}
The \textit{Painlev\'e property} (PP) of an ODE is the 
uniformizability of its general solution.
\index{Painlev\'e!property}
\end{definition}

In the above example, the uniformization is achieved by the change of 
independent variable $x=X^2$.
This definition is equivalent to the more familiar one.

\begin{definition}
The \textit{Painlev\'e property} (PP) of an ODE is 
the absence of movable critical singularities in its general solution.
\index{Painlev\'e!property}
\end{definition}

\begin{definition}
The \textit{Painlev\'e property} (PP) of a PDE is 
its integrability (Definition \ref{definitionIntegrable})
and the absence of movable critical singularities in its general solution.
\index{Painlev\'e!property}
\end{definition}

Let us recall that a singularity is said \textit{movable} 
(as opposed to \textit{fixed})
if its location depends on the initial conditions,
and \textit{critical} if multivaluedness takes place around it.
Indeed, out of the four configurations of singularities 
(critical or noncritical) and (fixed or movable),
only the configuration (critical and movable) prevents uniformizability~:
one does not know where to put the cut since the point is movable.

Wrong definitions of the PP, alas repeatedly published,
consist in replacing in the definition
``movable critical singularities''
by ``movable singularities other than poles'',
or ``its general solution'' by ``all its solutions''.
Even worse definitions only refer to Laurent series.
See Ref.~\cite{Cargese96Conte}, Section 2.6,
for the arguments of Painlev\'e himself.

The mathematicians like Painlev\'e want to integrate whole classes of
ODEs (e.g.~second order algebraic ODEs).
We will only use their methods for a given ODE or PDE,
with the aim of deriving the elements of integrability described in
Section \ref{sectionVariousLevels} (exact solutions, \dots).
This \textit{Painlev\'e analysis} is twofold 
(``double m\'ethode'', says Painlev\'e).
      \index{Painlev\'e!analysis}
\begin{enumerate}
\item
Build necessary conditions for an ODE or a PDE to have the PP
(this is called the \textit{Painlev\'e test}).
                     \index{Painlev\'e!test}

\item
When all these conditions are satisfied, or at least some of them,
find the global elements of integrability.
In the integrable case
this is achieved either (ODE case) by explicitly integrating or (PDE case)
by finding 
an auto-BT 
(like equations (\ref{eqSGBTx})--(\ref{eqSGBTt}) for sine--Gordon)
\index{sine-Gordon equation}
or a BT towards another PDE with the PP 
(like (\ref{eqLiouHeteroBTx})--(\ref{eqLiouHeteroBTt}) between the d'Alembert
and Liouville equations).
\index{B\"acklund transformation!hetero--}
\index{d'Alembert equation}
\index{Liouville equation}
In the partially integrable case, only degenerate forms of the above can be
expected, as described in Section \ref{sectionVariousLevels}.

\end{enumerate}

%\vfill \eject

% ==========================================================================
\section{The Painlev\'e test for PDEs in its invariant version}
\label{sectionPDEPtest}
\indent

When the PDE reduces to an ODE,
the Painlev\'e test (for shortness we will simply say the test)
reduces by construction to the test for ODEs,
presented in detail elsewhere \cite{Cargese96Conte}
and assumed known here.
\index{Painlev\'e!test}

We will skip those steps of the test which are the same
for ODEs and for PDEs (e.g., diophantine conditions that all the leading
powers and all the Fuchs indices be integral),
and we will concentrate on the features which are specific to PDEs,
namely the description of the movable singularities,
the optimal choice of the expansion variable for the Laurent series,
the advantage of the homographic invariance.

% ==========================================================================
\subsection{Singular manifold variable ${\bf \varphi}$
 and expansion variable ${\bf \chi}$}
\label{sectionSMVandEV}
\indent

Consider a nonlinear PDE
\begin{eqnarray}
& &
E(u,x,t,\dots)=0.
\label{eqPDE}
\end{eqnarray}

To test movable singularities for multivaluedness without integrating,
which is the essence of the test,
one must first describe them,
then,
among other steps,
check the existence near each movable singularity of a Laurent series which
represents the general solution.

For PDEs,
the singularities are not isolated in the space of the independent variables
$(x,t,\dots)$,
but they lay on a codimension one manifold
\begin{eqnarray}
& &
\varphi(x,t,\dots) - \varphi_0=0,
\label{eqPDEManifold}
\end{eqnarray}
in which the \textit{singular manifold variable}
              \index{singular manifold!variable}
$\varphi$ is an arbitrary function of the independent variables 
and $\varphi_0$ an arbitrary movable constant.
Even in the ODE case,
the movable singularity can be defined as $\varphi(x) - \varphi_0=0$,
since the implicit functions theorem allows this to be locally inverted to
$x-x_0=0$;
the arbitrary function $\varphi$ thus introduced may then be used to construct
exact solutions which would be impossible to find with the restriction
$\varphi(x)=x$ \cite{Weiss1984HHa,NTZ}.

One must then define from $\varphi - \varphi_0$ 
an \textit{expansion variable} $\chi$ for the Laurent series,
    \index{expansion variable}
for there is no reason to confuse the roles of
the singular manifold variable and the expansion variable.
Two requirements must be respected:
firstly, 
$\chi$ must vanish as $\varphi - \varphi_0$ when $\varphi \to \varphi_0$;
secondly,
the structure of singularities in the $\varphi$ complex plane must be in 
a one-to-one correspondence with that in the $\chi$ complex plane,
so $\chi$ must be a homographic transform of $\varphi - \varphi_0$
(with coefficients depending on the derivatives of $\varphi$).

The Laurent series for $u$ and $E$ involved in the Kowalevski-Gambier part
of the test are defined as
\begin{eqnarray}
& &
u = \sum^{ +\infty}_{j=0} u_j \chi^{j+p},\ -p \in {\mathcal N},\
E = \sum^{ +\infty}_{j=0} E_j \chi^{j+q},\ -q \in {\mathcal N}^*
\label{eqLaurentSeries}
\end{eqnarray}
with coefficients $u_j,E_j$ independent of $\chi$ and only depending on the
derivatives of $\varphi$.

To illustrate our point, let us take as an example the \KdV\ equation
\index{Korteweg-de Vries equation}
\begin{eqnarray}
E
& \equiv & b u_t + u_{xxx} - (6/a) u u_x=0
\label{eqKdVcons}
\end{eqnarray}
(this is one of the very rare locations where this equation can be taken as
an example;
indeed, usually, things work so nicely for KdV that it is hazardous to draw 
general conclusions from its single study).

The choice $\chi=\varphi - \varphi_0$ originally made by Weiss \etal\
\cite{WTC}
makes the coefficients $u_j,E_j$ invariant under the two-parameter group
of translations $\varphi \mapsto \varphi + b'$,
with $b'$ an arbitrary complex constant
and therefore they only depend on the differential invariant
$\grad \varphi$
of this group and its derivatives~:
\begin{equation}
\label{eqKdVLaura}
 u=2 a \varphi_x^2 \chi^{-2}
-2 a \varphi_{xx} \chi^{-1}
+ a b          {\varphi_t \over 6 \varphi_x}
+ {2 a \over 3} {\varphi_{xxx} \over \varphi_x}
- {a \over 2} \left[{\varphi_{xx} \over \varphi_x}\right]^2 + O(\chi),\
\chi=\varphi - \varphi_0.
\end{equation}

There exists a choice of $\chi$ for which the coefficients exhibit the highest
invariance and therefore are the shortest possible
(all details are in Section 6.4 of Ref.~\cite{Cargese96Conte}), 
this best choice is \cite{Conte1989} 
\begin{equation}
 \chi
={\varphi-\varphi_0 \over \varphi_x 
- \displaystyle{{\varphi_{xx} \over 2 \varphi_x}}
(\varphi-\varphi_0)}
=\left[{\varphi_x \over \varphi-\varphi_0}
- {\varphi_{xx} \over 2 \varphi_x}\right]^{-1},\
\varphi_x \not=0,
\label{eqchi}
\end{equation}
in which $x$ denotes one of the independent variables whose component of
$\grad \varphi$ does not vanish.
The expansion coefficients $u_j,E_j$ are then invariant under the
six-parameter group of homographic transformations
\index{homographic!group}
\begin{eqnarray}
& &
\varphi \mapsto {a' \varphi + b' \over c' \varphi + d'},\
a'd'-b'c' \not=0,
\end{eqnarray}
in which $a',b',c',d'$ are arbitrary complex constants.
Accordingly,
these coefficients only depend on the following elementary differential
invariants and their derivatives:
\index{homographic!invariants}
the \textit{Schwarzian}
     \index{Schwarzian}
\begin{eqnarray}
S=\lbrace \varphi;x \rbrace
={\varphi_{xxx} \over \varphi_x} - {3 \over 2} \left({\varphi_{xx} 
\over \varphi_x} \right)^2,
\label{eqS}
\end{eqnarray}
and one other invariant per independent variable $t,y,\dots$
\begin{eqnarray}
C=- \varphi_t / \varphi_x,\
K=- \varphi_y / \varphi_x,\ \dots
\label{eqC}
\end{eqnarray}
The reason for the minus sign in the definition of $C$ is that,
under the travelling wave reduction $\xi=x-ct$,
the variable $C$ becomes the constant $c$.
These two invariants are linked by the cross-derivative condition
\begin{eqnarray}
X & \equiv &
((\varphi_{xxx})_t - (\varphi_t)_{xxx})/ \varphi_x
= S_t + C_{xxx} + 2 C_x S + C S_x = 0,
\label{eqCrossXT}
\end{eqnarray}
identically satisfied in terms of $\varphi$.

For our KdV example, the final Laurent series,
as compared with the initial one (\ref{eqKdVLaura}),
is remarkably simple~:
\begin{equation}
 u=2 a \chi^{-2} - a b {C \over 6} + {2 a S \over 3} - 2 a (b C - S)_x \chi
+ O(\chi^2),\
\chi=(\ref{eqchi}).
\label{eqKdVLaurent}
\end{equation}

For the practical computation of $(u_j,E_j)$ as functions of $(S,C)$ only,
\ie\ what is called the invariant Painlev\'e analysis,
the above explicit expressions of $(S,C,\chi)$ in terms of $\varphi$
are \textit{not} required,
the variable $\varphi$ completely disappears,
and the only necessary information is the gradient
of the expansion variable $\chi$ defined by Eq.~(\ref{eqchi}).
This gradient is a polynomial of degree two in $\chi$
(this is a property of homographic transformations),
whose coefficients only depend on $S,C$:
\begin{eqnarray}
\chi_x 
& = &
 1 + {S \over 2} \chi^2,
%& & (\chi^{-1})_x= - \chi^{-2} - {S \over 2},
\label{eqChix}
\\
\chi_t 
& = &
 - C + C_x \chi  - {1 \over 2} (C S + C_{xx}) \chi^2.
%& & (\chi^{-1})_t=  C \chi^{-2} - C_x \chi^{-1} + {C S + C_{xx} \over 2}.
\label{eqChit}
\end{eqnarray}

The above choice (\ref{eqchi}) of $\chi$ which generates homographically
invariant coefficients is the simplest one, but it is only particular.
The general solution to the above two requirements
which also generates homographically invariant coefficients is 
defined by an affine transformation on the inverse of $\chi$
\cite{MC1991}
\begin{eqnarray}
& &
Y^{-1}=B(\chi^{-1} + A),\
B \not=0.
\label{eqMostGeneralY}
\end{eqnarray}
Since a homography conserves the Riccati nature of an ODE,
the variable $Y$ satisfies a Riccati system,
easily deduced from the canonical one (\ref{eqChix})--(\ref{eqChit}) 
satisfied by $\chi$,
see (\ref{eqLaxMatrix2xG})--(\ref{eqLaxMatrix2tG}).

A frequent worry is~:
is there any restriction (or advantage, or inconvenient) 
to perform the test with $\chi$ or $Y$ rather than with $\varphi-\varphi_0$?
The precise answer is : the three Laurent series are equivalent
(their set of coefficients are in a one-to-one correspondence,
only their radii of convergence are different).
As a consequence, the Painlev\'e test,
which involves the \textit{infinite} series,
is insensitive to the choice,
and the costless choice (the one which minimizes the computations)
is undoubtedly $\chi$ defined by its gradient 
(\ref{eqChix})--(\ref{eqChit})
(to perform the test, one can even set, following Kruskal \cite{JKM},
 $S=0,C_x=0$).
If the same question were asked not about the test but about 
the second stage of Painlev\'e analysis as formulated at the end of
Section \ref{sectionImportance},
the answer would be quite different, 
and it is given in Section \ref{sectionWhere}.

Finally, let us mention a useful technical simplification.
{}From its definition (\ref{eqchi}),
the variable $\chi^{-1}$ is a logarithmic derivative,
so the system (\ref{eqChix})--(\ref{eqChit}) can be integrated once
\begin{eqnarray}
\Psi & = & (\varphi - \varphi_0) \varphi_x^{-1/2},
\label{eqpsiphi}
\\
(\Log \Psi)_x 
& = &
\chi^{-1},
\label{eqPsiX}
\\
(\Log \Psi)_t 
& = &
-C \chi^{-1} + {1 \over 2} C_x.
\label{eqPsiT} % Do not change to eqPsit
\end{eqnarray}
This feature helps to process PDEs which
can be defined in either conservative or potential form
when the conservative field $u$ has a simple pole,
such as the Burgers equation
\begin{eqnarray}
& &
E(u) \equiv b u_t +  (u^2 / a + u_x)_x = 0,
\\
& &
F(v) \equiv b v_t + v_x^2 / a + v_{xx} + G(t) = 0,\ u=v_x,\ E=F_x.
\label{eqBurgersPot}
\end{eqnarray}
Despite its (unique) logarithmic term,
   the $\psi$-series for $v$
\index{$\psi$-series}
\begin{eqnarray}
v & = &
a \Log \Psi + v_0 + (2 v_{0,x} - a b C) \chi 
\nonumber
\\
& &
+(F(v_0) - aS/2 + a b C_x/2) \chi^2 + O(\chi^3),
\label{eqBurgersPotLaurent}
\end{eqnarray}
in which $v_0$ is arbitrary,
is ``shorter'' than the Laurent series for $u$
\begin{eqnarray}
u & = &
a \chi^{-1} + (a b / 2) C + u_2 \chi
\nonumber
\\
& &
+\left[(a/4) (b^2 (C_t + C C_x) + 2 b C_{xx} - S_x - u_{2,x})\right] \chi^2
+ O(\chi^3),
\label{eqBurgersConsLaurent}
\end{eqnarray}
in which $u_2$ is arbitrary,
and the resulting series for $F(v)$, 
which is not a $\psi$-series but a Laurent series,
is \textit{much} shorter than the Laurent series for $E(u)$.
See Section \ref{sectionKS} for an application.

% ==========================================================================
\subsection{The WTC part of the Painlev\'e test for PDEs}
\label{sectionPDEPtestSteps}
\indent

As mentioned at the beginning of Section \ref{sectionPDEPtest},
we do not give here all the detailed steps of the test
nor all the necessary conditions which it generates
(this is done in Section 6.6 of Ref.~\cite{Cargese96Conte}).
We mainly state the notation to be extensively used throughout next sections.

The WTC part \cite{WTC} of the full test,
when rephrased in the equivalent invariant formalism \cite{Conte1992a},
consists in checking the existence of all Laurent series 
(\ref{eqLaurentSeries})
able to represent the general solution,
maybe after suitable perturbations \cite{CFP1993,MC1995} not describe here.

The gradient of the expansion variable $\chi$ is given by
(\ref{eqChix})--(\ref{eqChit}),
with the cross-derivative condition (\ref{eqCrossXT}).
This condition may be used to eliminate, depending on the PDE,
either derivatives $S_{mx,nt}$, with $n\ge 1$,
or     derivatives $C_{mx,nt}$, with $m\ge 3$,
and all equations later written are already simplified in either way.

The \textit{first step} is to find all the admissible values $(p,u_0)$
which define the leading term of the series for $u$.
Such an admissible couple is called a \textit{family of movable singularities}
(the term \textit{branch} should be avoided for the confusion which it
induces with branching, \ie\ multivaluedness).
\index{family of movable singularities}
\index{branch}

The recurrence relation for the next coefficients $u_j$,
after replacement of $(p,u_0)$,
\begin{equation}
\forall j\ge 1:\
E_j \equiv P(j) u_j + Q_j(\{u_l\ \vert \ l<j \}) = 0
\label{eqMethodPole5}
\end{equation}
depends linearly on $u_j$
and nonlinearly on the previously computed coefficients $u_l$.

The \textit{second step} is to compute the \textit{indicial equation}
                                            \index{indicial equation}
\begin{equation}
P(i)=0
\label{eqIndicialEquation}
\end{equation}
(a determinant in the multidimensional case of a system of PDEs).
Its roots are called the \textit{Fuchs indices}
                          \index{Fuchs indices}
of the family because they are indeed the characteristic indices of a linear
differential equation near a Fuchsian singularity
(the name \textit{resonances} sometimes given to these indices
refers to no resonance phenomenon and should also be avoided).
\index{resonances}
One then requires that all indices be integral
and obey a rank condition which, for a single PDE,
reduces to the condition that all indices be distinct.
The value $i=-1$ is always a Fuchs index.

The \textit{third and last step} is to require that,
for any admissible family and any Fuchs index $i$ (a signed integer),
the \textit{no-logarithm condition}
     \index{no-logarithm condition}
\begin{eqnarray}
& &
\forall i \in {\mathcal Z},\
P(i)=0 :\
Q_i=0
\label{eqConditionQ}
\end{eqnarray}
holds true,
so that the coefficient $u_i$ is an arbitrary function
of the independent variables.
In the multidimensional case, this is the condition of orthogonality
between the vector $\bfQ_i$ and the adjoint of the linear operator $\bfP(i)$.
Whenever there exist negative integers in addition to the ever present
value $-1$ counted with multiplicity one,
the condition (\ref{eqConditionQ}) can only be tested by a perturbation
\cite{CFP1993}.

This ends this subset of the test which, let us insist on the terminology,
is only aimed at building \textit{necessary} conditions for the PP.

The Laurent series for $u$ built in this way depends on at most $N$ arbitrary
functions (if $N$ denotes the differential order),
namely the coefficients $u_i$ introduced at the $N$ Fuchs indices,
including $\varphi$ for the index $-1$.

Any item $u_j,E_j,Q_j$ depends, through the elementary invariants $(S,C)$,
on the derivatives of $\varphi$ up to the order $j+1$, 
so the dependences are as follows:
$u_0=f(C)$,
$u_1=f(C,C_x,C_t)$,
$u_2=f(C,C_x,C_t,C_{xx},C_{xt},C_{tt},S)$,
\dots

Let us take an example.

\begin{example}
The Kolmogorov-Petrovskii-Piskunov (KPP) equation 
\cite{KPP,NewellWhitehead}
\index{KPP equation}
\begin{eqnarray}
& & E(u) \equiv 
b u_t - u_{xx} + 2 d^{-2} (u-e_1)(u-e_2)(u-e_3)=0,\ 
e_j \hbox{ distinct},
\label{eqKPP}
\end{eqnarray}
encountered in reaction-diffusion systems
(an additional convection term $u u_x$ is quite important in physical
applications to prey-predator models \cite{Satsuma1987})
possesses the two families
($d$ denotes any square root of $d^2$)
\begin{eqnarray}
& & 
p=-1,\
u_0=d,\
% {\mathcal D}=d\partial_x 
\label{eqKPPFamilies}
\end{eqnarray}
each family has the same two indices $(-1,4)$,
and the Laurent series for each family reads
\begin{eqnarray}
& & 
{\hskip -7.0 truemm}
u=d \chi^{-1} + \left(s_1/3 - (b d / 6) C\right)
\\
& & 
{\hskip -7.0 truemm}
\phantom{u=}
-(d/6)\left((b^2 / 6) C^2 - 6 a_2 - S - b C_x\right) \chi
+ O(\chi^2),
\label{eqLaurentKPP}
\end{eqnarray}
with the notation
\begin{eqnarray}
{\hskip -5.0 truemm}
& & 
s_1=e_1 + e_2 + e_3,\ 
a_2 % =2(s_1^2 - 3 s_2) 
=\left( (e_2-e_3)^2 + (e_3-e_1)^2 + (e_1-e_2)^2\right) /(18 d^2).
\label{eqKPPNotation}
\end{eqnarray}
At index $i=4$, the two no-log conditions,
one for each sign of $d$ \cite{Conte1988},
\begin{eqnarray}
Q_4 & \equiv &
 C
 [(b d C + s_1 - 3 e_1)(b d C + s_1 - 3 e_2) (b d C + s_1 - 3 e_3)
\nonumber
\\
& &
\phantom{xxx} - 3 b^2 d^3 (C_t + C C_x)]
=0
\label{eqKPPQ4}
\end{eqnarray}
are not identically satisfied, so the PDE fails the test.
\end{example}

It is time to define a quantity which,
although useless for the test itself,
is of first importance at the second stage of Painlev\'e analysis,
which will be developed in Sections \ref{sectionTruncationPrinciples}
to \ref{sectionTruncationTwoFamilies}.
This quantity is defined from the finite subset of 
nonpositive powers of the Laurent series for $u$.

\begin{definition}
Given a family $(p,u_0)$,
the \textit{singular part operator} ${\mathcal D}$ is defined as
     \index{singular part operator}
\begin{equation}
\Log \varphi \mapsto {\mathcal D} \Log \varphi=u_T(0)-u_T(\infty),
\end{equation}
in which the notation $u_T(\varphi_0)$, 
which emphasizes the dependence on the movable constant $\varphi_0$,
stands for the principal part ($T$ like truncation) of the Laurent series
(\ref{eqLaurentSeries}),
\ie\ the finite subset of its nonpositive powers
\begin{eqnarray}
& &
u_T(\varphi_0) = \sum^{-p}_{j=0} u_j \chi^{j+p}.
\label{eqLaurentSeriesTruncated}
\end{eqnarray}
\end{definition}
For most PDEs, this operator is linear.
For the Laurent series already considered
(\ref{eqKdVLaurent}),
(\ref{eqBurgersPotLaurent}),
(\ref{eqBurgersConsLaurent}),
(\ref{eqLaurentKPP}),
the operator is, respectively,
${\mathcal D}=-2 a \partial_x^2, a, a \partial_x, d \partial_x$.
For the Kuramoto-Sivashinsky equation (\ref{eqKS}),
\index{Kuramoto-Sivashinsky equation}
there exists a unique Laurent series (\ref{eqLaurentSeries}) with $p=-3$
(given by (\ref{eqKSLaurent}) for a particular value of $\chi$,
and by the derivative of (\ref{eqKS4}) for any $\chi$),
with a singular part operator equal to
\begin{eqnarray}
& & 
{\mathcal D}=60 \nu \partial_x^3 + (60/19) \mu \partial_x.
\end{eqnarray}
This is precisely the third order linear operator on the \rhs\ of 
(\ref{eqKSSolitaryWithD}).

% ==========================================================================
\subsection{The various ways to pass or fail the Painlev\'e test for PDEs}
\label{sectionPDEPtestFailure}
\indent

If one processes a multidimensional PDE the coefficients of which 
depend on some parameters $\mu$,
\begin{eqnarray}
& & \bfE(\bfu,\bfx;\mu)=0,
\label{eqPDEbf}
\end{eqnarray}
(boldface means multicomponent),
the Painlev\'e test generates the following output :

\begin{enumerate}
\item
leading order $(p,u_0)$,
Fuchs indices $i$
and singular part operator ${\mathcal D}$ for each admissible family,

\item
diophantine conditions
that all singularity orders $p$ and all Fuchs indices $i$ be integral,
conditions whose solution creates constraints of the type
\begin{eqnarray}
& & F(\mu,C)=0,
\end{eqnarray}
\index{diophantine conditions}

\item
no-log conditions
\begin{eqnarray}
& & 
\forall i\
\forall n\
\forall u_j\
 :\
\bfQ_i^{n}(\mu,S,C,u_j)=0,
\end{eqnarray}
arising from any integral Fuchs index $i$,
in which $n$ is the Fuchsian perturbation order \cite{CFP1993} if necessary,
$u_j$ are the arbitrary coefficients $u_{\rm arb}$
introduced at earlier Fuchs indices $j$.

\end{enumerate}

In particular, the Laurent series (\ref{eqLaurentSeries})
are of no use and should not be computed beyond the highest Fuchs integer.
All this output (items 1 and 3)
is easily produced with a computer algebra program
and, in all further examples,
we will simply list these results without any more detail.

Strictly speaking, the answer provided by the test 
to the question ``Has the PDE the PP?''
is either \textit{no} (at least one of the necessary conditions fails)
or \textit{maybe} 
(all necessary conditions are satisfied, and the PDE may possess
the PP but this still has to be proven).
It is never \textit{yes},
as shown by the famous counterexample of Painlev\'e
(the second order ODE with the general solution 
$\wp(\lambda \Log (c_1 x + c_2),g_2,g_3)$,
which therefore has the PP iff $2 \pi i \lambda$ is a period of the
Weierstrass elliptic function $\wp$,
a transcendental condition impossible to generate by a finite
algebraic procedure).
\index{elliptic function}

Now that the necessary part (\ie\ the Painlev\'e test) of Painlev\'e analysis
is finished, let us turn to the question of sufficiency.

To reach our goal which is to obtain as many analytic results as possible,
we do not adopt such a drastic point of view,
but the opposite one.
Instead of the logical \textit{and} performed by the mathematician on all the
necessary conditions generated by the test,
we perform a logical \textit{or} operation on these conditions.
Therefore 
the above Painlev\'e test must be performed to its end, \ie without stopping
even in case of failure of some condition,
so as to collect all the necessary conditions.
Turning to sufficiency, 
these conditions have to be examined independently in the hope of finding
some global element of integrability.
An application of this point of view to the Lorenz model,
a third order ODE,
can be found in Section 6.7 of Ref.~\cite{Cargese96Conte}.

If the PDE under study possesses a singlevalued exact solution,
\index{solution!exact}
there must exist a Laurent series 
(\ref{eqLaurentSeries})
which represents it locally.
Therefore the practical criterium to be implemented deals with the existence
of \textit{particular} Laurent series, and the result of the test belongs to
one of the following mutually exclusive situations.

\begin{enumerate}
\item
(The best situation)
Success of the test, at least for some values of $\mu$ selected by the test.
The PDE may have the PP,
and one must look for its BT;

\item
There exists at least one value of $(\mu, \varphi, u_{\rm arb})$
which ensures the existence of a particular Laurent series.
For these values,
an exact solution may exist;

\item
There exists at least one value of $(\mu, \varphi, u_{\rm arb})$
enforcing some of, but not all,
the no-log conditions of at least one particular Laurent series.
Quite probably no exact solution exists,
but there may exist a conservation law
(a first integral for an ODE);

\item
(The worst situation)
There is no value of $(\mu, \varphi, u_{\rm arb})$
enforcing at least one of the no-log conditions of the various series.
Quite probably the PDE is chaotic and possesses no exact solution at all.

\end{enumerate}

Examples of these various situations are, respectively~:

\begin{enumerate}
\item
All the PDEs which have the PP
(sine-Gordon, Korteweg-de Vries, \dots),
but also the counterexample of Painlev\'e quoted above;
\index{sine-Gordon equation}
\index{Korteweg-de Vries equation}

\item
The equation of Kuramoto and Sivashinsky (\ref{eqKS}),
with the particular Laurent series (\ref{eqKSLaurent});
\index{Kuramoto-Sivashinsky equation}

\item
The Lorenz model for $b=2 \sigma$,
for which the no-log condition at $i=4$ is violated
and there exists a first integral;
\index{Lorenz model}

\item
The R\"ossler dynamical system
for which the unique family has the never satisfied condition 
$Q_2 \equiv 16 =0$.
\index{R\"ossler system}

\end{enumerate}

%\vfill \eject

% ==========================================================================
\section{Ingredients of the ``singular manifold method''}
\label{sectionTruncationPrinciples}
\indent

The methods to handle the integrable and nonintegrable situations are the 
same,
simply a more or less important result is obtained.

The goal is to find a (possibly degenerate) couple 
(Darboux transformation, Lax pair)
\index{Darboux transformation}
\index{Lax pair}
in order to deduce the B\"acklund transformation or,
\index{B\"acklund transformation}
if a BT does not exist,
to generate some exact solutions.

The full Laurent series is of no help,
for this is not an exact solution according to the definition in Section
\ref{sectionVariousLevels}.
Since this is the only available piece of information
and since a finite (closed form) expression is required to represent an
exact solution,
let us represent,
and this is the idea of Weiss, Tabor and Carnevale \cite{WTC},
an unknown exact solution $u$
as the sum of a singular part,
built from the finite principal part of the Laurent series
(i.e.~the finite number of terms with negative powers),
and of a regular part made of one term denoted $U$.
This assumption is identical to that of a Darboux transformation
(\ref{eqDTGeneral}),
\index{Darboux transformation}
in which nothing would be specified on $U$.

This method is widely known as the 
\textit{singular manifold method} or \textit{truncation method}
because it selects
the beginning of the Laurent series and discards (``truncates'')
the remaining infinite part.
\index{truncation!method}
\index{singular manifold!method}

Since its introduction by WTC \cite{WTC},
it has been improved in many directions
\cite
{MC1991,EstevezEtAl1993,Garagash1993,MC1994,CMGalli95,Pickering1996,MC1998},
and we present below the current status of the method.

% ==========================================================================
\subsection{The ODE situation}
\label{sectionODESituation}
\indent

For the six ordinary differential equations (ODE) (P1)--(P6) 
which bear his name,
Painlev\'e proved the PP by showing \cite{PaiActa,PaiCRAS1906}
the existence of one (case of (P1)) or two ((P2)--(P6)) 
function(s) $\tau=\tau_1, \tau_2$,
called tau-functions,
linked to the general solution $u$ by logarithmic derivatives
\index{Painlev\'e!equations}
\index{tau-function}
\begin{eqnarray}
\hbox{(P1)}:\
& u= & {\mathcal D}_1 \Log \tau 
\label{eqlogPone}
\\
(\hbox{P}n),\ n=2,\dots,6 :\
& u= & {\mathcal D}_n (\Log \tau_1 - \Log \tau_2)
\label{eqlogPtwo}
\end{eqnarray}
where the operators ${\mathcal D}_n$ are linear:
\begin{eqnarray}
& &
{\mathcal D}_1= - \partial_x^2,\
{\mathcal D}_2={\mathcal D}_4=\pm \partial_x,\
{\mathcal D}_3=\pm e^{-x} \partial_x,\
\\
& &
{\mathcal D}_5=\pm x e^{-x} (2 \alpha)^{-1/2} \partial_x,\
{\mathcal D}_6=\pm x (x-1) e^{-x} (2 \alpha)^{-1/2} \partial_x.
\end{eqnarray}
These functions $\tau_1,\tau_2$ satisfy third order nonlinear ODEs
and they have the same kind of singularities than
solutions of \textit{linear} ODEs, namely
they have no movable singularities at all;
they are entire functions for (P1)--(P5), 
and their only singularities for (P6) are the three fixed critical points
$(\infty, 0, 1)$.

ODEs cannot possess an auto-BT,
since the number of independent arbitrary coefficients in a solution
cannot exceed the order of the ODE.
They can however possess a Schlesinger transformation
(see definition Section \ref{sectionCJP}).
\index{Schlesinger transformation}

% ==========================================================================
\subsection{Transposition of the ODE situation to PDEs}
\label{sectionTransposition}
\indent

For PDEs, similar ideas prevail.
The analogue of (\ref{eqlogPone})--(\ref{eqlogPtwo}),
with an additional \rhs\ $U$,
is now the Darboux transformation (\ref{eqDTGeneral}),
\index{Darboux transformation}
and the scalar(s) $\psi$ to which the scalar(s) $\tau$ are linked by
(\ref{eqLinkTauPsi})
are assumed to satisfy a linear system, the Lax pair.

Another interesting observation must be made.
There seems to exist two and only two classes of integrable $1+1$-dimensional
PDEs,
at least at the level of the base member of a hierarchy~:
\index{hierarchy}
those which have only one family of movable singularities,
and those which have only pairs of families with opposite principal parts,
similarly to the distinction between (P1) on one side
and (P2)--(P6) on the other side.
Among the $1+1$-dimensional integrable equations,
those with one family include 
KdV, 
the AKNS, Hirota-Satsuma and Boussinesq equations;
they also include
the Sawada-Kotera, 
Kaup-Kupershmidt and Tzitz\'eica equations
because only one of their two families is relevant
\cite{MC1998,CMG1999}.
Equations with pairs of opposite families
include sine-Gordon, mKdV and Broer-Kaup (two families each),
NLS (four families).
\index{sine-Gordon equation}
\index{Boussinesq equation}
\index{Hirota-Satsuma equation}
\index{Sawada-Kotera equation}
\index{Kaup-Kupershmidt equation}
\index{Tzitz\'eica equation}
\index{Broer-Kaup equation}

%\vfill \eject

% ==========================================================================
\subsection{The singular manifold method as a Darboux transformation}
\label{sectionTruncationDarboux}
\indent

As qualitatively described in Section \ref{sectionTruncationPrinciples},
the singular manifold method looks very much like a resummation of the Laurent
series,
just like the geometric series
\index{truncation!method}
\index{singular manifold!method}
\index{resummation}
\begin{eqnarray}
& &
\sum_{j=0}^{+ \infty} x^j,\ x \to 0,
\end{eqnarray}
becomes a finite sum in the resummation variable $X=x/(1-x)$
\begin{eqnarray}
& &
\sum_{J=0}^{1} X^J,\ X \to 0.
\end{eqnarray}

The principle of the method is the following \cite{WTC}.
One first notices that the (infinite) Laurent series (\ref{eqLaurentSeries}) 
in the variable $\varphi - \varphi_0$ 
can be rewritten as the sum of two terms
\begin{eqnarray}
& & 
u={\mathcal D} \Log \tau + \hbox{ regular part}.
\label{eqLaurentDtau}
\end{eqnarray}
The first term 
${\mathcal D} \Log \tau$,
built from the singular part operator defined 
in Section \ref{sectionPDEPtestSteps},
is a finite Laurent series
and,
if $\tau$ is any variable fulfilling the two requirements
for an expansion variable enunciated
in Section \ref{sectionSMVandEV},
it captures all the singularities of $u$ when $\varphi \to \varphi_0$.
The second term,
temporarily called ``regular part'' for this reason,
is yet unspecified.
The sum of these two terms is therefore a finite Laurent series
(hence the name \textit{truncated series}),
and the variable $\tau$ is a \textit{resummation variable}
which has made the former infinite series in $\varphi-\varphi_0$ a finite one.
\index{resummation}
One then tries to identify 
this resummation (\ref{eqLaurentDtau}) with the definition
of a Darboux transformation (\ref{eqDTGeneral}).
This involves two features.
The first feature is to uncover a link (\ref{eqLinkTauPsi})
between $\tau$ and a scalar component $\psi$ of a Lax pair.
The second feature is to prove that the left over
``regular part'' is indeed a second solution to the PDE under study.

% ==========================================================================
\subsection{The degenerate case of linearizable equations}
\label{sectionLinearizable}
\indent

The Burgers equation (\ref{eqBurgersPot}),
under the transformation of Forsyth (Ref.~\cite{Forsyth1906} p.~106),
\begin{eqnarray}
& & 
v=a \Log \tau,\
\tau=\psi,\
\end{eqnarray}
is linearized into the heat equation
\begin{eqnarray}
& & 
b \psi_t + \psi_{xx} + G(t) \psi=0.
\end{eqnarray}
This can be considered as a degenerate Darboux transformation 
(\ref{eqDTGeneral}),
in which $U$ is identically zero
and $\psi$ satisfies a single linear equation,
not a pair of linear equations,
so this fits the general scheme.

Another classical example is the second order particular Monge-Amp\`ere
equation $s+pq=0$,
linearized into the d'Alembert equation $s=0$ :
\index{d'Alembert equation}
\begin{eqnarray}
& & 
s+pq \equiv u_{xt} + u_x u_t=0,
\\
& & 
u=\Log \tau,\
\tau=\psi,\
\psi_{xt}=0.
\end{eqnarray}

%\vfill \eject

% ==========================================================================
\subsection{Choices of  Lax pairs and equivalent Riccati pseudopotentials}
\label{sectionAFewLaxPairs}
\indent

To fix the ideas, 
we list here a few usual second order and third order 
Lax pairs depending on undetermined coefficients,
together with the constraints imposed on these coefficients by the
commutativity condition.

It is sometimes appropriate to represent an $n$-th order Lax pair
by the $2(n-1)$ equations satisfied by
an equivalent $(n-1)$-component pseudopotential $\bfY$ of Riccati type,
the first component of which is chosen as
\begin{eqnarray}
Y_1=\psi_x / \psi,
\label{eqY1}
\end{eqnarray}
in which $\psi$ is a scalar component of the Lax pair.

% ==========================================================================
\subsubsection{Second-order Lax pairs and their privilege}
\indent

The general second-order scalar Lax pair reads,
in the case of two independent variables $(x,t)$,
\begin{eqnarray}
L_1 \psi & \equiv & 
\psi_{xx} -d \psi_{x} - a \psi=0,
\label{eqLaxScalar2xG}
\\
L_2 \psi & \equiv & 
\psi_{t} - b \psi_{x} - c \psi=0,
\label{eqLaxScalar2tG}
\\
\lbrack L_1,L_2 \rbrack & \equiv & X_0 + X_1 \partial_x,
\\
\\
X_0 & \equiv & - a_t + a_x b + 2 a b_x + c_{xx} - c_x d =0,
\\
X_1 & \equiv & - d_t + (b_x + 2 c - b d)_x =0.
\label{eqOrder2GDX1}
\end{eqnarray}
For the inverse scattering method to apply,
the coefficients $(d,a)$ of the $x$-part (\ref{eqLaxScalar2xG})
are required to depend linearly on the field $U$ of the PDE.

The Lax pair (\ref{eqLaxScalar2xG})--(\ref{eqLaxScalar2tG})
is identical to a 
linearized version of the Riccati system 
satisfied by the most general expansion variable $Y$ 
defined by (\ref{eqMostGeneralY}),
under the correspondence
\begin{eqnarray}
{\hskip -3.0 truemm}
& & 
Y= B^{-1} {\psi \over \psi_x},\ B\not=0,
\label{eqYpsi}
\\
{\hskip -3.0 truemm}
& & 
d=2 A,\
a= A_x - A^2 -S/2,\
b=-C,\
c=C_x/2 + A C + \int A_t \D x,
\end{eqnarray}
and the commutator of the Lax pair is (\ref{eqCrossXT}).

In particular, 
when the coefficient $d$ is zero or when, 
by a linear change 
$\psi \mapsto e^{\displaystyle{\int d \D x/2}} \psi$,
it can be set to zero without altering the linearity of $a$ on $U$,
the correspondence is \cite{MC1991}
\begin{eqnarray}
   & & \chi= {\psi \over \psi_x},\ B=1,\ A=0,
\label{eqchipsi}
\\ & & 
d=0,\
a= -S/2,\
b=-C,\
c=C_x/2,
\\ L_1 \psi & \equiv & \psi_{xx} + {S \over 2} \psi=0,
\label{eqLaxScalar2x}
\label{eqPsixx}
\\ L_2 \psi & \equiv & \psi_{t} + C \psi_{x} - {C_x \over 2} \psi=0,
\label{eqLaxScalar2t}
\label{eqPsit}
\\
2 \lbrack L_1,L_2 \rbrack & \equiv & X = S_t + C_{xxx} + C S_x + 2 C_x S=0.
\label{eqCrossSC}
\end{eqnarray}

Therefore second order Lax pairs are privileged in the singularity approach,
in the sense that their coefficients can be identified with the elementary
homographic invariants $S,C$ of the invariant Painlev\'e analysis and, 
if appropriate, $A,B$.
Conversely,
and this has historically been the reason of some errors described
in Section \ref{sectionWTCSuitable},
at the stage of searching for the BT,
these homographic invariants $S,C$ are useless 
when the Lax order is higher than two
and they should not be considered.

As explained in Section \ref{sectionTruncationDarboux},
given a Lax pair,
one should define from it either one or two scalars $\psi_f$.
Consider the second order Lax pair defined by the gradient of $Y$.
Then, for one-family PDEs, this unique scalar $\psi$ is 
defined by (\ref{eqYpsi}). % Not by (\ref{eqchipsi}).
For two-family PDEs, the two scalars $\psi_f$ are defined by 
\begin{eqnarray}
   & & Y={\psi_1 \over \psi_2},\
\label{eqYpsi1psi2}
\end{eqnarray}
which leads to the zero-curvature representation of the Lax pair
\begin{eqnarray}
{\hskip -2.0truemm}
& &
(\partial_x - L) \pmatrix{\psi_1 \cr \psi_2 \cr}=0,\
L=
\pmatrix{-A-B^{-1}B_x/2 & B^{-1} \cr B(A_x - A^2 -S/2) & A+B^{-1}B_x/2 \cr},\
\label{eqLaxMatrix2xG}
\\
{\hskip -2.0truemm}
& &
(\partial_t - M) \pmatrix{\psi_1 \cr \psi_2 \cr}=0,\
\label{eqLaxMatrix2tG}
\\
{\hskip -2.0truemm}
& &
M=
\pmatrix{A C + C_x/2 - B^{-1} B_t/2 &
        -C B^{-1} \cr
        B((C S + C_{xx})/2 +A_t + C A^2 + C_x A) & 
        -A C - C_x/2 + B^{-1} B_t/2 \cr}.
\nonumber
\end{eqnarray}
The reason why the Riccati form is the most suitable characterization
of the Lax pair is that it allows two linearizations
\cite{MC1994,Pickering1996},
namely (\ref{eqYpsi}) and (\ref{eqYpsi1psi2}),
depending on whether the PDE has one family or two opposite families.

% ==========================================================================
\subsubsection{Third-order Lax pairs}
\indent

The general third-order scalar Lax pair is defined as
\begin{eqnarray}
{\hskip - 6.0 truemm}
L_1 \psi & \equiv &
\psi_{xxx} - f \psi_{xx} - a \psi_x - b \psi=0,
\label{eqLaxScalar3xG}
\\
{\hskip - 6.0 truemm}
L_2 \psi & \equiv &
\psi_t  - c\psi_{xx} - d \psi_x - e \psi=0,
\label{eqLaxScalar3tG}
\\
{\hskip - 6.0 truemm}
\lbrack L_1,L_2 \rbrack & \equiv &
X_0 + X_1 \partial_x + X_2 \partial_{x}^2,
\\
{\hskip - 6.0 truemm}
X_0  & \equiv &
 - b_t - a e_x + e_{xxx} + b_{xx} c 
 + 2 b c f_x + b c_x f - e_{xx} f
\nonumber
\\ 
{\hskip - 6.0 truemm}
& & + 3 b c_{xx} + 3 b_x c_x + 3 b d_x + b_x d = 0,
\label{eqX0}
\\ 
{\hskip - 6.0 truemm}
X_1  & \equiv &
 - a_t + 3 e_{xx} + 2 b_xc + a_{xx} c + d_{xxx} + 3 a c_{xx}+ 2 a d_x
\nonumber
\\ 
{\hskip - 6.0 truemm}
& & + 3 a_x c_x + 3 b c_x + a_x d 
+ 2 a c f_x + a c_x f - 2 e_x f - d_{xx} f
= 0,
\label{eqX1}
\\ 
{\hskip - 6.0 truemm}
X_2  & \equiv &
- f_t + (c f^2 + c f_x + 2 c_x f + d f + 2 a c + c_{xx} + 3 d_x + 3 e)_x=0.
\label{eqX2}
\end{eqnarray}

An equivalent two-component pseudopotential is the projective Riccati one
$\bfY=(Y_1,Y_2)$
\cite{MC1991,MCGallipoli1991}
(written below, for simplicity, in the case $f=0$)
\begin{eqnarray}
Y_1 & = & {\psi_{x}  \over \psi},\
Y_2={\psi_{xx} \over \psi},\
\\
Y_{1,x} & = & - Y_1^2 + Y_2,
\label{eqProjRiccatiY1x}
\\ 
Y_{2,x} & = & -Y_1 Y_2 + a Y_1 + b,
\label{eqProjRiccatiY2x}
\\
Y_{1,t} & = & - (d Y_1 + c Y_2) Y_1 + (a c+d_x)Y_1 + (c_x+d)Y_2 +e_x+b c
\label{eqProjRiccatiY1t}
\\ 
        & = & (c Y_2 + d Y_1 + e)_x,
\label{eqProjRiccatiY1tCons}
\\ 
Y_{2,t} & = & - (d Y_1 + c Y_2) Y_2 +(2 a c_x+a_x c+b c+d_{xx}+a d+2e_x)Y_1
\nonumber
\\ & &
 +(c_{xx}+2 d_x+a c) Y_2 +2 b c_x+b_x c+b d+e_{xx},
\label{eqProjRiccatiY2t}
\\
Y_{1,tx}-Y_{1,xt} & = &  X_1 + X_2 Y_1,
\\
Y_{2,tx}-Y_{2,xt} & = & -X_0 + X_2 Y_2.
\end{eqnarray}

When there is no reason to distinguish between $x$ and $t$,
for instance because the PDE is 
invariant under the permutation (Lorentz transformation)
\begin{eqnarray}
& &
{\mathcal P}:\ (\partial_x,\partial_t) \to (\partial_t,\partial_x),
\label{eqPermutation}
\end{eqnarray}
it is natural to consider the following third-order matrix Lax pair,
invariant under (\ref{eqPermutation}),
defined in the basis $(\psi_x, \psi_t, \psi)$ \cite{CMG1999},
\begin{eqnarray}
& &
(\partial_x - L) \pmatrix{\psi_x \cr \psi_t \cr \psi \cr}=0,\
L=
\pmatrix{f_1 & f_2 & f_3 \cr g_1 & g_2 & g_3 \cr 1 & 0 & 0 \cr},\
\label{eqLaxMatrix3Invarx}
\\
& &
(\partial_t - M) \pmatrix{\psi_x \cr \psi_t \cr \psi \cr}=0,\
M=
\pmatrix{g_1 & g_2 & g_3 \cr h_1 & h_2 & h_3 \cr 0 & 1 & 0 \cr}.
\label{eqLaxMatrix3Invart}
\end{eqnarray}
In the equivalent projective Riccati components
$(Y_1,Y_2)$
\begin{eqnarray}
Y_1 & = & {\psi_{x} \over \psi},\
Y_2   =   {\psi_{t} \over \psi},\
\end{eqnarray}
with the property $Y_{1,t}=Y_{2,x}$,
it is defined as
\begin{eqnarray}
Y_{1,x} & = & - Y_1^2   + f_1 Y_1 + f_2 Y_2 + f_3,
\label{eqProjRiccatiY1xc}
\\ 
Y_{2,x} & = & - Y_1 Y_2 + g_1 Y_1 + g_2 Y_2 + g_3,
\label{eqProjRiccatiY2xc}
\\
Y_{1,t} & = & - Y_1 Y_2 + g_1 Y_1 + g_2 Y_2 + g_3,
\label{eqProjRiccatiY1tc}
\\ 
Y_{2,t} & = & - Y_2^2   + h_1 Y_1 + h_2 Y_2 + h_3.
\label{eqProjRiccatiY2tc}
\end{eqnarray}
The nine functions $f_j,g_j,h_j, j=1,2,3,$ must satisfy six cross-derivative
conditions $X_j=0$
\begin{eqnarray}
(Y_{1,x})_t - (Y_{1,t})_x & = & X_0 + X_1 Y_1 + X_2 Y_2=0,
\label{eqProjRiccatiCross012}
\\ 
(Y_{2,x})_t - (Y_{2,t})_x & = & X_3 + X_4 Y_1 + X_5 Y_2=0,
\label{eqProjRiccatiCross345}
\end{eqnarray}
easy to write explicitly.
It is worth noticing that there exists no such invariant
second-order matrix Lax pair.

% ==========================================================================
\subsection{The admissible relations between $\tau$ and $\psi$}
\label{sectionGambier}
\indent

By elimination of $\partial_t$, 
one of the two PDEs defining the BT to be found can be made an ODE,
e.g.~(\ref{eqChix}) or (\ref{eqBTxOrder3}).
This nonlinear ODE,
with coefficients depending on $U$ and,
in the $1+1$-dimensional case, on an arbitrary constant $\lambda$,
has the property \cite{MC1998} of being linearizable.
This very strong property restricts the admissible choices 
(\ref{eqLinkTauPsi}) to a finite number of possibilities,
and full details can be found in Musette lecture \cite{CetraroMusette}.

%\vfill \eject

% ==========================================================================
\section{The algorithm of the singular manifold method}
\label{sectionAlgorithmTruncation}
\indent

We now have all the ingredients
to give a general exposition of the method
in the form of an algorithm.
The present exposition largely follows the lines of Ref.~\cite{MC1998}.

The various situations thus implemented are :
one-family and two-opposite-family PDEs,
second or higher order Lax pair,
various allowed links between the two sets of functions $(\tau,\psi)$.

Consider a PDE (\ref{eqPDE}) with only one family of movable singularities
or exactly two families of movable singularities with opposite values of
$u_0$,
and denote ${\mathcal D}$ the \textit{singular part operator} 
of either the unique family or anyone of the two opposite families.

\textit{First step}.
Assume a Darboux transformation defined as
\begin{eqnarray}
& &
u=U + {\mathcal D} (\Log \tau_1 - \Log \tau_2),\
E(u)=0,
\label{eqDT}
\end{eqnarray}
with $u$ a solution of the PDE under consideration,
$U$ an unspecified field which most of the time will be found to be
a second solution of the PDE,
$\tau_f$ the ``entire'' function
(or more precisely ratio of entire functions)
attached to each family $f$.
For one-family PDEs, one denotes $\tau_1=\tau, \tau_2=1$,
so the DT assumption (\ref{eqDT}) becomes
\begin{eqnarray}
& &
u=U + {\mathcal D} \Log \tau,\
E(u)=0.
\label{eqDTOneFamily}
\end{eqnarray}

A consequence of the assumption (\ref{eqDT}) is the existence of the 
involution
\begin{eqnarray}
& &
\forall f :\
(u,U,\tau_f) \mapsto (U,u,\tau_f^{-1}),
\end{eqnarray}
since the operator ${\mathcal D}$ is linear,
and, for two-family PDEs, of the involution
\begin{eqnarray}
& &
\forall (u,U):\ ({\mathcal D},\tau_1,\tau_2) \mapsto (-{\mathcal D},\tau_2,\tau_1).
\end{eqnarray}

\textit{Second step}.
Choose the order two, then three, then \dots,
for the unknown Lax pair,
and define one or two (as many as the number of families)
\textit{scalars} $\psi_f$ from the component(s) of its wave vector
(e.g.~the scalar wave vector if the PDE has one family
and the pair is defined in scalar form).
Such sample Lax pairs and scalars can be found in
Section \ref{sectionAFewLaxPairs}. 

\textit{Third step}.
Choose an explicit link $F$
\begin{eqnarray}
& &
\forall f:\
{\mathcal D} \Log \tau_f=F(\psi_f),
\label{eqTauPsiLink}
\end{eqnarray}
the same for each family $f$,
between the functions $\tau_f$
and the scalars $\psi_f$ defined from the Lax pair.
According to Section \ref{sectionGambier},
at each scattering order,
there exists only a finite number of choices (\ref{eqTauPsiLink}),
among them the most frequent one
\begin{eqnarray}
& &
\forall f:\
\tau_f=\psi_f.
\label{eqTauPsiLink1}
\end{eqnarray}

\textit{Fourth step}.
Define the ``truncation'' and solve it,
that is to say~:
with the assumptions 
(\ref{eqDT}) for a DT,
(\ref{eqTauPsiLink}) for a link between $\tau_f$ and $\psi_f$,
(\ref{eqLaxScalar2xG})--(\ref{eqLaxScalar2tG})
or
(\ref{eqLaxScalar3xG})--(\ref{eqLaxScalar3tG})
or other
for the Lax pair in $\psi$,
express $E(u)$ as a polynomial in the derivatives of $\psi_f$
which is irreducible \textit{modulo} the Lax pair.
For the above pairs and a one-family PDE,
this amounts to eliminate any derivative of $\psi$ of order in $(x,t)$ 
higher than or equal 
   to $(2,0)$ or $(0,1)$ (second order case)
or to $(3,0)$ or $(0,1)$ (third order),
thus resulting in a polynomial of one variable $\psi_x / \psi$ (second order)
or two variables $\psi_x / \psi, \psi_{xx} / \psi$ (third order)
\begin{eqnarray}
{\hskip -4.0 truemm}
& &
E(u)= \sum_{j=0}^{-q} E_j(S,C,U) (\psi / \psi_x)^{j+q}
\hbox{ (one-family PDE, second order)},
\label{eqTruncationOrder2}
\\
{\hskip -4.0 truemm}
& &
E(u)= \sum_{k \ge 0} \sum_{l \ge 0} E_{k,l}(a,b,c,d,e,U) 
(\psi_x / \psi)^k (\psi_{xx} / \psi)^l
\nonumber
\\
{\hskip -4.0 truemm}
& &
\hbox{         (one-family PDE, third order)}.
\label{eqTruncationOrder3}
\end{eqnarray}
Since one has no more information on this polynomial $E(u)$ except the fact
that it must vanish,
one requests that it identically vanishes,
by solving the set of \textit{determining equations}
\begin{eqnarray}
\forall j\
& &
E_{j}(S,C,U) =0
\hbox{ (one-family PDE, second order)}
\label{eqDetermining2}
\\
\forall k\
\forall l\
& &
E_{k,l}(a,b,c,d,e,U) =0
\hbox{ (one-family PDE, third order)}
\label{eqDetermining3}
\end{eqnarray}
for the unknown coefficients $(S,C)$ or $(a,b,c,d,e)$ as functions of $U$,
and one establishes the constraint(s) on $U$ by eliminating
$(S,C)$ or $(a,b,c,d,e)$.
The strategy of resolution is developed in Section \ref{sectionKS}.

The constraints on $U$ reflect the integrability level of the PDE.
If the only constraint on $U$ is to satisfy some PDE,
one is on the way to an auto-BT 
if the PDE for $U$ is the same as the PDE for $u$,
or to a remarkable correspondence (hetero-BT)
between the two PDEs.
\index{B\"acklund transformation!hetero--}

The second, third and fourth steps must be repeated until a success occurs.
The process is successful if and only if all the following conditions are met
\begin{enumerate}
\item
$U$ comes out with one constraint exactly,
namely~: to be a solution of some PDE,
\item
(if an auto-BT is desired) the PDE satisfied by $U$ is identical to 
(\ref{eqPDE}),
\item
the vanishing of the commutator $[L_1,L_2]$ is equivalent to the vanishing
of the PDE satisfied by $U$,
\item
in the $1+1$-dimensional case only
and if the PDE satisfied by $U$ is identical to (\ref{eqPDE}),
the coefficients depend on an arbitrary constant $\lambda$,
the spectral or B\"acklund parameter.
\index{spectral parameter}
\index{B\"acklund parameter}
\end{enumerate}

At this stage, one has obtained the DT and the Lax pair.

\textit{Fifth step}.
Obtain the two equations for the BT by eliminating $\psi_f$ \cite{Chen1974}
between the DT and the Lax pair.
This sometimes uneasy operation when the order $n$ of the Lax pair
is too high may become elementary by 
considering the equivalent Riccati representation of the Lax pair and
eliminating the appropriate components of $\bfY$ rather than $\psi$.
Assume for instance that $\tau=\psi$, ${\mathcal D}=\partial_x$,
and the PDE has only one family.
Then Eq.~(\ref{eqDT}) reads
\begin{eqnarray}
& &
Y_1=u-U
\label{eqY1uU}
\end{eqnarray}
with $Y_1$ defined in (\ref{eqY1}),
and the BT is computed as follows~:
eliminate all the components of ${\bf Y}$ but $Y_1$ between the equations for 
the gradient of ${\bf Y}$,
then in the resulting equations substitute $Y_1$ as defined in (\ref{eqY1uU}).

If the computation of the BT requires the elimination of $Y_2$
between (\ref{eqProjRiccatiY1x})--(\ref{eqProjRiccatiY2t}),
this BT is 
\begin{eqnarray}
& &
Y_{1,xx} + 3 Y_1 Y_{1,x} + Y_1^3 - a Y_1 - b =0,
\label{eqBTxOrder3}
\\
& &
Y_{1,t} - (c Y_{1,x} + c Y_1^2 + d Y_1 + e)_x=0,
\label{eqBTtOrder3}
\\
& &
(Y_{1,xx})_t-(Y_{1,t})_{xx}=X_0 + X_1 Y_1 + X_2 Y_1^2=0,
\end{eqnarray}
in which $Y_1$ is replaced by an expression of $u-U$, e.g.~(\ref{eqY1uU}).

Although, let us repeat it,
the method equally applies to integrable as well as nonintegrable PDEs,
examples are split according to that distinction,
to help the reader to choose his/her field of interest.

%\vfill \eject

% ==========================================================================
\subsection{Where to truncate, and with which variable?}
\label{sectionWhere}
\indent

This Section is self-contained,
and mainly destinated to persons accustomed to perform the WTC truncation.
Although some paragraphs might be redundant with 
Section \ref{sectionAlgorithmTruncation},
it may help the reader by presenting a complementary point of view.

Let us assume in this Section that the unknown Lax pair is second order.
Then the truncation defined in the fourth step of
Section \ref{sectionAlgorithmTruncation}
is performed in the style of Weiss \etal\ \cite{WTC},
\ie\ with a single variable.
This WTC truncation
consists in forcing the series (\ref{eqLaurentSeries}) to terminate;
let us denote
$p$ and $q$ the singularity orders of $u$ and $E(u)$,
$-p'$ the rank at which the series for $u$ stops,
and $-q'$ the corresponding rank of the series for $E$
\index{truncation!method}
\index{singular manifold!method}
\begin{eqnarray}
& &
u=\sum_{j=0}^{-p'} u_j Z^{j+p},\
u_0 u_{-p'} \not=0,\
E=\sum_{j=0}^{-q'} E_j Z^{j+q},
\end{eqnarray}
in which the \textit{truncation variable}
              \index{truncation!variable}
$Z$ chosen by WTC is $Z=\varphi - \varphi_0$.
Since one has no more information on $Z$,
the method of WTC is to require the separate satisfaction of each of the
\textit{truncation equations}
 \index{truncation!equations}
\begin{eqnarray}
& &
\forall j =0,\dots,-q'\ : E_j=0.
\end{eqnarray}

In earlier presentations of the method,
one had to prove by recurrence that,
assuming that enough consecutive coefficients $u_j$ vanish beyond $j=-p'$,
then all further coefficients $u_j$ would vanish.
This painful task is useless if one defines the process as done above.

The first question to be solved is :
what are the admissible values of $p'$,
\ie\ those which respect the condition $u_{-p'} \not=0$?

The answer depends on the choice of the truncation variable $Z$.
In Section \ref{sectionSMVandEV},
three choices were presented,
$Z=$ either $\varphi - \varphi_0,\ \chi$ or $Y$,
respectively defined by
equations (\ref{eqPDEManifold}), (\ref{eqchi}), (\ref{eqMostGeneralY}),
with the property that any two of their inverse are linearly dependent.

The advantage of $\chi$ or $Y$ over $\varphi - \varphi_0$ is the following.
The gradient of $\chi$ (resp.~$Y$) is a polynomial of degree two in
$\chi$ (resp.~$Y$),
so each derivation of a monomial $a Z^k$ increases the degree by one,
while the gradient of $\varphi - \varphi_0$ is a polynomial of degree zero
in $\varphi - \varphi_0$,
so each derivation decreases the degree by one.
Consequently,
one finds two solutions and only two to the condition $u_{-p'} \not=0$ 
\cite{Pickering1993}
\begin{enumerate}
\item
$p'=p,q'=q$,
in which case the three truncations are identical,
since the three sets of equations $E_j=0$ are equivalent
(the finite sum $\sum E_j Z^{j+q}$ is just the same polynomial of $Z^{-1}$
written with three choices for its base variable),

\item
for $\chi$ and $Y$ only,
$p'=2 p,q'=2 q$,
in which case the two truncations are different
since the two sets of equations $E_j=0$ are inequivalent
(they are equivalent only if $A=0$).

\end{enumerate}

To perform the first truncation $p'=p,q'=q$,
one must then choose $Z=\chi$
since $Y$ brings no more information
and $\varphi - \varphi_0$ creates equivalent but lengthier expressions.

To perform the second truncation $p'=2 p,q'=2 q$,
one must choose $Z=Y$,
since $\chi$ would create the \textit{a priori} constraint $A=0$.

The second question to be solved is :
given some PDE with such and such structure of singularities,
and assuming that one of the above two truncations is relevant
(which is a separate topic),
which one should be selected?

The answer lies in the two elementary identities \cite{CM1993}
\begin{equation} 
   \tanh z - {1 \over \tanh z}= -2 i \sech\left[2 z + i {\pi \over 2}\right],\
   \tanh z + {1 \over \tanh z}=  2   \tanh\left[2 z + i {\pi \over 2}\right].
\label{eqIdentitiestanhsech}
\end{equation}
Let us explain why on two examples,
the ODEs whose general solution is $\tanh(x-x_0)$ and $\sech(x-x_0)$,
namely
% (\ref{eqsech}) by $u'^2+e^2(u-a)(u-b)(u-c)(u-d)=0$.
\begin{eqnarray}
& &
E \equiv u'+u^2-1=0,\ u=\tanh(x-x_0),
\label{eqtanh}
\\
& &
E \equiv v'^2 + a^{-2} v^4 -v^2=0,\ v= a \sech(x-x_0),
\label{eqsech}
\end{eqnarray}
(this is just for convenience that we do not set $a=1$).
Equation (\ref{eqtanh}) has the single family
\begin{eqnarray}
& &
p=-1, q=-2, u_0=1, 
\hbox{Fuchs indices}=(-1),
\end{eqnarray}
and equation (\ref{eqsech}) has the two opposite families
\begin{eqnarray}
& &
p=-1, q=-4, v_0= i a,
\hbox{Fuchs indices}=(-1),
\end{eqnarray}
in which $i a$ denotes any square root of $-a^2$.
The first truncation
\begin{eqnarray}
& &
u = \sum^{-p}_{j=0} u_j \chi^{j+p},\ 
E = \sum^{-q}_{j=0} E_j \chi^{j+q},\
\forall j\ : E_j=0,
\end{eqnarray}
generates the respective results 
\begin{eqnarray}
& &
u=\chi^{-1},\ S=-2,
\\
& &
v= i a \chi^{-1},\
E_2\equiv a^{2}(1-S)=0,\
E_3\equiv 0,\
E_4\equiv -a^{2} S^2/4,
\end{eqnarray}
thus providing (after integration of the Riccati ODE (\ref{eqChix}))
the general solution of equation (\ref{eqtanh}),
and no solution at all for equation (\ref{eqsech}).

The second truncation
\begin{eqnarray}
& &
u = \sum^{-2 p}_{j=0} u_j Y^{j+p},\ 
E = \sum^{-2 q}_{j=0} E_j Y^{j+q},\
\forall j\ : E_j=0,
\label{eqTruncation2p}
\end{eqnarray}
generates the respective results
% Exact expr to be checked
\begin{eqnarray}
& &
u=B^{-1} Y^{-1} + (1/4) B Y,\
A=0,\
S=-1/2,\
B \hbox{ arbitrary},
\\
& &
v=i a B^{-1} Y^{-1} - (1/4) i a B Y,\
A=0,\
S=-1/2,\
B \hbox{ arbitrary},
\end{eqnarray}
thus providing, thanks to the identities (\ref{eqIdentitiestanhsech}),
the general solution for both equations.

The conclusions from this exercise which can be generalized are :
\begin{enumerate}
\item
for PDEs with only one family,
the second truncation brings no additional information as compared to
the first one and is always useless;

\item
for PDEs with two opposite families 
(two opposite values of $u_0$ for a same value of $p$),
the first truncation can never provide the general solution
and can only provide particular solutions,
while the second one may provide the general solution.

\end{enumerate}

This defines the guideline to be followed in the respective Sections
\ref{sectionTruncationOneFamily}
and
\ref{sectionTruncationTwoFamilies}.
The question of the relevance of the parameter $B$,
which seems useless in the above two examples,
is addressed in Section \ref{sectionTruncationTwoFamilies}.

% ==========================================================================
\section{The singular manifold method applied to one-family PDEs}
\label{sectionTruncationOneFamily}
\indent

% ==========================================================================
\subsection{Integrable equations with a second order Lax pair}
\label{sectionOneFamSecondOrderIntegrable}
\indent

There is only one truncation variable,
which must be chosen as $\chi$.

Weiss introduced a nice notion,
initially for one-family integrable equations with a second order Lax pair,
later extended to two-family such equations by Pickering \cite{Pickering1996}.
This is the following.

\begin{definition}
(\cite{Conte1990})
Consider the set of $-q+p$ \textit{determining equations} 
(\ref{eqDetermining2})
$E_j=0$, which depend on $(S,C,U)$.
One calls
\textbf{singular manifold equation} (SME)
 \index{singular manifold!equation}
the result of the elimination of $U$ between them.
\end{definition}

In the two-family situation, these determining equations also depend
on $(A,B)$, see (\ref{eqTruncation2p}),
and the extension of this definition \cite{Pickering1996} is to also
require the elimination of $(A,B)$.

Despite its name, originally restricted to integrable equations,
the SME can be made of several equations in the nonintegrable case.

The SME has the following properties.
\begin{enumerate}
\item
unicity, whatever be the integrability of the PDE,

\item
invariance under homography by construction \cite{Conte1989},
i.e. dependence only on one Schwar\-zian $S$ and as many $C$ quantities as
independent variables other than the one in the Schwarzian,

\item
the SME set is made of one and only one equation if and only if the PDE is 
integrable.

\end{enumerate}

Although one can define a SME whatever be the order of the Lax pair,
it is inconsistent,
as will be explained in Section \ref{sectionWTCSuitable},
to do so whenever this order is higher than two.

% ==========================================================================
\subsubsection{The Liouville equation}
\label{sectionLiouville}
\indent

\index{Liouville equation}

It is convenient to consider, 
following Zhiber and Shabat \cite{ZhiberShabat1979}, the equation
\begin{eqnarray}
& &
E(u) \equiv u_{xt} + \alpha e^u + a_1 e^{-u} + a_2 e^{-2 u}=0,\ \alpha \not=0,
\label{eqZS}
\end{eqnarray}
which has the advantage to include
the Liouville   equation $a_1=a_2=0$,
the sine-Gordon equation $(a_1 \not=0,a_2=0)$
and
the Tzitz\'eica equation $(a_1 =0,a_2 \not=0)$.
As to the case $a_1 a_2 \not=0$, it fails the test.
Let us consider here the Liouville case.
The results to be found are
its auto-BT \cite{McLaughlinScott}
and its hetero-BT to the d'Alembert equation.
This will be achieved with two different truncations.
\index{d'Alembert equation}

Although not algebraic in $u$,
the PDE is algebraic in either $e^u$ or $e^{-u}$.

Equation (\ref{eqZS}) always possesses the family
\begin{eqnarray}
& &
 e^u \sim -(2/ \alpha) \varphi_x \varphi_t (\varphi - \varphi_0)^{-2},\
\hbox{indices } (-1,2),\
{\mathcal D}=(2/ \alpha) \partial_x \partial_t.
\label{eqFamilyOne}
\end{eqnarray}
For Liouville, this is the only family.

The special form of Liouville equation allows the assumption 
\begin{eqnarray}
& &
e^u = {\mathcal D} \Log \tau + e^U,\
E(u)=0,\
{\mathcal D}=(2/ \alpha) \partial_x \partial_t,
\label{eqTruncationGeneral}
\end{eqnarray}
to be integrated twice to yield
\begin{eqnarray}
& &
u=-2 \Log \tau + V,\
E(u)=\sum_{j=0}^{2} E_j \tau^{j-2}=0,\
\label{eqTruncationLiouville}
\end{eqnarray}
in which nothing is assumed on $V$.

The Liouville equation is nongeneric for the singular manifold method
in the sense that it is linearizable into another equation
(thus, it should even \textit{not} be part of the 
Section \ref{sectionOneFamSecondOrderIntegrable}).

Therefore we define the first truncation in an exceptional way, 
namely we do not assume any linear relations on $\tau \equiv \psi$
and just treat $\tau$ as the truncation variable.
The three determining equations are then quite simple
\cite{CMG1999}
\begin{eqnarray}
& & E_0 \equiv 2 \tau_x \tau_t + \alpha e^V=0,\
\\
& & E_1 \equiv \tau_{xt}=0,\
\\
& & E_2 \equiv V_{xt}=0,
\end{eqnarray}
and their general solution depends on two arbitrary functions of one variable
\begin{eqnarray}
& &
\tau=f(x) + g(t),\
\\
& &
e^V=-{2 \over \alpha} \tau_x \tau_t =- {2 \over \alpha} f'(x) g'(t),\
\label{eqeV}
\\
& &
e^u=-{2 \over \alpha} {\tau_x \tau_t \over \tau^2}
=- {2 \over \alpha} {f'(x) g'(t) \over (f(x)+g(t))^2},
\label{eqeu}
\\
& &
e^U=\tau^{-2} e^V +{2 \over \alpha} {\tau_x \tau_t \over \tau^2}=0.
\end{eqnarray}
Thus, the two fields $u$ and $V$ are the general solution of, respectively,
the Liouville and d'Alembert equations.
The hetero-BT between these two equations
is provided by
the elimination of $f$ and $g$ % more precisely $f'/(f+g),g'/(f+g)$ 
between (\ref{eqeV}), (\ref{eqeu})
and the $x-$ and $t-$derivatives of (\ref{eqTruncationLiouville})
\index{d'Alembert equation}
\index{B\"acklund transformation!hetero--}
\begin{eqnarray}
& &
(u-v)_x = \alpha \lambda e^{(u+v)/2},\
\\
& &
(u+v)_t = - 2 \lambda^{-1} e^{(u-v)/2},\
\end{eqnarray}
in which $v$ is another solution of d'Alembert equation defined by
\begin{eqnarray}
& &
e^v=(\lambda \tau_t)^{-2} e^V 
= - {2 \over \alpha} \lambda^{-2} {f'(x) \over g'(t)}.
%v=V - 2 \Log (\lambda \tau_t).
\end{eqnarray}

\bigskip

{\it Remark}.
When performing the truncation (\ref{eqTruncationGeneral}),
Tamizhmani and Lakshmanan \cite{TamLak}
already found $e^U=0, \tau_{xt}=0$ as a {\it particular} solution,
while the above truncation (\ref{eqTruncationLiouville}) proves it to be the
{\it general} solution.
Another difference between the two truncations is the presence of a field $V$
in (\ref{eqTruncationLiouville}),
which allows us to find in addition the hetero-BT between
the Liouville and d'Alembert equations.

Let us now define the second truncation, by the assumption
% The equation (55) in \cite{CMG1999} is misleading,
% for the PDE has only one family.
\begin{eqnarray}
& &
u=-2 \Log \tau + \tilde W,\
\end{eqnarray}
and the link (\ref{eqTauPsiLink1}),
with $\psi$ solution of 
the Lax pair (\ref{eqLaxScalar2xG})--(\ref{eqLaxScalar2tG}).
Introducing the Riccati variable $Y$ defined by (\ref{eqYpsi}),
this second truncation is equivalent to \cite{CMG1999} 
\begin{eqnarray}
& &
u=-2 \Log Y + W,\
Y^{-1}=B(\chi^{-1} + A),\
\nonumber
\\
& &
E(u)=\sum_{j=0}^{4} E_j(S,C,A,B,W) Y^{j-2},\
\forall j:\ E_j=0,
\label{eqTruncation2Liouville}
\end{eqnarray}
and its result is recovered from the truncation of sine-Gordon
in Section \ref{sectionSG} by simply setting $a_1=0$.

% ==========================================================================
\subsubsection{The AKNS equation}
\label{sectionAKNSPDE}
\indent

The AKNS equation \cite{AKNS}
\index{AKNS equation}
\begin{eqnarray}
& &
E(u) \equiv u_{xxxt} 
+ 4 \alpha^{-1} (2 (u_x - \beta) u_{xt} + (u_t - \gamma) u_{xx})= 0
\end{eqnarray}
admits the single family
\begin{eqnarray}
& &
p=-1,\
q=-5,\
u_0=\alpha,\
\hbox{indices } (-1,1,4,6),\
{\mathcal D}= \alpha \partial_x,
\end{eqnarray}
so the assumption for the DT is (\ref{eqDTOneFamily}).
Let us choose
at second step the scalar Lax pair 
(\ref{eqLaxScalar2x})--(\ref{eqLaxScalar2t}) for $\psi$,
at third step the link (\ref{eqTauPsiLink1}) between $\tau$ and $\psi$.
Then there are only three non identically zero determining equations 
(\ref{eqDetermining2})
\cite{MusetteSainteAdele}
\begin{eqnarray}
 E_2 & \equiv & 4 \alpha S C + 8 (U_{t} - \gamma) - 16 C (U_{x} - \beta) = 0,
\\
 E_3 & \equiv & - \alpha (C S_x + 4 S C_x) + 16 C_x (U_{x} - \beta) 
- 8 U_{xt} + 4 C U_{xx} = 0,
\\
 E_5 & \equiv & E(U) + (\alpha/2) (2 S S_t - C S S_x - S_{xxt} - S_x C_{xx})
- 2 S_x (U_{t} - \gamma) 
\nonumber
\\
& &
- 4 S_t (U_{x} - \beta) - 4 S U_{xt} 
+ 2 (S C + C_{xx}) U_{xx} = 0,
\end{eqnarray}
plus the ever present condition $X=0$, Eqn.~(\ref{eqCrossSC}).
Their detailed resolution for $(U_x,U_t)$ is as follows.
One eliminates $U_t$ between $E_2$ and $E_3$
\begin{eqnarray}
E_3 + E_{2,x} & \equiv & 3 C (-4 U_{xx} + \alpha S_x) =0,
\end{eqnarray}
discards the nongeneric solution $C=0, U_t=\gamma, S_t=0$,
introduces an arbitrary function of $t$ after one integration,
and solves for $U_{x}$ 
\begin{eqnarray}
U_x - \beta & = & (\alpha/4)  (S + 2 \lambda(t)).
\end{eqnarray}
Then $E_2$ is solved for $U_t$
\begin{eqnarray}
U_t - \gamma & = & \alpha \lambda(t) C.
\end{eqnarray}
The cross-derivative condition $U_{xt}=U_{tx}$ is solved for $S_t$
\begin{eqnarray}
S_t & = & 4 \lambda(t) C_x -2 \lambda'(t).
\end{eqnarray}
Substituting $U_x,U_t,S_t$ and $C_{xxx}$ taken from (\ref{eqCrossXT})
in $E_5$, one obtains the condition
\begin{eqnarray}
E_5 & = & 2 \alpha \lambda(t) \lambda'(t) =0,
\end{eqnarray}
which introduces the spectral parameter as the arbitrary constant $\lambda$.

The solution for $(U_x,U_t)$ is
\begin{eqnarray}
& & 
U_x - \beta  = (\alpha/4) (S + 2 \lambda),\
U_t - \gamma = \alpha \lambda C,\
\end{eqnarray}
and the elimination of $U$ defines the SME
 \index{singular manifold!equation}
\begin{eqnarray}
{S_t \over C_x} - 4 \lambda =0.
\end{eqnarray}

The solution for $(S,C)$ is
\begin{eqnarray}
& &
S=(4/ \alpha) (U_{x} - \beta) - 2 \lambda,\ 
C = (U_{t} - \gamma)/(\alpha \lambda),\
\end{eqnarray}
and its cross-derivative condition
\begin{eqnarray}
& &
X \equiv E(U)/(\alpha \lambda)=0
\end{eqnarray}
creates on the field $U$ the only constraint that $U$ satisfy the AKNS PDE.

The BT is the result of the substitution $\chi^{-1}=(u-U)/ \alpha$ in 
(\ref{eqChix})--(\ref{eqChit}).

% ==========================================================================
\subsubsection{The KdV equation}
\label{sectionKdV}
\indent

\index{Korteweg-de Vries equation}

The \KdV\ equation for $u$ (\ref{eqKdVcons}) is defined in conservative form,
so it is cheaper to process the potential form
\begin{eqnarray}
& &
E(v) \equiv b v_t + v_{xxx} - (3/a) v_x^2 + F(t) = 0,\
v=u_x.
\label{eqKdVpot}
\end{eqnarray}
Its unique family is
\begin{eqnarray}
& &
p=-1,\
q=-4,\
v_0=-2 a,\
\hbox{indices } (-1,1,6),\
{\mathcal D}= - 2 a \partial_x.
\end{eqnarray}

With the assumption
\begin{eqnarray}
& &
v=V + {\mathcal D} \Log \tau,\
E(v)=0,
\label{eqDTpKdV}
\end{eqnarray}
for the DT,
the choice of the second-order scalar
(\ref{eqLaxScalar2x})--(\ref{eqLaxScalar2t}) for $\psi$,
the link (\ref{eqTauPsiLink1}) between $\tau$ and $\psi$,
one generates the three determining equations
\begin{eqnarray}
E_2 & \equiv & - 2 a (b C + 2 S) - 12 V_x = 0,
\\
E_3 & \equiv & 2 a (b C - S)_x = 0,
\\
E_4 & \equiv & E(V) + {S \over 2} E_2 - {1 \over 2} E_{3,x} = 0.
\end{eqnarray}
After one integration of $E_3$,
the system $(E_2,E_3)$ is solved for $(S,C)$
\begin{eqnarray}
& &   S = - 2 \lambda(t) - (2/a) V_x,\
    b C =   4 \lambda(t) - (2/a) V_x,
\end{eqnarray}
in which $\lambda(t)$ is an arbitrary integration function.
Then $E_4$, as seen from its above written compacted expression,
expresses that $V$ satisfies the PDE.
Last, the cross-derivative condition (\ref{eqCrossXT})
\begin{eqnarray}
& & 
X \equiv -2 \lambda'(t) - 2 (E(V))_x / (a b)=0
\end{eqnarray}
introduces the spectral parameter as an arbitrary complex constant
and proves that a Lax pair has been obtained for the conservative
(not the potential) equation.
This Lax pair can be written, at the reader's taste, 
either in the scalar representation
(\ref{eqLaxScalar2x})--(\ref{eqLaxScalar2t}), with $U=V_x$,
\begin{eqnarray}
& & 
L_1 \equiv   \partial_x^2 - U / a - \lambda,\
\\
& &
L_2 \equiv b \partial_t
+ ( 4 \lambda - 2 U / a) \partial_x
+ U_x / a,\
\\
& &
a \lbrack L_1,L_2 \rbrack = b U_t + U_{xxx} - (6/a) U U_x,
\end{eqnarray}
or in the zero-curvature representation
(\ref{eqLaxMatrix2xG})--(\ref{eqLaxMatrix2tG})
\begin{eqnarray}
& &
L= \pmatrix{0 & 1 \cr U/a + \lambda & 0 \cr},\
\\
& &
M= b^{-1} \pmatrix{-U_x/a & 2 U/a - 4 \lambda \cr
               -U_{xx}/a + 2 (U/a + \lambda) (U/a - 2 \lambda) & U_x/a \cr},\
\end{eqnarray}
or in the Riccati representation for $\omega=\chi^{-1}$
(see (\ref{eqChix})--(\ref{eqChit}) and (\ref{eqPsiT}))
\begin{eqnarray}
& &
\omega_x= - {S \over 2} - \omega^2
        = \left({U \over a} + \lambda \right) - \omega^2,
\label{eqKdVomegax}
\\
& &
\omega_t=(-C \omega + C_x/2)_x 
        =b^{-1} ((2 U/a - 4 \lambda) \omega - U_x/a)_x.
\label{eqKdVomegat}
\end{eqnarray}
This last representation is by far the best one,
for it allows one to deduce immediately two quite important informations, 
namely the auto-B\"acklund transformation of KdV
and the hetero-B\"acklund transformation between KdV and mKdV.
Firstly, the substitution of the inverse relation of (\ref{eqDTpKdV})
\begin{eqnarray}
& &
\omega= (v-V)/(-2 a)
\end{eqnarray}
in (\ref{eqKdVomegax})--(\ref{eqKdVomegat})
provides the auto-BT for the conservative form of KdV
\begin{eqnarray}
 a (v + V)_x            &=& - 2 a^2 \lambda + (v-V)^2 / 2,\
\\
a(b(v + V)_t - 2 F'(t)) &=& -(v - V)(v - V)_{xx} + 2 (V_x^2 + v_x V_x + v_x^2),
\label{eqKdVBTx}
\end{eqnarray}
after suitable differential consequences of the $x$-part have been added
to the $t$-part in order to suppress $\lambda$ and cubic terms
in (\ref{eqKdVBTx}).

Secondly,
the elimination of $U$ between (\ref{eqKdVomegax})--(\ref{eqKdVomegat})
leads to the mKdV equation (\ref{eqmKdV}) for $w$,
with the identification $w=\alpha \omega, \nu=\lambda$;
since conversely the elimination of $\omega$ leads to the KdV equation for $U$,
the system (\ref{eqKdVomegax})--(\ref{eqKdVomegat}) also represents
the hetero-BT between KdV and mKdV (\cite{Lamb1974} Eq.~(5.16), \cite{WE1975}).
\index{B\"acklund transformation!hetero--}

As to the SME, it results from the elimination of $V$ between $(E_2,E_3,E_4)$
 \index{singular manifold!equation}
\begin{eqnarray}
& &
b C -S - 6 \lambda=0.
\label{eqSMEKdV}
\end{eqnarray}
Most of these results for KdV were found in the original paper of WTC 
\cite{WTC}.

{\it Remark}.
The transformation (\ref{eqKdVomegax}) between $w=\alpha \omega$ and $U$ is
often called a \textit{Miura transformation},
\index{Miura transformation}
but it is really just one half of the hetero-BT.
The advantage of the hetero-BT it that it is invertible,
while the Miura transformation as defined in the previous sentence is not.

%\vfill \eject

% ==========================================================================
\subsection{Integrable equations with a third order Lax pair}
\label{sectionOneFamThirdOrderIntegrable}
\indent

Let us process a few PDEs which possess a third order Lax pair,
and let us first perform their one-family truncation
with the (wrong) assumption of a second order Lax pair,
because this often provides interesting results.

% ==========================================================================
\subsubsection{The Boussinesq equation}
\label{sectionBoussinesq}
\indent

\index{Boussinesq equation}

The Boussinesq equation (Bq) is often defined 
in a two-component evolution form
\cite{ZS1974}
\begin{equation}
\hbox{sBq}(u,r) \equiv
 \left\{\matrix{
 u_t- r_x=0,\ (\alpha,\beta,\varepsilon) \hbox{ constant},
 \cr
 r_t+ \varepsilon^2 ((u+\alpha)^2 + (\beta^2/3) u_{xx})_{x})=0. \hfill
 \cr}
  \right.
\label{eqsBq}
\end{equation}
Let us consider its one-component ``potential'' form
\begin{eqnarray}
{\hskip -4.0truemm}
& &
\hbox{pBq}(v) \equiv
 v_{tt}
+\varepsilon^2 \left((v_x+\alpha)^2 + (\beta^2/3) v_{xxx} \right)_{x}=0,\
u=v_x,\
r=v_t.
\label{eqpBq}
\end{eqnarray}
Equation (\ref{eqpBq}) has only one family of movable singularities
\begin{equation}
p=-1,\ q=-5,\ \hbox{indices } (-1,1,4,6),\ {\mathcal D}=2 \beta^2 \partial_x,
\end{equation}
and it passes the Painlev\'e test \cite{Weiss1985Bq}.
Since (\ref{eqpBq}) is a conservation law,
the computations can be reduced by considering the 
``second potential Bq'' equation
\begin{eqnarray}
& &
{\hskip -9.0 truemm}
\hbox{ppBq}(w) \equiv
w_{tt}+\varepsilon^2\left((w_{xx}+\alpha)^2+(\beta^2/3) w_{xxxx}\right)=0,\
u=v_x=w_{xx},
\end{eqnarray} 
whose single family is of the logarithmic type $w \sim 2 \beta^2 \Log \chi$
\begin{equation}
p=0^{-},\ q=-4,\ \hbox{indices} (-1,0,1,6),\ {\mathcal D}=2 \beta^2.
\end{equation}

Let us assume for the would-be DT the relation
\begin{eqnarray}
& &
w = 2\beta^2 \Log \tau + W,\
\hbox{ppBq}(w)=0,\
\label{eqDTppBq}
\end{eqnarray}
and for the link between $\tau$ and $\psi$ the identity (\ref{eqTauPsiLink1}).

Let us first assume that $\psi$ satisfies the second-order scalar Lax pair 
(\ref{eqLaxScalar2x})--(\ref{eqLaxScalar2t}).
This is equivalent to the usual WTC truncation
in the invariant formalism \cite{Conte1989}
\begin{eqnarray}
& &
\hbox{ppBq}(w) \equiv \sum_{j=0}^{4} E_j \chi^{j-4}=0,
\end{eqnarray}
and this generates the three determining equations
\begin{eqnarray}
& &
E_2 \equiv
(4/3) \beta^2 \varepsilon^2 S - 2 C^2 - 4 \varepsilon^2 (W_{xx} + \alpha)=0,
\\
& &
E_3 \equiv -2(C_t - C C_x -(\beta^2 \varepsilon^2/3) S_x) = 0,
\\
& &
E_4 \equiv (S E_2 - E_{3,x})/2 + C_x^2 + \beta^2 \hbox{ppBq}(W)=0.
\end{eqnarray}
{}From the last equation $E_4=0$,
the desired solution $\hbox{ppBq}(W)=0$ cannot be generic, 
so this second-order assumption fails to provide the auto-BT.
However, it does provide another information,
namely a hetero-BT between the Boussinesq PDE and another PDE.
Indeed, under the natural parametric representation of $E_3$
(which, by the way, would be the SME if the second order were the correct one),
 \index{singular manifold!equation}
\index{B\"acklund transformation!hetero--}
\begin{eqnarray}
& &
S=3 z_t - 3 (\beta\varepsilon)^2 z_x^2/2,\
C=(\beta \varepsilon)^2 z_x,\
\end{eqnarray}
the field $z$, by the cross-derivative condition (\ref{eqCrossXT}),
satisfies the modified Boussinesq equation \cite{HS1977}
\begin{eqnarray}
{\hskip -4.0 truemm}
& &
\hbox{MBq}(z) \equiv
z_{tt}
 +((\beta\varepsilon)^2 / 3) z_{xxxx}
+2 (\beta\varepsilon)^2 z_t z_{xx}
-2 (\beta\varepsilon)^4 z_x^2 z_{xx}=0.
\end{eqnarray}

Just like for the KdV equation (Section \ref{sectionKdV}),
this leads, after a short computation left to the reader,
to the hetero-BT between the Boussinesq and the modified Boussinesq equations.

Going to third order, 
the assumption (\ref{eqDTppBq}) and (\ref{eqTauPsiLink1}),
with $\psi$ solution of the scalar Lax pair 
(\ref{eqLaxScalar3xG})--(\ref{eqLaxScalar3tG}),
generates
\begin{eqnarray}
& &
\hbox{ppBq}(w) \equiv
\sum_{k=0}^{2} \sum_{l=0}^{2} E_{k,l} Y_1^k Y_2^l,\ k+l \le 2.
\end{eqnarray}
These six determining equations $E_{k,l}=0$,
plus the three cross-derivative conditions $X_j=0,j=0,1,2,$ 
are solved as follows
in the Gel'fand-Dikii case $f=0$: \cite{MCGallipoli1991,MC1998}

\begin{equation} 
\begin{array}{lclcl}
E_{02} & \equiv & (\beta \varepsilon)^2-c^2=0
& \Rightarrow &
c=\beta \varepsilon,
\\ 
E_{11} & \equiv & d=0
& \Rightarrow &
d=0,
\\ 
E_{20} & \equiv & 3 (V_{x}+ \alpha) + 2 \beta^2 a=0
& \Rightarrow &
a=-3 (V_{x}+ \alpha)/(2 \beta^2),
\\ 
E_{10} & \equiv &  \varepsilon V_{xx} - \beta e_x =0
& \Rightarrow &
e_x=\beta^{-1} \varepsilon V_{xx},
\\ 
X_1 & \equiv & 
3 V_{xt} + 3 \beta \varepsilon V_{xxx} + 4 \beta^3 \varepsilon b_x =0
& \Rightarrow &
b=g(t)-3 (\beta^{-2} V_{xx}+ \beta^{-3} \varepsilon^{-1} V_{t})/4,
\\ 
X_0 & \equiv & 
(3/(4 \varepsilon \beta^2)) \hbox{pBq}(V)=0
& \Rightarrow &
V \hbox{ satisfies the PDE (\ref{eqpBq})},
\\ 
E_{00} & \equiv & 2 \beta^2 g'(t) =0
& \Rightarrow &
g(t)=\lambda,
\end{array}
\end{equation}
in which $\lambda$ is an arbitrary constant.
The coefficients $a,b,c,d,e$ are  
\begin{eqnarray} 
& &
a=-(3/2) \beta^{-2}(V_x+\alpha),\  
b=\lambda -(3/4)\beta^{-2} V_{xx} -(3/4)\beta^{-3} \varepsilon^{-1} V_{t},\
\nonumber 
\\ 
& &
c=\beta\varepsilon,\ 
d=0,\
e= \beta^{-1} \varepsilon (V_{x} + \alpha),
\\
& &
X_0=(3/(4 \varepsilon \beta^2)) \hbox{pBq}(V),
X_1=0,
X_2=0,
\end{eqnarray} 
and they define a third-order Lax pair of the potential Boussinesq equation
(\ref{eqpBq}) \cite{Zakharov1973,ZS1974,Morris1976}.

The BT is just (\ref{eqBTxOrder3})--(\ref{eqBTtOrder3})
or equivalently, after substitution of $Y_1=(v-V)/(2 \beta^2)$,
\begin{eqnarray}
& &
(v-V)_{xx} 
+ 3 \beta^{-1} \varepsilon^{-1} (v+V)_t 
+ 3 \beta^{-2} (v-V)((v+V)_x + 2 \alpha)
\nonumber 
\\
& &
+ \beta^{-4} (v-V)^3 
-8 \beta^2 \lambda =0,
\\ 
& & 
(v+V)_{xx} 
- \beta^{-1} \varepsilon^{-1} (v-V)_t 
+ \beta^{-2} (v-V)(v-V)_x 
=0.
\end{eqnarray}

% ==========================================================================
\subsubsection{The Hirota-Satsuma equation}
\label{sectionHirotaSatsuma}
\indent

\index{Hirota-Satsuma equation}

Defined as
\cite{HS1976a}
\begin{eqnarray}
& &
\hbox{HS}(w) \equiv [w_{xxt} + (6/a) w_x w_t]_x =0,\ a \not=0,
\label{eqHS}
\end{eqnarray}
it is better processed on its potential form
\begin{eqnarray}
& &
\hbox{pHS}(w) \equiv w_{xxt} + (6/a) w_x w_t + F(t)=0,\ a \not=0.
\label{eqpHS}
\end{eqnarray}

The second order assumption 
(\ref{eqLaxScalar2x})--(\ref{eqLaxScalar2t}) 
generates the three determining equations
\begin{eqnarray}
{\hskip -7.0truemm}
& & E_2 \equiv - 2 a S C - 6 W_t + 6 C W_x = 0,
\nonumber
\\
{\hskip -7.0truemm}
& & E_3 \equiv a S_t + 2 a S C_x - 6 C_x W_x = 0,
\\
{\hskip -7.0truemm}
& & E_4 \equiv \hbox{pHS}(W)
 - a (S^2 C + S_{xt}/2 + S C_{xx}) - 3 S W_t + 3 (S C + C_{xx}) W_x=0.
\nonumber
\end{eqnarray}
In the generic case $C_x \not=0$,
their general solution is unknown,
in particular we have not succeeded to perform the elimination of $(S,C)$
to find the constraint(s) satisfied by $W$.
It is easy to eliminate $W$ but this gives rise to two equations for $(S,C)$
\begin{eqnarray}
& & 
6 W_x=2 a S + a {S_t \over C_x},\
6 W_t=  a {C S_t \over C_x},
\\
& &
M_{23} \equiv
\left({C S_t \over C_x}\right)_x - \left(2 S + {S_t \over C_x}\right)_t = 0,
\\
& &
M_{4} \equiv
1 - 6 a^{-1} F(t)
-   C_x^{-1} (4 C S S_t + C S_{xxt} + 2 C_{xx} S_t)
\nonumber
\\
& &
\phantom{xxxx}
-   C_x^{-2} (2 C S_t^2 + C^2 S_t S_x - 2 C C_{xx} S_{xt})
- 2 C_x^{-3} C C_{xx}^2 S_t=0,
\end{eqnarray}
and their possible functional dependence is unsettled.
Anyhow, the field $W$ cannot be a second solution of (\ref{eqHS})
\cite{MC1991}.

The third order assumption 
(\ref{eqLaxScalar3xG})--(\ref{eqLaxScalar3tG}),
with the link (\ref{eqTauPsiLink1}) and the truncated expansion
\begin{eqnarray}
& & w=W + a \partial_x \Log \tau,
\end{eqnarray}
generates seven determining equations (\ref{eqDetermining3}).
They are easily solved \cite{MC1991} and their unique solution defines
the Lax pair (\ref{eqHSLaxx})--(\ref{eqHSLaxt}),
with $W$ a second solution of (\ref{eqpHS}).

% ==========================================================================
\subsubsection{The Tzitz\'eica equation}
\label{sectionTzi}
\indent

\index{Tzitz\'eica equation}

The equation is defined by (\ref{eqZS}), in the case $(a_1 =0,a_2 \not=0)$.
It posseses two families, the first one defined by (\ref{eqFamilyOne}),
the second one by 
\begin{eqnarray}
& &
e^{-u} \sim \sqrt{(1 / a_2) \varphi_x \varphi_t} (\varphi- \varphi_0)^{-1},\
\hbox{indices } (-1,2).
\label{eqFamilyTwoTzi}
\end{eqnarray}
These two families are not opposite,
but the second family is irrelevant because the Tzitz\'eica equation 
has a one-to-one correspondence \cite{HR1980} 
with a one-family equation,
namely the potential form (\ref{eqpHS}) 
of the Hirota-Satsuma PDE 
in the particular case $F(t)=0$.
This correspondence is obtained by the 
elimination of $a_2$ in equation (\ref{eqZS})
\begin{eqnarray}
& &
{\hskip -5.0 truemm}
\left(F(t)=0, e^u={2 \over a \alpha} w_t\right)
\Longrightarrow
\left(
e^{-2u} (e^{2 u} \hbox{Tzi}(u))_x
 = \left({\hbox{pHS}(w) \over w_t}\right)_t
\right).
\label{eqTziandpHS}
\end{eqnarray}
The irrelevance of the second family is confirmed
by the negative result of Weiss \cite{Weiss1986,Conte1989} 
obtained when performing a truncation on $e^{-u}$.

All the truncations will accordingly take the same form 
(\ref{eqTruncationGeneral}) as for the Liouville equation,
which implies that $\tau$ is an object invariant under the 
permutation (\ref{eqPermutation}).
Depending on the Lax pair assumption,
the link between $\tau$ and $\psi$ will be
either the identity (case of a scalar $\psi$ invariant under the
permutation (\ref{eqPermutation}))
or not (if the scalar $\psi$ is not invariant,
e.g.~because the Lax pair itself is not invariant),
as detailed below.

Let us first assume a second order Lax pair.
To the author's knowledge,
one cannot define a scalar $\psi$,
linked to such a Lax pair,
which, like $\tau$, would be invariant under (\ref{eqPermutation}).
This is probably the reason why
the assumption $\tau=\psi$ with $\psi$ solution of
the noninvariant Lax pair (\ref{eqLaxScalar2x})--(\ref{eqLaxScalar2t}) 
generates so intricate determining equations
that their general solution has not yet been obtained
\cite{MC1994};
these equations are however consistent in the sense that
one easily finds the particular exact solution
\begin{eqnarray}
{\hskip -9.0 truemm}
& &
  \alpha e^u=2 c \wp(x-ct-x_1,g_2,A+{a_2 \alpha^2 \over 8 c^3})
            -2 c \wp(x+ct-x_2,g_2,A-{a_2 \alpha^2 \over 8 c^3}),\
\end{eqnarray}
depending on five arbitrary constants $(x_1,x_2,c,g_2,A)$
and representing the superposition of two traveling waves of opposite
velocities.

{}From this second-order WTC truncation,
and with appropriate assumptions,
one can also find a particular solution which represents
a binary Darboux transformation \cite{Schief1996a}.
\index{Darboux transformation!binary}

%The general solution of the four truncation equations of $e^u$ is
%currently under investigation \cite{Meleshko},
%using sophisticated elimination techniques \cite{Janet1929,Pommaret1978}.

Let us now turn to the third order assumption.
One can postulate 
either a Lax pair invariant under (\ref{eqPermutation}),
such as the matrix pair (\ref{eqLaxMatrix3Invarx})--(\ref{eqLaxMatrix3Invart}),
or a noninvariant Lax pair 
such as the scalar pair (\ref{eqLaxScalar3xG})--(\ref{eqLaxScalar3tG}).
In the first case,
one must assume the identity link $\tau=\psi$,
while in the second case the assumed link must be noninvariant.
Both assumptions lead to a success \cite{CMG1999}.
Let us detail here the invariant assumption,
\ie\ \textit{a priori} the simpler one.

The truncation is defined by (\ref{eqTruncationGeneral}),
the link (\ref{eqTauPsiLink1}),
and the matrix Lax pair (\ref{eqLaxMatrix3Invarx})--(\ref{eqLaxMatrix3Invart})
\begin{eqnarray}
& &
E(u) \equiv \sum_{k=0}^{3} \sum_{l=0}^{3-k}
 E_{kl}(f_j,g_j,h_j,U) Y_1^k Y_2^l,\
\forall k,l~:~
 E_{kl}=0,
\end{eqnarray}
in which $(Y_1,Y_2)$ 
are the two components of the projective Riccati pseudopotential
(\ref{eqProjRiccatiY1xc})--(\ref{eqProjRiccatiY2tc})
equivalent to the Lax pair.
To these ten determining equations in $U$ and the nine unknown coefficients,
one must add the six cross-derivative conditions $X_j=0$
(\ref{eqProjRiccatiCross012})--(\ref{eqProjRiccatiCross345}).

During their resolution, one first proves that the product $f_2 h_1$
cannot vanish (otherwise $a_2$ would be zero).
This makes the sixteen equations algebraically independent
and equivalent to the fifteen differential relations
\begin{eqnarray}
& &
f_{j,t},g_{j,x},g_{j,t},h_{j,x},g_{j,xt}
=P(\lbrace f_k,g_k,h_k \rbrace, k=1,2,3),\
j=1,2,3,
\label{eqNotation1}
\end{eqnarray}
with $P$ polynomials whose coefficients depend on $U,U_x,U_t,U_{xt}$,
plus the single algebraic relation
\begin{eqnarray}
& &
E_{00} \equiv a_2 - {4 \over \alpha^2} 
\left(g_3 + g_1 g_2 + (\alpha/2) e^U\right)^2=0.
\label{eqNotation2}
\end{eqnarray}
They are solved successively as
[equations are referenced as in (\ref{eqNotation1})--(\ref{eqNotation2})]
\begin{equation} 
\begin{array}{lcl}
g_{3,xt}-(g_{3,x})_t & : & E(U)=0,\ 
\\ 
g_{1,x}-g_{2,t} & : & \exists\ g_0(x,t) : g_1=g_{0,t},\ g_2=g_{0,x},
\\ 
g_{2,t} & : & g_3=- \alpha e^U - g_{0,x} g_{0,t} - g_{0,xt},
\\ 
E_{00} & : & \exists\ f_0(x,t)\not=0 : f_2 = \sqrt{a_2} W^{-1} f_0,\
                                      h_1 = \sqrt{a_2} W^{-1} f_0^{-1},\
\\ 
       & & \hbox{notation } W=e^U + (2 / \alpha) g_{0,xt},
\\ 
g_{2,x} & : & f_3=- \sqrt{a_2} W^{-1} f_0 g_{0,t} - f_1 g_{0,x} - g_{0,x}^2
                  + g_{0,xx},\
\\ 
g_{3,x} & : & f_1= W_x/W + 2 g_{0,x},\
\\ 
f_{2,t} & : & h_2= W_t/W + 2 g_{0,t} - f_{0,t}/f_0,\
\\ 
h_{1,x} & : & f_{0,x}=0,\
\\ 
g_{1,t} & : & h_3=g_{0,t} (f_{0,t}/f_0 -W_t/W - g_{0,t})
                    + g_{0,tt} - \sqrt{a_2} W^{-1} g_{0,x}/f_0,\
\\ 
g_{3,t} & : & f_{0,t}=0,\ 
\\ 
h_{2,x} & : & g_{0,xt}=0.
\end{array}
\end{equation}
The irrelevant arbitrary function $g_0$ reflects the freedom in the definition 
(\ref{eqTruncationGeneral})
of $\tau$ and can be absorbed by redefining $\tau$ as $\tau e^{-g_0}$.
Thus the solution is unique~:
the field $U$ must satisfy the Tzitz\'eica PDE,
and $f_0$ is an arbitrary nonzero complex constant $\lambda$.
Accordingly,
one has obtained a Lax pair and a Darboux transformation.
The equivalent projective Riccati representation of the matrix Lax pair is
\begin{eqnarray}
Y_{1,x} & = & - Y_1^2 + U_x Y_1 + \sqrt{a_2} \lambda      e^{-U} Y_2,
\label{eqProjRiccatiY1x0}
\\ 
Y_{2,x} & = & - Y_1 Y_2 - \alpha e^U,
\label{eqProjRiccatiY2x0}
\\
Y_{1,t} & = & - Y_1 Y_2 - \alpha e^U,
\label{eqProjRiccatiY1t0}
\\ 
Y_{2,t} & = & - Y_2^2 + U_t Y_2 + \sqrt{a_2} \lambda^{-1} e^{-U} Y_1,
\label{eqProjRiccatiY2t0}
\end{eqnarray}
with cross-derivative conditions proportional to the Tzitz\'eica equation
\begin{eqnarray}
& &
(Y_{1,x})_t - (Y_{1,t})_x = Y_1 E(U),\
(Y_{2,x})_t - (Y_{2,t})_x = Y_2 E(U).
\end{eqnarray}
This Lax pair is the rewriting in matrix form of the scalar triplet
 given by Tzitz\'eica \cite{Tzitzeica1908b}
\begin{eqnarray}
& &
- \tau_{xx} + U_x \tau_x + \sqrt{a_2} \lambda e^{-U} \tau_t =0,
\label{eqTziScalarLax1}
\\
& &
-\tau_{tt} +U_t \tau_t + \sqrt{a_2} \lambda^{-1} e^{-U} \tau_x=0,
\label{eqTziScalarLax2}
\\
& &
- \tau_{xt} - \alpha e^U \tau =0.
\label{eqTziScalarLax3}
\end{eqnarray}
The Lax pair admits by construction the involution 
\cite{Tzitzeica1910b,Gaffet1987} 
\begin{eqnarray}
& &
(\tau, e^U, \lambda) \to 
\left({1 \over \tau}, - e^U - {2 \over \alpha} {\tau_x \tau_t \over \tau^2},
- \lambda\right),
\label{eqInvolution}
\end{eqnarray}
equivalent to 
\begin{eqnarray}
& &
(\tau, e^U, \lambda) \to 
(1/\tau, e^U + {\mathcal D} \Log \tau, - \lambda),\
\end{eqnarray}
which defines another, equivalent, writing of the Darboux transformation
\begin{eqnarray}
& &
e^u = - e^U - {2 \over \alpha} {\tau_x \tau_t \over \tau^2}.
\label{eqDT1910}
\end{eqnarray}

{\it Remark}.
Knowing these results,
one can also write this DT \cite{YangLi,AG1998}
as a difference of the two fields $u-U$ in terms of the two components of
a projective Riccati pseudopotential
\begin{eqnarray}
& &
u=U + \Log (-2 \lambda^2 y_1 y_2 -1),\
y_j=\alpha^{-1/2} \lambda^{-1} e^{-U/2} Y_j,\
\label{eqTziDTForm2}
\end{eqnarray}
in a quite similar manner to the DT of 
Liouville and sine-Gordon (\ref{eqDTLiouvilleandSG}).
However, the field $u$ is multivalued.

In order to find the BT, one must now eliminate one of the two equivalent
projective components,
and this defines two possible, different, eliminations.

\bigskip

In the first elimination,
one takes $Y_2$ from (\ref{eqProjRiccatiY1x0})
and substitutes it into the three remaining equations, which results in
\begin{eqnarray}
& &
Y_2=(Y_{1,x} + Y_1^2 - U_x Y_1) e^U/(\sqrt{a_2} \lambda),
\\ 
& &
\hbox{ODE} \equiv
Y_{1,xx} + 3 Y_1 Y_{1,x} + Y_1^3 - e^{-U} (e^U)_{xx} Y_1 
+ \alpha \sqrt{a_2} \lambda=0,
\label{eqODEG5HS}
\\ 
& &
\hbox{PDE} \equiv
Y_{1,t}
 + e^U \left((Y_1 Y_{1,x}+ Y_1^3) - Y_1^2 U_x \right)/(\sqrt{a_2} \lambda)
 + \alpha e^U=0,
\label{eqPDEBTHS}
\\ 
& &
(\ref{eqProjRiccatiY2t0}) \equiv
- Y_1 E(U) 
- {e^U Y_1 \over \sqrt{a_2} \lambda} \hbox{ODE} 
+ (2 Y_1 - U_x + \partial_x) \hbox{PDE} =0,
\label{eqBTThirdHS}
\\ 
& &
\hbox{[ODE,PDE]} =(Y_{1,xx})_t - (Y_{1,t})_{xx} = Y_1 (e^{2 U} E(U))_x.
\label{eqBadCommutator}
\end{eqnarray}
Only two of them are functionally independent, as shown by relation
(\ref{eqBTThirdHS}), 
but the commutator (\ref{eqBadCommutator}) of equations
(\ref{eqODEG5HS})--(\ref{eqPDEBTHS}) 
shows that this elimination fails to generate the auto-BT of Tzitz\'eica 
equation. 

However, it does provide another result,
which we now derive.
The ODE (\ref{eqODEG5HS}) belongs to the classification of Gambier
-- this is the number 5, see Section \ref{sectionGambier} --,
it is linearizable by the transformation
$ Y_1 = \partial_x\Log\psi$
into a third-order linear ODE,
with the relation $\tau=\psi$ between the two functions.
This transformation also linearizes the PDE (\ref{eqPDEBTHS}),
and the resulting linear system 
\begin{eqnarray}
& &
\tau_{xxx} - (U_{xx} + U_x^2) \tau_x + \sqrt{a_2} \alpha \lambda \tau=0,
\\
& &
- \sqrt{a_2} \lambda \tau_t + e^{U} \tau_{xx} - U_x e^U \tau_x =0,
\end{eqnarray}
which cannot be a scalar Lax pair of the Tzitz\'eica equation,
is, in fact, the scalar Lax pair of the Hirota-Satsuma equation
(\ref{eqHS}), see Section \ref{sectionHirotaSatsuma},
\index{Hirota-Satsuma equation}
\begin{eqnarray}
& &
\tau_{xxx} -(6/a) w_x \tau_x + \Lambda \tau =0,
\label{eqHSLaxx}
\\
& &
 \Lambda \tau_t -(2/a) w_t \tau_{xx} + (2/a) w_{xt} \tau_x =0,
\label{eqHSLaxt}
\end{eqnarray}
under the change of variables (\ref{eqTziandpHS}).

\bigskip

In the second elimination,
one takes $Y_1$ from (\ref{eqProjRiccatiY2x0}) 
and substitutes it into the three remaining equations
\begin{eqnarray}
Y_1 & = &
-(Y_{2,x} + \alpha e^U)/Y_2,\
\\ 
\hbox{ODE} & \equiv &
Y_2 Y_{2,xx}- 2 Y_{2,x}^2 - (U_x Y_2 + 3 \alpha e^U) Y_{2,x}
\nonumber 
\\ & &
+ \sqrt{a_2} \lambda e^{-U} Y_2^3 - \alpha^2 e^{2 U}=0,
\label{eqODEBTTzi}
\\ 
\hbox{PDE} & \equiv &
Y_2 Y_{2,t} + Y_2^3 - U_t Y_2^2
 + \sqrt{a_2} \lambda^{-1}(\alpha + e^{-U} Y_{2,x})=0,
\label{eqPDEBTTzi}
\\ 
(\ref{eqProjRiccatiY1t0}) & \equiv &
 E(U) 
+ (\partial_x - \alpha e^U Y_2^{-1}) \hbox{PDE}
\nonumber 
\\ & &
- \sqrt{a_2} \lambda^{-1} e^{-U} Y_2^{-2} \hbox{ODE} =0,
\label{eqBTThirdTzi}
\\ 
\hbox{[ODE,PDE]} & = &
(Y_{2,xx})_t - (Y_{2,t})_{xx}
\nonumber 
\\
& = & (3 \alpha e^U + U_x Y_2 + 3 Y_{2,x} - Y_2 \partial_x) E(U).
\label{eqGoodCommutator}
\end{eqnarray}
Only two of them are functionally independent,
as shown by the relation (\ref{eqBTThirdTzi}),
and the vanishing of the commutator (\ref{eqGoodCommutator}) of equations
(\ref{eqODEBTTzi})--(\ref{eqPDEBTTzi})
is equivalent to the vanishing of the Tzitz\'eica equation for $U$. 
This elimination therefore generates the auto-BT of Tzitz\'eica equation,
by the substitution
\begin{eqnarray}
& &
Y_2 = (\alpha/2) \int \left(e^u - e^U \right) \D x
\end{eqnarray}
into (\ref{eqODEBTTzi})--(\ref{eqPDEBTTzi}).

The ODE part (\ref{eqODEBTTzi}) of the BT is equivalently written as
\cite{MCV2000}
\begin{eqnarray}
& &
{w_{xx} \over w_{x}} - {W_{xx} \over W_{x}}
- 2 {w_{x}+ W_{x} \over w - W}
+ \alpha \sqrt{a_2} \lambda {(w-W)^2 \over 2 w_{x} W_{x}}=0,
\end{eqnarray}
with the notation $Y_2=(\alpha/2) (w-W), e^u=w_{x},e^U=W_{x}$.

The nonlinear ODE (\ref{eqODEBTTzi}) again belongs to the equivalence class 
of the fifth Gambier equation (G5), see section \ref{sectionGambier},
and its linearization
\begin{eqnarray}
& &
Y_2^{-1} = - \alpha^{-1} e^{-U} \partial_x \Log(e^U \psi)
\label{eqY2FunctionOfPsi}
\end{eqnarray}
transforms the two equations (\ref{eqODEBTTzi})--(\ref{eqPDEBTTzi}) into
the third-order scalar Lax pair of the Gel'fand and Dikii type
(\ie\ $f=0$ in (\ref{eqLaxScalar3xG})--(\ref{eqLaxScalar3tG}))
\begin{eqnarray}
{\hskip -5.0 truemm}
{\mathcal L} \psi & \equiv &
\psi_{xxx} + (2 U_{xx} -U_x^2) \psi_x
+((2 U_{xx} -U_x^2)_x/2 + \sqrt{a_2} \alpha \lambda) \psi
=0,
\label{eqScalarLaxTzix}
\\
{\hskip -5.0 truemm}
{\mathcal M} \psi & \equiv &
\psi_t
+\sqrt{a_2}(\alpha \lambda)^{-1} e^{-2 U}(\psi_{xx} + U_x \psi_x + U_{xx} \psi)
\nonumber
\\
{\hskip -5.0 truemm}
& &
 + (U_t + \int \left(\alpha e^U + a_2 e^{-2U}\right) \D x) \psi
=0,
\label{eqScalarLaxTzit}
\\
{\hskip -5.0 truemm}
\lbrack {\mathcal L}, {\mathcal M} \rbrack & = &
3 E \partial_x^2 
+(2 (e^U)_x E + E_x) \partial_x
\nonumber
\\
{\hskip -5.0 truemm}
& &
+ (e^U)_x E_x + (3 U_{xx} - U_x^2) E,\
E=E(U).
\end{eqnarray}

Thus, the noninvariant (under (\ref{eqPermutation}))
link between $\tau$ and $\psi$ that one would have had to postulate
if one had chosen the scalar Lax pair 
(\ref{eqLaxScalar3xG})--(\ref{eqLaxScalar3tG})
is {\it a posteriori}
provided by the linearizing formula (\ref{eqY2FunctionOfPsi})
and the Riccati equation (\ref{eqProjRiccatiY1t0}),
this is the invertible transformation
\begin{eqnarray}
& &
e^U \tau= (e^U \psi)_x,\
e^U \psi=- \alpha^{-1} \tau_t,
\end{eqnarray}
and it clearly breaks the invariance under (\ref{eqPermutation}).

%\vfill \eject

% ==========================================================================
\subsubsection{The Sawada-Kotera and Kaup-Kupershmidt equations}
\label{sectionKKSK}
\indent

\index{Sawada-Kotera equation}
\index{Kaup-Kupershmidt equation}

Because of their duality \cite{Kaup1980,Weiss1984KKSK},
it is convenient to introduce simultaneously
the Sawada-Kotera equation (SK)
and the Kaup-Kupershmidt equation (KK).
These are defined as
\begin{eqnarray}
{\rm SK}(u)
& \equiv &
\beta u_t 
+ \left(u_{xxxx}+{30 \over \alpha} u u_{xx}+{60 \over \alpha^2} u^3\right)_x=0,
\label{eqSKcons}
\\
{\rm pSK}(v)
& \equiv &
\beta v_t 
+ v_{xxxxx}+{30 \over \alpha} v_x v_{xxx}+{60 \over \alpha^2} v_x^3 =0,
\label{eqSKpot}
\\
{\rm KK}(u)
& \equiv &
\beta u_t + \left(
u_{xxxx} + {30 \over \alpha} u u_{xx} + {45 \over 2 \alpha} u_x^2
 + {60 \over \alpha^2} u^3 \right)_x = 0,
\label{eqKKcons}
\\
{\rm pKK}(v)
& \equiv &
\beta v_t + 
v_{xxxxx} + {30 \over \alpha} v_{x} v_{xxx} + {45 \over 2 \alpha} v_{xx}^2
 + {60 \over \alpha^2} v_{x}^3 = 0,
\label{eqKKpot}
\end{eqnarray}
in which $u$ denotes the conservative field and $v$ the potential one,
with $u=v_x$.

Both equations have the Painlev\'e property \cite{Weiss1984KKSK}.
Each of them has two families \cite{Weiss1984KKSK}
\begin{eqnarray}
{\rm pSK, F1} & : &
p=-1,\ v_0=  \alpha,\ \hbox{indices } -1,1,2,3,10,
\label{eqpSKF1}
\\
{\rm pSK, F2} & : &
p=-1,\ v_0=2 \alpha,\ \hbox{indices } -2,-1,1,5,12,
\\
{\rm pKK, F1} & : &
p=-1,\ v_0=\alpha/2,\ \hbox{indices } -1,1,3,5,7,
\label{eqpKKF1}
\\
{\rm pKK, F2} & : &
p=-1,\ v_0=4 \alpha,\ \hbox{indices } -7,-1,1,10,12.
\end{eqnarray}
The singular part operator ${\mathcal D}$ attached to a given family is
${\mathcal D}= v_0 \partial_x$.
The two families have residues which are not opposite,
but fortunately each potential equation
possesses in its hierarchy a ``minus-one'' equation \cite{Willox}
\begin{eqnarray}
{\rm pSK}_{-1} & : &
v_{xxt} + {6 \over \alpha} v_x v_t=0,
\label{eqSKMinus1}
\\
{\rm pKK}_{-1} & : &
v_t v_{xxt} -{3 \over 4} v_{xt}^2 + {6 \over \alpha} v_x v_t^2=0,
\label{eqKKMinus1}
\end{eqnarray}
which has only one family
(the first one is nothing else than the Hirota-Satsuma PDE,
already processed in Section \ref{sectionHirotaSatsuma}).
The equations SK and KK are therefore to be considered as 
possessing the single family F1, Eqs.~(\ref{eqpSKF1}) and (\ref{eqpKKF1}).

Let us assume the one-family DT (\ref{eqDTpKdV})
and, succesively,
the second-order scalar Lax pair 
(\ref{eqLaxScalar2x})--(\ref{eqLaxScalar2t}),
then the third-order scalar one 
(\ref{eqLaxScalar3xG})--(\ref{eqLaxScalar3tG}) with the
Gel'fand-Dikii simplification $f=0$.
As to the link between $\tau$ and $\psi$,
at second order this is the identity,
while at third order it can be,
as outlined in Section \ref{sectionGambier} 
and detailed in Musette lecture \cite{CetraroMusette},
either the linearizing transformation of the fifth Gambier equation
or that of the twenty-fifth Gambier equation.

Therefore, at the fourth step of the singular manifold method,
for each PDE,
one has only three possibilities to examine~:
order two and Riccati,
order three and (G5),
order three and (G25).
This is done in the next two Sections.

% ==========================================================================
\subsubsection{The Sawada-Kotera equation}
\label{sectionSK}
\indent

\index{Sawada-Kotera equation}
\index{Kaup-Kupershmidt equation}

\underbar{First truncation} (order two and Riccati).
The one-family truncation (\ref{eqTruncationOrder2}) with $\tau=\psi$
generates the three equations \cite{Conte1990}
\begin{eqnarray}
E_4 & \equiv &
 \beta C - 4 S^2 + 9 S_{xx} + 60 S V_x / \alpha - 180 (V_x / \alpha)^2
 - 30 V_{xxx} / \alpha = 0,
\\
E_5 & \equiv &
- \beta C_x - 2 S S_x + S_{xxx} + 30 S_x V_{x} / \alpha = 0,
\\
E_6 & \equiv &
{\rm pSK}(V)+ (S E_4 - E_{5,x})/2 + 5 S_x (3 V_{xx}/ \alpha - S_x/2) = 0.
\end{eqnarray}
These equations possess two solutions \cite{Conte1990},
a nongeneric one $S_x=0$
\begin{eqnarray}
& &
S=- k^2 / 2,\
C=c+c_0,\
c=k^4 / \beta + 2 c_0/3,\
\nonumber
\\
& &
V/ \alpha =\zeta(x-(c-c_0)t,k^4/12+ \beta c_0/9,g_3)- k^2 x / 12
\nonumber
\\
& & \phantom{xxxxx}   +((5(k^4 + \beta c_0) k^2/36 - 12 g_3) t)/ \beta,
\end{eqnarray}
in which $\zeta$ is the Weierstrass function
and $(k,c_0,g_3)$ are arbitrary constants,
and a generic one $S_x \not=0$ defined by the four equations
\begin{eqnarray}
& &
V_{x} = \alpha (\beta C_x + 2 S S_x - S_{xxx})/(30 S_x),\
\label{eqSKTrunc2Start}
\\
& &
V_{t} = \dots
\\
& &
M_1 \equiv
(G_{x} / S_x)_{xx} - G - S_x^2 G_x^2 / 5 + 2 S G_x / S_x = 0,\
\\
& &
M_2 \equiv 29 \hbox{ terms}=0, \hbox{ also vanishing if } G=0,
\label{eqSKTrunc2End}
\end{eqnarray}
in which $G$ is defined by
\begin{eqnarray}
& &
G \equiv S_{xx} + 4 S^2 - \beta C.
\label{eqSKSubSME}
\end{eqnarray}
The general solution $(S,C)$ of the system $M_1=0,M_2=0$
has not yet been obtained,
for the elimination of $S$ or $C$ is difficult.
This difficulty reflects the fact that second order is not the correct order.
Nevertheless,
these complicated equations admit the very simple \textit{particular}
solution \cite{Weiss1984KKSK}
(it would be interesting to prove that this is the \textit{general} solution),
\begin{eqnarray}
& &
G=0,\
V_x=\alpha S/3,\
\hbox{pKK}(V)=0,
\label{eqSKSolHetero}
\end{eqnarray}
so the field $V$ in the DT assumption (\ref{eqDTpKdV}) satisfies a different
PDE, namely the potential KK equation.
This defines a hetero-BT between the conservative forms of SK and KK
\cite{FG1980,HR1980,Hirota1985a}
\index{Kaup-Kupershmidt equation}
\index{B\"acklund transformation!hetero--}
\begin{eqnarray}
& &
\alpha (v+V/2)_x+(v-V)^2=0,\
\beta  (v+V/2)_t+\dots=0,\
\nonumber
\\
& &
\hbox{pSK}(v)=0,\
\hbox{pKK}(V)=0
\end{eqnarray}
(see Ref.~\cite{Hirota1985a} for the exact expression of the $t$-part).

With $G=0$, the linear system (\ref{eqLaxScalar2x})--(\ref{eqLaxScalar2t})
is a degenerate Lax pair for KK,
since it lacks a spectral parameter.

Still when (\ref{eqSKSolHetero}) holds,
the field $\chi^{-1}$ satisfies a fifth order PDE, 
the Fordy-Gibbons equation,
and the explicit writing of its hetero-BT with the SK equation is left to
the reader.
\index{Fordy-Gibbons equation}

\underbar{Second truncation} (order three and (G5)).
This assumption creates no \textit{a priori} constraint on the coefficients
$(a,b)$ of the spectral problem (\ref{eqLaxScalar3xG}),
and the linearizing transformation of (G5) is just the identity $\tau=\psi$.
This generates six determining equations (\ref{eqDetermining3}).
The process is successful \cite{Weiss1984KKSK,Weiss1986,Conte1991}
and $V$ is found to be a second solution of pSK
(notation $U=V_x$ as usual)
\begin{eqnarray}
b
& = &
\lambda,
\\
a
& = &
- 6 U / \alpha,
\\
L_1
& = &
\partial_x^3
+ 6 {U \over \alpha} \partial_x - \lambda,
\\
L_2
& = &
\beta \partial_t
+ \left(18 {U_x \over \alpha} - 9 \lambda\right) \partial_x^2
+ \left(36 {U^2 \over \alpha^2} - 6 {U_{xx} \over \alpha}\right) \partial_x
- 36 \lambda {U \over \alpha},
\\
\lbrack L_1,L_2 \rbrack & = & 6 \beta^{-1} \alpha^{-1} \hbox{SK}(U).
\end{eqnarray}
This is the Lax pair given by Satsuma and Kaup~\cite{SK1977}.

The BT results from the elimination of $Y_2$,
which provides Eqs.~(\ref{eqBTxOrder3})--(\ref{eqBTtOrder3})
for $Y_1=Y$,
\begin{eqnarray}
{\hskip -14.0 truemm}
& &
Y_{xx} + 3 Y Y_x + Y^3 + 6 (U/ \alpha) Y - \lambda=0,\
\\
{\hskip -14.0 truemm}
& &
\beta Y_t
-9[(\lambda-2 U_{x}/ \alpha)(Y_x +Y^2)
\\
{\hskip -14.0 truemm}
& &
\phantom{xxxxxx}
+4(\lambda U/ \alpha-(U/ \alpha)^2 Y)+(2/3) (U_{xx}/ \alpha) Y]_x
=0,
\nonumber
\\
{\hskip -14.0 truemm}
& &
\beta((Y_{xx})_t-(Y_t)_{xx})/Y=-(6 / \alpha) \hbox{SK}(U),
\end{eqnarray}
followed by the substitution $Y=(v-V) / \alpha$,
\begin{eqnarray}
& &
(v-V)_{xx} / \alpha +3 (v-V) (v+V)_x / \alpha^2 +(v-V)^3/ \alpha^3 - \lambda=0,
\label{eqSKBTxvV}
\\
& &
\beta (v-V)_t/ \alpha -(3/2)[
(v-V)_{xxxx} / \alpha 
\nonumber
\\
& &
+ (5 (v-V) (v+V)_{xxx} + 15 (v+V)_x (v-V)_{xx}) / \alpha^2
\nonumber
\\
& &
+ (15 (v-V)^2 (v-V)_{xx} + 30 (v-V) (v+V)_x^2) / \alpha^3
\nonumber
\\
& &
+ 30 (v-V)^3 (v+V)_x / \alpha^4
+ 6 (v+V)^5/ \alpha^5
]_x=0,
\label{eqSKBTtvV}
\end{eqnarray}
a result due to Satsuma and Kaup~\cite{SK1977}.

% ==========================================================================
\subsubsection{The Kaup-Kupershmidt equation}
\label{sectionKK}
\indent

\index{Kaup-Kupershmidt equation}

\underbar{First truncation} (order two and Riccati).
The one-family truncation (\ref{eqTruncationOrder2}) with $\tau=\psi$
generates the three equations \cite{MC1998}
\begin{eqnarray}
E_2 & \equiv &
15 (S/4 - 3 V_{x}/ \alpha) = 0,
\\
E_4 & \equiv &
\beta C/2 + 7 S^2/4 + 3 S_{xx}/4
- 15 (S V_{x} + V_{xxx}) / \alpha - 90 (V_{x} / \alpha)^2 = 0,
\\
E_6 & \equiv &
(S/2) E_4/ \alpha
+ (4 \beta (C S + C_{xx}) - S^3 + (21/2) S_x^2 + 14 S S_{xx} - 4 S_{xxxx})/16
\nonumber
\\
& &
+ (15/4) (3 S^2 - 2 S_{xx}) V_{x}/ \alpha
- 45 S (V_{x} / \alpha)^2 + {\rm pKK}(V)
=0.
\end{eqnarray}
As opposed to the (difficult) SK case,
these equations are easy to solve and possess the unique solution
\cite{Weiss1984KKSK}
\begin{eqnarray}
& &
V_x=\alpha S/12,\
\hbox{pSK}(V)=0,\
S_{xx} + S^2/4 - \beta C=0.
\end{eqnarray}
This is a strong indication that the particular solution 
(\ref{eqSKSolHetero}) of (\ref{eqSKTrunc2Start})--(\ref{eqSKTrunc2End})
should be the general one.
One again recovers, by a nice duality,
the hetero-BT between KK and SK.
\index{Sawada-Kotera equation}
\index{B\"acklund transformation!hetero--}

\underbar{Second truncation} (order three and (G5)).
This generates thirteen determining equations (\ref{eqDetermining3}).
This truncation fails and provides no solution at all
(one determining equation is $E_{2,2} \equiv 45/8 =0$),
not even the one-soliton solution.
Indeed, the one-soliton solution of Kaup
corresponds to constant coefficients for the scalar Lax pair
(\ref{eqLaxScalar3xG})--(\ref{eqLaxScalar3tG}) with $f=0$,
and,
with the above procedure,
the only way to obtain it \cite{CM1993,Pickering1996}
is to enforce the two first integrals $K_1$ and $K_2$
which result from the zero value of $b$,
\begin{eqnarray}
K_1
& = &
\psi_{xx} - a \psi,
K_2=\psi_x^2 - a \psi^2 - 2 (\psi_{xx} - a \psi) \psi.
\end{eqnarray}

\underbar{Third truncation} (order three and (G25)).
This assumption implies, 
among the coefficients $(a,b)$ of the spectral problem (\ref{eqLaxScalar3xG}),
the \textit{a priori} constraint \cite{CetraroMusette}
\begin{eqnarray}
& &
b - a_x/2 = \lambda(t),
\end{eqnarray}
and the linearizing transformation of (G25)
defines the link between $\tau$ and $\psi$
\begin{eqnarray}
& &
{\tau_x \over \tau}={\lambda(t) \over Y_{1,x} + (1/2)Y_1^2 - a/2},\
Y_1={\psi_x \over \psi}.
\end{eqnarray}
This generates fourteen determining equations (\ref{eqDetermining3})
(i.e.~the same order of magnitude as for the (G5) assumption)
in the basis
$(\psi_x / \psi,\psi_{xx}/ \psi-((\psi_x/ \psi)^2 +b_x \psi_x /(b \psi)-a)/2)$;
they are solved as follows
($g_k$ denotes an arbitrary integration function, 
$\lambda$ an arbitrary integration constant)
\cite{MC1998}
\begin{equation} 
\begin{array}{lcl}
E_{4,4} & : & c = 9 \lambda(t) / \beta,
\\ 
E_{3,5} & : & a = -6 V_x / \alpha,
\\ 
E_{1,5} & : & d = (3 V_{xxx} / \alpha - (6 V_x / \alpha)^2)/ \beta,
\\ 
E_{0,5} & : & \lambda(t) = \lambda \hbox{ independent of } t,
\\ 
E_{0,6} & : & \hbox{pKK}(V)=0,
\\ 
E_{1,4} & : & e = g_2(t) + (36 \lambda V_x / \alpha + 72 V_x V_{xx} / \alpha^2
 + 3 V_{xxxx} / \alpha)/ \beta,
\\ 
X_1 & : & b = g_4(t) - 3 V_{xx} / \alpha +V_{xxx} / (3 \alpha)),
\\ 
X_0 & : & g_4=\lambda,
\end{array}
\end{equation}
and the result is, with $U=V_x$,
\begin{eqnarray}
{\hskip -5.0truemm}
b
& = &
\lambda - 3 U_x / \alpha,
\\
{\hskip -5.0truemm}
a
& = &
-6 U / \alpha,
\\
{\hskip -5.0truemm}
\partial_x \Log \tau & = &
{\lambda \over \psi_{xx} / \psi -(1/2) (\psi_{x} / \psi)^2 + 3 (U / \alpha)}.
\label{eqKKDLogtau}
\\
{\hskip -5.0truemm}
L_1
& = &
\partial_x^3 + 6 {U \over \alpha} \partial_x + 3 {U_x \over \alpha} - \lambda,
\\
{\hskip -5.0truemm}
L_2
& = &
\beta \partial_t
- 9 \lambda \partial_x^2
+\left(3 {U_{xx} \over \alpha} + 36 {U^2 \over \alpha^2}\right) \partial_x
-3 {U_{xxx} \over \alpha}
\nonumber
\\
{\hskip -5.0truemm}
& &
 - 72 {U U_x \over \alpha^2}
 - 36 \lambda {U \over \alpha},
\\
{\hskip -5.0truemm}
\beta [L_1,L_2]
& = &
(6/ \alpha) \hbox{KK}(U) \partial_x +(3/ \alpha) \hbox{KK}(U)_x.
\end{eqnarray}
This is the Lax pair given by Kaup \cite{Kaup1980}.
The integration of the first-order ODE (\ref{eqKKDLogtau}) 
{\it modulo} the Lax pair yields
the DT given by Levi and Ragnisco \cite{LR1988}~:
\begin{eqnarray}
\tau & = & \psi \psi_{xx} -(1/2) \psi_x^2 + 3 (U / \alpha) \psi^2,\
\tau_x = \lambda \psi^2.
\label{eqTauPsiLink2}
\end{eqnarray}

Although the relation
$\tau_x / \psi^2=$ constant 
is the same as in the case of KdV (see Eq.~(4.14) in Ref.~\cite{WTC}),
it cannot be taken as an {\it a priori} assumption,
it is the result of the method.

Starting from the {\it vacuum} solution $U=0$,
the general solution $\psi$ of $L_1 \psi=0, L_2 \psi=0$,
\begin{eqnarray}
& &
\psi=
 c_1 e^{    K x + 9     K^5 t / \beta}
+c_2 e^{j   K x + 9 j^2 K^5 t / \beta}
+c_3 e^{j^2 K x + 9 j   K^5 t / \beta},\
\nonumber
\\
& &
j^3=1,\
K^3=\lambda,
\end{eqnarray}
in which $c_1,c_2,c_3,K$ are arbitrary complex constants,
leads by (\ref{eqDT}) to the one-soliton solution of Kaup \cite{Kaup1980}
\begin{eqnarray}
& &
u=(\alpha/2) \partial_x^2 \Log (2 + \cosh (k/2) (x-(k/2)^4 t/ \beta)),\
k \in {\mathcal R}
\end{eqnarray}
for the choice
$(c_1,c_2,c_3)=(0,j^2,-j), K^2=-k^2/12$,
which corresponds to the entire function
\begin{eqnarray}
& &
\tau=-(k^2/12)
 (2 + \cosh (k/2) (x-(k/2)^4 t/ \beta)) e^{(k/2)(x + (k/2)^4 t / \beta)}.
\end{eqnarray}

Let us now obtain the auto-BT of KK, by an elimination.
In order to perform this elimination easily,
it is convenient to choose one of the two components 
of the pseudopotential $\bfY$
so as to characterize the DT,
\begin{eqnarray}
{\rm KK} &:& 
{2(v-V) \over \alpha}={\tau_x \over \tau}=Z.
\label{eqKKDT}
\end{eqnarray}
The chosen equivalent system
is the system  satisfied by $(Y_1,Z)$
\begin{eqnarray}
{\hskip -10.0 truemm}
Y_1 &=& {\psi_x \over \psi},\
Z={\tau_x \over \tau},
\\
{\hskip -10.0 truemm}
Y_{1,x} & = & - Y_1^2/2 + \lambda Z^{-1} - 3 U / \alpha,
\label{eqKKgradx1}
\\
{\hskip -10.0 truemm}
Z_{x} & = & 2 Y_1 Z - Z^2,
\label{eqKKgradx2}
\\
{\hskip -10.0 truemm}
\beta Y_{1,t} & = &
[9 \lambda Y_1^2/2
-(3 U_{xx} / \alpha + 36 (U/ \alpha)^2) Y_1
 + 9 \lambda^2 Z^{-1}
\nonumber
\\
{\hskip -10.0 truemm}
& &
+ 3 U_{xxx}/ \alpha + 72 U U_x / \alpha^2 + 9 \lambda U / \alpha]_x,
\label{eqKKgradt1}
\\
{\hskip -10.0 truemm}
\beta Z_{t} & = &
\big[
18 \lambda U / \alpha
+ 9 \lambda^2 Z^{-1}
+ 9 \lambda Y_1^2
\nonumber
\\
{\hskip -10.0 truemm}
& &
+ (45 (U/ \alpha)^2 
+ 6 (U_{xx} / \alpha)
- 18 (U_x / \alpha) Y_1
\nonumber
\\
{\hskip -10.0 truemm}
& &
+ 27 (U/ \alpha) Y_1^2 
+(9/4) Y_1^4) Z
\big]_x.
\label{eqKKgradt2}
\end{eqnarray}

The BT then arises from the elimination of $Y_1$ between 
(\ref{eqKKgradx1}), (\ref{eqKKgradx2}) and (\ref{eqKKgradt2})
(Eq.~(\ref{eqKKgradt1}) must be discarded),
which results in the two equations for $Z=Y$,
\begin{eqnarray}
& &
Y_{xx} - (3/4) Y_x^2/Y + 3 Y Y_x/2 + Y^3/4 + 6 (U / \alpha) Y - 2 \lambda=0,
\\
& &
\beta Y_t -(3/16)
[3 Y^5 + 15 Y Y_x^2 + 30 Y^2 Y_{xx} + 8 Y_{xxxx}
\nonumber
\\
& &
+ 30 (Y^3 + 2 Y_{xx})(Y_x + 4 V_x/ \alpha)
+ 60 Y (Y_x + 4 V_x / \alpha)^2
\nonumber
\\
& &
+ 30 Y_x (Y_{xx} + 4 V_{xx} / \alpha)
+ 20 Y (Y_{xxx} + 4 V_{xxx}/ \alpha)
]_x=0,
\\
& &
\beta((Y_{xx})_t-(Y_t)_{xx})/Y=-(6 / \alpha) \hbox{KK}(U),
\end{eqnarray}
followed by the substitution $Y=2 (v-V) / \alpha$,
\cite{MC1998}
\begin{eqnarray}
& &
{\hskip -11.0 truemm}
(v-V)_{xx} / \alpha -(3/4) (v-V)_x^2 / (\alpha (v-V))
\nonumber
\\
& &
{\hskip -11.0 truemm}
\phantom{xxx}
+ 3 (v-V) (v+V)_x / \alpha^2 + (v-V)^3/ \alpha^3 - \lambda=0,
\label{eqKKBTxvV}
\\
& &
{\hskip -11.0 truemm}
\beta (v-V)_t/ \alpha -(3/2)[
2 (v-V)_{xxxx} / \alpha 
+ 60 (v-V)^3 (v+V)_x / \alpha^4
\nonumber
\\
& &
{\hskip -11.0 truemm}
\phantom{xxx}
+ 12 (v-V)^5/ \alpha^5
+(10(v-V) (v+V)_{xxx} +30 (v+V)_x (v-V)_{xx}
\nonumber
\\
& &
{\hskip -11.0 truemm}
\phantom{xxx}
+15 (v-V)_x (v+V)_{xx})/ \alpha^2
+ (30 (v-V)^2 (v-V)_{xx}
\nonumber
\\
& &
{\hskip -11.0 truemm}
\phantom{xxx}
+ 60 (v-V) (v+V)_x^2 + 15 (v-V) (v-V)_x^2) / \alpha^3
]_x=0.
\label{eqKKBTtvV}
\end{eqnarray}
The simple form of the conservative equations
(\ref{eqSKBTtvV}) and (\ref{eqKKBTtvV})
results from the addition of suitable differential consequences of
(\ref{eqSKBTxvV}) and (\ref{eqKKBTxvV}).
The $x$-part of the BT has already be given by Rogers and Carillo
\cite{RogersCarillo} for $\lambda=0$.

If we write 
(\ref{eqKKgradx1})--(\ref{eqKKgradx2})
in the variables
$(Y_1,Z_2=Z^{-1})$,
\begin{eqnarray}
Y_{1,x} & = & - Y_1^2/2 + \lambda Z_2 - 3 U / \alpha,
\label{eqKKgradx1bis}
\\
Z_{2,x} & = & - 2 Y_1 Z_2 + 1,
\label{eqKKgradx2bis}
\end{eqnarray}
both systems
(\ref{eqProjRiccatiY1x})--(\ref{eqProjRiccatiY2x})
and
(\ref{eqKKgradx1bis})--(\ref{eqKKgradx2bis})
are coupled Riccati systems,
with the difference that
the transformation from $Y_1$ to $\psi_x / \psi$ is a point transformation
while the one      from $Z_2$ to $\psi_x / \psi$ is a contact one.
Thus,
the Riccati system (\ref{eqProjRiccatiY1x})--(\ref{eqProjRiccatiY2x}) is in
the classification of linearizable coupled Riccati systems given by Lie 
(this is the projective one),
while
the Riccati system (\ref{eqKKgradx1bis})--(\ref{eqKKgradx2bis}) is outside it.

A minor open problem is to find a bilinear BT for SK equation
(the one given in Ref.~\cite{MC1998} contains a nonbilinear term).
\index{B\"acklund transformation!bilinear}

%\vfill \eject

% ==========================================================================
\subsection{Nonintegrable equations, second scattering order}
\label{sectionOneFamSecondOrderNonintegrable}
\indent

Strictly speaking, nonintegrable equations have no associated scattering
order. 
What is meant in the title of this section is that one assumes a given
scattering order to process some nonintegrable PDEs.

For algebraic PDEs in two variables, 
particular solutions in which $(S,C)$ are constant are quite easy to find.
They correspond to solutions $u$ polynomial in
$\tanh {k \over 2} (x-ct-x_0)$.
The privilege of $\tanh$ is to be the general solution
of the \textit{unique} first order first degree nonlinear ODE with the PP, 
namely the Riccati equation
$\tanh' + \tanh^2 + S/2$,
in the particular case $S=$ constant.

A characteristic feature of nonintegrable equations is the absence of a BT.
Therefore, the iteration of Section \ref{sectionVariousLevels}
can only generate a finite number of new solutions 
\cite{CM1989,CT1989},
this will be seen on examples.
\index{B\"acklund transformation}

% ==========================================================================
\subsubsection{The Kuramoto-Sivashinsky equation}
\label{sectionKS}
\indent

It is worth to handle this example in detail because it exhibits 
all the features of what should be done and more importantly 
of what should \textit{not} be done when solving truncation equations.

The equation of Kuramoto and Sivashinsky (\ref{eqKS})
(notation $\mu = 19 \mu'$) 
possesses a single family \cite{Nozaki1987,FournierSpiegelThual}
\index{Kuramoto-Sivashinsky equation}
\begin{equation}
p=-3,\ q=-7,\ u_0=120 \nu,\ 
\hbox{ indices } -1,6,{13 \pm i \sqrt{71} \over 2},
{\mathcal D}=60 \nu \partial_x^3 + 60 \mu' \partial_x.
\end{equation}
and the orthogonality condition at index $6$ is satisfied.
Since equation (\ref{eqKS}) is a conservation law,
we therefore study it on its potential form
\begin{equation}
\label{eqKS1}
E \equiv v_t+{v_x^2 \over 2} + \mu v_{xx} + \nu v_{xxxx} + G(t)=0,\ u=v_x,
\end{equation}
which has the unique family
\begin{equation}
p=-2,\ q=-6,\ v_0=-60 \nu,\ 
\hbox{ indices } -1,2,{13 \pm i \sqrt{71} \over 2},\
{\mathcal D}=60 \nu \partial_x^2 + 60 \mu'.
\end{equation}
Although the no-log condition at index $i=2$ is not satisfied,
the $\psi$-series
\index{$\psi$-series}
\begin{equation}
\label{eqKS4}
v=-60 \nu \chi^{-2} + 60 \mu' \Log \psi + v_2 + 0(\chi),\
v_2 \hbox{ arbitrary function},
\end{equation}
in which the gradients of $\psi$ and $\chi$ are given by
(\ref{eqPsiX})--(\ref{eqPsiT}) and (\ref{eqChix})--(\ref{eqChit}),
contains one logarithm only,
which cancels by derivation.

The one-family truncation assumption is
\begin{equation}
\label{eqKS2}
v_T=v_0 \chi^{-2} + v_1 \chi^{-1} + v_{02} \Log \psi + v_2,\
v_{02} \hbox{ constant}
\end{equation}
equivalent to a truncated series $(-3:0)$ for $u$.
Substituting (\ref{eqKS2}) in (\ref{eqKS1}) and eliminating any derivative of
$\chi$ and $\psi$, one obtains
\begin{equation}
\label{eqKS3}
E = \sum_{j=0}^6 E_j \chi^{j-6}.
\end{equation}
Together with the identity (\ref{eqCrossXT}),
this defines a system of eight equations in the six unknowns 
$(v_0,v_1,v_{02},v_2,S,C)$.

Equations $j=0,1,2$ 
are solved for $v_0, v_1, v_{02}, v_2$ exactly as in the Painlev\'e
test and yield
the values in (\ref{eqKS4}).
The next five equations ($j=3,4,5,6$ and (\ref{eqCrossXT}))
now read \cite{CM1989}
\begin{eqnarray}
E_3
& \equiv &
120 \nu \left(- C + 15 \nu S_x + v_{2,x} \right) =0,
\\
E_4
& \equiv &
60 \left(
- 6 \nu^2 S_{xx} - 4 \nu^2 S^2 - 20 \mu' \nu S + 2 \nu C_x
+ 11 \mu'^2 \right) =0,
\\
E_5
& \equiv &
{S \over 2} E_3
+ 60 \left(
- \mu' C 
+ 20 \mu' \nu S_x 
- 2 \nu^2 S S_{x} 
+ \nu^2 S_{xxx}
\right.
\nonumber
\\
& &
\left.
- \nu C_{xx}
+ \mu' v_{2,x} \right) =0,
\\
E_6
& \equiv &
E(v_2)
+ 30 \left(
     \mu' C_x 
- 19 \mu'^2 S 
-    \mu' \nu (20 S^2 + S_{xx})
\right.
\nonumber
\\
& &
\left.
-         \nu^2 (4 S^3 + 3 S_x^2 + 4 S S_{xx})
\right) =0,
\nonumber
\\
X
& \equiv &
S_t + C_{xxx} + 2 C_x S + C S_x=0.
\end{eqnarray}
The principles to be obeyed during the resolution are the following.
\begin{enumerate}
\item
Never increase the differential order of a given variable.
On the contrary,
solve for the higher derivatives in terms of the lower ones,
and substitute the result, as well as its differential consequences,
in the remaining equations.

\item
Never integrate a differential equation,
unless it is just a total derivative.
On the contrary, perform an {\it algebraic} resolution.

\item
Never solve for a function of, say, one variable as an expression in
several variables.

\item
Close the solution by exhausting all Schwarz cross-derivative conditions.

\end{enumerate}

This computation is systematic,
and its algorithmic version is known as 
the construction of a differential Groebner basis 
\cite{MansfieldThesis,BoulierThese}.
\index{differential Groebner basis}

The full system is split into $(E_3,E_4,E_5)$, independent of $\partial_t$,
and $(E_6,X)$, explicitly depending on $\partial_t$.
The subsystem $(E_3,E_4)$ is first solved, according to rule 1,
as a Cramer system for $(v_{2,x},S_{xx})$.
After substitution of  $(v_{2,x},S_{xx})$ and their derivatives in 
all the other equations,
equation $E_5$ is solved, according to rule 1, for $C_{xx}$
and the result is recognized as being an $x-$derivative.
This allows us to solve for $C_x$ after the introduction of an arbitrary 
integration function $\lambda$ of $t$ only.
As to $(E_6,X)$,
they are solved, according to rule 1,
as a Cramer system in variables involving only $t$-derivatives,
namely $(v_{2,t},S_t)$,
for expressions independent of $\partial_t$.
To summarize this first stage,
the original system is now equivalent to 
\begin{eqnarray}
S_{xx}
& = &
- {3 \over 2} S^2 
- {5 \mu' \over 2 \nu} S 
+ {\mu'^2 \over 8 \nu^2} (\lambda + 22),
\label{eqKS03}
\\
C_x
& = &
- {5 \nu \over 2} S^2 
+ {5 \mu' \over 2} S 
+ {\mu'^2 \over 8 \nu} (3 \lambda + 22),
\label{eqKS04}
\\
v_{2,x}
& = &
 C - 15 \nu S_x,
\\
{v_{2,t} \over \nu}
& = &
- {   1         \over   \nu  } G(t)
+ {1243  \mu'^3 \over 2 \nu^2}
+    16 {\mu'^3 \over   \nu^2} \lambda
+    10 {\mu'^2 \over   \nu  } (\lambda - 2) S
\nonumber
\\
& &
+ 110 \mu' S^2
 - {1 \over 2 \nu} C^2
+ 15 C S_x 
- {125 \nu \over 2} S_x^2,
\\
S_t
& = &
- {5 \mu'^3 \over 16 \nu^2} (\lambda + 22)
- {  \mu'^2 \over  8 \nu  } (\lambda - 116) S
- {55 \mu' \over 4} S^2
- {5  \nu \over 2} S^3
\nonumber
\\
& &
- C S_x + 5 \nu S_x^2.
\end{eqnarray}
One equation, and only one, namely (\ref{eqKS03}), is an ODE.
Integrating it as an elliptic ODE for $S$ \cite{CM1989}
would create useless subcases and complications
and should, according to rule 2, \textit{not} be done.
This ODE should also \textit{not} be replaced by its first integral,
because the integrating factor $S_x$ could be, and will indeed be, zero.
According to rule 3, it is also forbidden to eliminate $\lambda(t)$
by solving e.g.~(\ref{eqKS04}) for it.
The only thing to do is (rule 4) 
to close this solution by cross-differentiation.
There are two such conditions:
\begin{eqnarray}
(S_{xx})_t-(S_t)_{xx}
& \equiv &
- {   \mu'^4 \over 64 \nu^3} (3 \lambda^2 + 308 \lambda + 5324)
+ { 5 \mu'^3 \over  8 \nu^2} (15 \lambda + 374) S
\nonumber
\\
& &
+ { 25 \mu'^2 \over 8 \nu} (\lambda - 16) S^2
- {165 \mu'   \over 2} S^3
- { 75 \nu    \over 4} S^4
\nonumber
\\
& &
+ {   \mu'^2 \over 8 \nu^2} \lambda'
+ {85 \mu'   \over 2}       S_x^2
+  25 \nu                   S S_x^2=0,
\label{eqKS5}
\\
(v_{2,t})_x-(v_{2,x})_t
& \equiv &
- C_t
-  { \mu'^2 \over 8 \nu} (3 \lambda + 22) C
- {5 \mu'   \over 2} S C
+ {5 \nu    \over 2} S^2 C
\nonumber
\\
& &
+ {15 \mu'^2 \over 4} (3 \lambda + 71) S_x
- 255 \mu' \nu S S_x
\nonumber
\\
& &
- 150 \nu^2 S^2 S_x = 0.
\end{eqnarray}

The latter is solved for $C_t$ and provides a third cross-derivative condition
\begin{eqnarray}
{(C_{x})_t-(C_t)_{x} \over \nu}
& \equiv &
- {  \mu'^4 \over 64 \nu^3} (81 \lambda^2 + 4028 \lambda + 47476)
\nonumber
\\
& &
+ {5 \mu'^3 \over  8 \nu^2} (101 \lambda + 2322) S
+ {   5 \mu'^2 \over 8 \nu  } (55  \lambda + 96) S^2
\nonumber
\\
& &
- {1415 \mu'   \over 2} S^3
- { 825 \nu    \over 4} S^4
\nonumber
\\
& &
+ {3 \mu'^2 \over 8 \nu^2} \lambda'
+ {535 \mu' \over 2} S_x^2
+ 275 \nu S S_x^2 = 0.
\label{eqKS6}
\end{eqnarray}

This ends the linear part of the resolution,
and now comes the nonlinear part (algebraic Groebner).
The two remaining equations (\ref{eqKS5}) and (\ref{eqKS6}),
considered as nonlinear in the two unknowns $(S_x,S)$,
imply without computation $S_x=0$,
which then allows one to solve (\ref{eqKS03}) for the monomial $S^2$
as a polynomial in $S$ of a smaller degree.
Equations (\ref{eqKS5}) and (\ref{eqKS6}) thus become linear in 
$S$ and $\lambda'$,
and their resultant in $\lambda'$ factorizes as a product of \textit{linear}
factors
\begin{equation}
(\lambda + 33)
(\lambda - 11 + {40 \nu \over \mu'} S) = 0.
\end{equation}
The first factor yields no solution.
The second one provides the unique solution
\begin{equation}
\left(S- {\mu \over 38 \nu} \right) \left(S+ {11 \mu \over 38 \nu} \right)=0,\
C=\hbox{ arbitrary constant } c
\label{eqKSSME}
\end{equation}
and it leads to the two-parameter $(c,x_0)$ solution
(\ref{eqKuramotoTsuzuki}).
The two equations (\ref{eqKSSME}) represent the SME.
 \index{singular manifold!equation}

If one performs the iteration of Section \ref{sectionVariousLevels},
starting from $u=c$, 
one generates the solitary wave (\ref{eqKuramotoTsuzuki}) 
and no more \cite{CM1989}.

The reduction $u(x,t) = c + U(\xi), \xi=x-ct $ of
the PDE (\ref{eqKS}) yields the ODE
\begin{equation}
 \nu U''' + \mu U' + U^2/2 + K = 0,\ K \hbox{ arbitrary},
\label{eqKSRed}
\end{equation}
with the indices $-1, (13 \pm i \sqrt{71})/2$.
Due to the two irrational indices,
the \textit{general analytic solution}
(see definition in Section \ref{sectionCGL3})
can only depend on one arbitrary constant.
This one-parameter solution,
whose local expansion contains no logarithm,
is known globally only for $ K= -450 \nu k^2/(19^2 \mu)$,
this is (\ref{eqKuramotoTsuzuki}),
but its closed form expression for any $K$ is still an open problem.
 \index{general analytic solution}

Being autonomous, the ODE (\ref{eqKSRed}) is equivalent to the nonautonomous
second order ODE for $V(U)$
\begin{equation}
V=\frac{\D U}{\D \xi}\ :\
 \nu \frac{\D^2 (V^2)}{\D U^2} + 2 \mu + \frac{U^2 + 2 K}{V}=0,
\end{equation}
an equation which has been studied from the Hamiltonian point of view
\cite{Bouquet1995}.

%\vfill \eject

% ==========================================================================
\subsection{Nonintegrable equations, third scattering order}
\label{sectionOneFamThirdOrderNonintegrable}
\indent

An example is given in the Section \ref{sectionKPP}.

% ==========================================================================
\section{Two common errors in the one-family truncation}
\indent

Two errors are
frequently made in the method of section \ref{sectionTruncationOneFamily}.

% ==========================================================================
\subsection{The constant level term does not define a BT}
\indent

Consider the one-family truncation as done by WTC
(the subscript $T$ means ``truncated'')
\begin{equation}
u_T^{\rm WTC}=\sum_{j=0}^{-p} u_j^{\rm WTC} \varphi^{j+p}
\label{eqWeissTruncationu}
\end{equation}
in which $\varphi$ is the function defining the singularity manifold.

In the WTC truncation,
one considers three solutions of the PDE
\begin{enumerate}
\item
the lhs $u_T^{\rm WTC}$ of the truncation (\ref{eqWeissTruncationu}),

\item
the ``constant level'' coefficient $u_{-p}^{\rm WTC}$,

\item
the field $U$ which appears in the Lax pair after the successful completion
of the method.

\end{enumerate}

The frequently encountered argument
``The constant level coefficient $u_{-p}^{\rm WTC}$ also satisfies the PDE,
therefore one has obtained a BT''
is wrong.
This is obvious, since nonintegrable PDEs, which have no BT,
nevertheless have this property.
One can check it by taking the explicit example of a nonintegrable
PDE \cite{CM1989}.

A hint that the above argument might be wrong is the fact,
observed on all successful truncations,
that the $U$ in the Lax pair is \textit{never} $u_{-p}^{\rm WTC}$.
Let us prove this fact, with the homographically invariant analysis 
\cite{Conte1989}.
The truncation of the same variable in the invariant formalism is
\begin{equation}
u_T=\sum_{j=0}^{-p} u_j \chi^{j+p},
\label{eqInvarTruncationu}
\end{equation}
in which $\chi$ is given by
(\ref{eqchi}).
This $u_T$ depends on the movable constant $\varphi_0$ and one has
\begin{eqnarray}
& &
\left\lbrace
\begin{array}{ll}
\displaystyle{
u_T^{\rm WTC} = u_T (\varphi_0=0)
}
\\
\displaystyle{
u_{-p}^{\rm WTC} = u_T (\varphi_0=\infty).
}
\end{array}
\right.
\end{eqnarray}
Since the results of the truncation do not depend on the movable constant
$\varphi_0$,
this proves that the lhs $u_T^{\rm WTC}$ of the truncation
and the constant level coefficient $u_{-p}^{\rm WTC}$
are not considered as distinct by the singular manifold method.
Since the $U$ in the Lax pair cannot be the truncated $u$
(otherwise one would not have a Darboux transformation),
this ends the proof.

% ==========================================================================
\subsection{The WTC truncation is suitable iff the Lax order is two}
\label{sectionWTCSuitable}
\indent

We mean the truncation as originally introduced,
not its updated version of Section \ref{sectionTruncationOneFamily}.

When the Lax pair has second order, everything is consistent.
When the Lax pair has a higher order, e.g.~three,
the original method,
as well as its original invariant version \cite{MC1991},
presents the following inconsistency.
In a first stage,
it generates the $-q+p$ equations $E_j(S,C,U)=0$
of formula (\ref{eqTruncationOrder2}),
which intrinsically correspond to a \textit{second}-order scattering problem
(and this is precisely the inconsistency),
and in a second stage it injects in each of these $-q+p$ equations
a link between $(S,C)$ and the scalar field $\psi$ of the Lax pair of
\textit{higher} order,
thus generating determining equations which are hybrid between the second order
and the higher one.
The first nearly correct treatment has been made in 
Ref.~\cite{MCGallipoli1991}.

For the same reason, 
in order to obtain the Lax pair when its order is higher than two,
it is also inconsistent to consider the so-called
\textit{singular manifold equation} (SME) \cite{WTC,Conte1989,Pickering1996},
defined in Section \ref{sectionOneFamSecondOrderIntegrable}.
 \index{singular manifold!equation}
When the Lax order is three,
the correct extension of this SME notion would be the set of
three relations on $(a,b,c,d)$ resulting from the 
elimination of $U$ between the four coefficients of the Lax pair
($e$ is derivable from (\ref{eqX2}) so we discard it),
but this seems of little interest.

Although these inconsistencies may still provide the full result
for some ``robust'' equations 
(Boussinesq \cite{Weiss1985Bq}, 
 Sawada-Kotera \cite{Weiss1984KKSK},
 Hirota-Satsuma \cite{MC1991}),
there do exist equations for which it leads to a failure,
and the Kaup-Kupershmidt equation \cite{MC1998} is one of them.

\index{Boussinesq equation}
\index{Hirota-Satsuma equation}
\index{Sawada-Kotera equation}
\index{Kaup-Kupershmidt equation}
% ==========================================================================
\section{The singular manifold method applied to two-family PDEs}
\label{sectionTruncationTwoFamilies}
\indent

By two-family, we mean two opposite families.

This includes also the one-family truncation as a particular case.

When the base member of the hierarchy of integrable equations has more than
a single family,
these families usually come by pairs of opposite singular part operators,
just like (P2)--(P6).
Examples are enumerated at the end of Section \ref{sectionTransposition}.
Then the sum of the two opposite singular parts
\begin{eqnarray}
& &
{\mathcal D} \Log \tau_1 - {\mathcal D} \Log \tau_2
\end{eqnarray}
only depends on the variable 
\begin{eqnarray}
& &
Y={\tau_1 \over \tau_2}.
\label{eqYtau12}
\end{eqnarray}

The current status of the method \cite{MC1994,Pickering1996},
which used to be called the two-singular manifold method \cite{MC1994},
is as follows.
Most of the method for one-family equations still applies,
with the difference that it is much more convenient to represent the
Lax pairs in a Riccati form than in a scalar linear form.
Let us restrict here to second-order scattering problems
(for the third order case, see Section \ref{sectionTwoFamThirdOrderIntegrable})
and to identity links (\ref{eqTauPsiLink1}) between the two $\tau$ and the
two $\psi$ functions.
Then $Y$ satisfies a Riccati system and, as explained in Section
\ref{sectionWhere},
its most general expression is given by (\ref{eqMostGeneralY}).

In the first step, $\tau$ is simply replaced by $Y$ 
in the assumption (\ref{eqDT}) for a DT.

In the second step, the scattering problem is represented by the Riccati
system satisfied by $Y$,
whose coefficients depend on $(S,C,A,B)$.

The fourth step contains the main difference.
Rather than truncating $u$ at the level $j=-p$,
one truncates it at the level $j=-2p$ 
\cite{MC1994,Pickering1996},
in order to implement the two movable singularities $\tau_1=0$ and $\tau_2=0$.
So the truncation is
\cite{Pickering1996}
(for second order Lax pairs only)
\begin{eqnarray}
& &
u={\mathcal D} \Log Y + U,\
\label{eqTruncationTwoFamily1}
\\
& &
Y^{-1}=B(\chi^{-1} + A),\
\\
& &
E(u)=\sum_{j=0}^{-2 q} E_j(S,C,A,B,U) Y^{j+q},\
\\
& &
\forall j\
E_{j}(S,C,A,B,U) =0,
\label{eqTruncationTwoFamily4}
\end{eqnarray}
in which nothing is imposed on $U$.

Let us remark that the relation $A\not=0$ does not characterize two-family
PDEs,
see the Liouville case in Section \ref{sectionSG}.

% ==========================================================================
\subsection{Integrable equations with a second order Lax pair}
\label{sectionTwoFamSecondOrderIntegrable}
\indent

% ==========================================================================
\subsubsection{The sine-Gordon equation}
\label{sectionSG}
\indent

The sine-Gordon equation is defined for convenience as the case 
$a_1 \not=0, a_2=0$ of the equation (\ref{eqZS}).
Although not algebraic in $u$,
it becomes algebraic in $e^u$
and it possesses
two opposite families (opposite in the field $u$),
both with $p=-2,q=-2$
\index{sine-Gordon equation}
\begin{eqnarray}
{\hskip -4.0 truemm}
& &
e^u \sim -(2/ \alpha) \varphi_x \varphi_t (\varphi - \varphi_0)^{-2},\
\hbox{indices } (-1,2),\
{\mathcal D}=(2/ \alpha) \partial_x \partial_t.
%\label{eqFamilyOne}
\\
{\hskip -4.0 truemm}
& &
e^{-u} \sim (2/ a_1)  \varphi_x \varphi_t (\varphi - \varphi_0)^{-2},\
\hbox{indices } (-1,2),\
{\mathcal D}=-(2/ a_1) \partial_x \partial_t.
%\label{eqFamilyTwoSG}
\end{eqnarray}
The resulting DT assumption
\begin{eqnarray}
& &
e^u  + (a_1 / \alpha) e^{-u}=(2/ \alpha) \partial_x \partial_t \Log Y 
+ \tilde{W},\
E(u)=0
\end{eqnarray}
with $Y$ defined by (\ref{eqYtau12}),
can be integrated twice due to the special form of the PDE, resulting in
\begin{eqnarray}
& &
u=-2 \Log Y + W,\ E(u)=0,
\end{eqnarray}
in which nothing is imposed on $W$ (we use $W$ to reserve the symbol $U$
for future use).
For $a_1=0$, 
this truncation is what was called in Section \ref{sectionLiouville}
the second truncation of Liouville equation.
\index{Liouville equation}

The five determining equations in the unknowns $(S,C,A,B,W)$ are
\cite{Pickering1996,CMG1999} % CMG1999 eqs (55)--(70)
\begin{eqnarray}
& &
E_0 \equiv \alpha B^2 e^W - 2 C=0,
\\
& &
E_1 \equiv 2 (C_x + 2 A C)=0,
\\
& &
E_2 \equiv 0, \hbox{ (Fuchs index)}
\\
& &
E_3 \equiv - \sigma_t - \sigma (C_x + 2 A C)=0,
\\
& &
E_4 \equiv \sigma \left(C \sigma + (C_x + 2 A C)_x \right)/2
 + a_1 B^{-2} e^{-W}=0,
\end{eqnarray}
with the abbreviation
\begin{eqnarray}
& &
\sigma=S + 2 A^2 - 2 A_x,
\label{eqsigma}
\end{eqnarray}
and, together with the cross-derivative condition (\ref{eqCrossXT}),
they are solved as usual by ascending values of $j$
\begin{eqnarray}
& &
E_0:\
B^2 e^W ={2 \over \alpha} C,
\\
& &
E_1:\
A=-{1 \over 2} (\Log C)_x,\
\\
& &
E_3:\
S=-F(x) + {C_x^2 \over 2 C^2} - {C_{xx} \over C},
\\
& &
E_4:\
C C_{xt} - C_x C_t + F(x) C^3 + a_1 \alpha F(x)^{-1} C=0,
\\
& &
X:\
a_1 F'(x)=0.
\end{eqnarray}
in which $F$ is a function of integration.
For sine-Gordon, $F(x)$ must be a constant
\begin{eqnarray}
& &
F(x)=2 \lambda^2.
\end{eqnarray}
In the Liouville case, 
for which the truncation imposes no restriction on $F(x)$,
let us also require that $F(x)$ be a constant.
Then, for both equations,
$\Log C$ is proportional to a second solution $U$ of the PDE
\begin{eqnarray}
& &
C={\alpha \over 2} \lambda^{-2} e^{U},\
E(U)=0,
\end{eqnarray}
and one has obtained the Darboux transformation 
\begin{eqnarray}
& &
u = -2 \Log y + U,\
y=\lambda B Y,
\label{eqDTLiouvilleandSG}
\end{eqnarray}
in which $y$ satisfies the Riccati system 
\begin{eqnarray}
& &
y_x = \lambda +  U_x y - \lambda y^{2},
\label{eqRiccatiZx}
\\
& &
y_t = -{\alpha \over 2} \lambda^{-1} (e^{U} + (a_1 / \alpha) e^{-U} y^2),
\label{eqRiccatiZt}
\\
& &
(\Log y)_{xt} - (\Log y)_{tx} = E(U).
\end{eqnarray}
The linearization 
\begin{eqnarray}
& &
y=\psi_1 / \psi_2
\label{eqytopsi}
\end{eqnarray}
yields the second-order matrix Lax pair 
\begin{eqnarray}
& &
(\partial_x - L) \pmatrix{\psi_1 \cr \psi_2 \cr}=0,\
L=
\pmatrix{U_x/2 & \lambda \cr \lambda & - U_x/2 \cr},\
\\
& &
(\partial_t - M) \pmatrix{\psi_1 \cr \psi_2 \cr}=0,\
M= - (\alpha/2) \lambda^{-1}
 \pmatrix{0 & e^{U} \cr - (a_1 / \alpha) e^{- U} & 0 \cr}.
\label{eqSGLaxMatricial}
\end{eqnarray}

The auto-BT 
(classical for sine-Gordon, Ref.~\cite{McLaughlinScott} for Liouville)
results from the substitution
$y= e^{-(u - U)/2}$ into (\ref{eqRiccatiZx})--(\ref{eqRiccatiZt})
\begin{eqnarray}
& &
(u+\tilde{U})_x = - 4 \lambda \sinh {u - \tilde{U} \over 2},\
\\
& &
(u-\tilde{U})_t = \lambda^{-1} 
\left(\alpha e^{(u + \tilde{U})/2} + a_1 e^{-(u + \tilde{U})/2}\right).
\end{eqnarray}
It coincides in the sine-Gordon case with the one given earlier,
equations (\ref{eqSGBTx})--(\ref{eqSGBTt}).
The ODE part (\ref{eqRiccatiZx}) of the BT is a Riccati equation.

The SME is \cite{Pickering1996}
 \index{singular manifold!equation}
\begin{eqnarray}
& &
S+C^{-1} C_{xx} - {1 \over 2} C^{-2} C_x^2 + 2 \lambda^2=0,
\end{eqnarray}
and it coincides, but this is not generic,
with the one \cite{Weiss1984c,Conte1989}
obtained from the (incorrect) truncation in $\chi$.

{\it Remarks}.
\begin{enumerate}
\item
The reason for the presence of the apparently useless parameter $B$ in the
definition (\ref{eqMostGeneralY}) is to allow the precise correspondence
(\ref{eqTauPsiLink1})
\begin{eqnarray}
& &
\tau_1=\psi_1,\
\tau_2=\psi_2
\end{eqnarray}
for some choice of $B$, namely
\begin{eqnarray}
& &
B=\lambda^{-1},\
y=Y,\
W=U.
\end{eqnarray}

\item
In the Liouville case $a_1=0$,
this is an example of a PDE with only one family and a nonzero value of
$A$.
\index{Liouville equation}

\end{enumerate}
%\vfill \eject

% ==========================================================================
\subsubsection{The modified Korteweg-de Vries equation}
\label{sectionmKdV}
\indent

This PDE has the same scattering problem as sine-Gordon,
so the computation should be, and indeed is,
quite similar to that for sine-Gordon.

\index{Korteweg-de Vries equation!modified}

Since this PDE has the conservative form
\begin{eqnarray}
& &
\hbox{mKdV}(w) \equiv 
b w_t + \left(w_{xx} -2 (w - \beta)^3 / \alpha^2 + 6 \nu w \right)_x=0,\
w=r_x,
\label{eqmKdV}
\end{eqnarray}
it is technically cheaper to process its potential form
\begin{eqnarray}
{\hskip -4.0 truemm}
& &
\hbox{p-mKdV}(r) \equiv 
b r_t + r_{xxx} - 2 (r_x - \beta)^3 / \alpha^2 + 6 \nu (r_x - \beta)
+ F(t) = 0.
\end{eqnarray}
Its invariance under the involution $w - \beta \mapsto -(w - \beta)$
provides an elegant way \cite{CetraroMagri}
to derive the BT of the KdV equation and its hierarchy.
Although the constants $\beta$ and $\nu$ could be set to zero by a 
transformation on $(r,x,t)$ preserving the PP,
it is convenient to keep them nonzero,
for reasons explained at the end of this Section.
This last PDE admits the two opposite families
($\alpha$ is any square root of $\alpha^2$)
\begin{eqnarray}
& &
p=0^{-},\
q=-3,\
r \sim \alpha \Log \psi,\
\hbox{indices } (-1,0,4),\
{\mathcal D} = \alpha.
\end{eqnarray}

The truncation is defined by
\begin{eqnarray}
& &
r=\alpha \Log Y + R,\
\end{eqnarray}
with (\ref{eqTruncationTwoFamily1})--(\ref{eqTruncationTwoFamily4}),
and this generates five equations $E_j=0$ \cite{Pickering1996},
with the notation (\ref{eqsigma}) for $\sigma$
\begin{eqnarray}
{\hskip -7.0 truemm}
E_1 & \equiv & 6 \alpha A - 6 ((R-\alpha \Log B)_x-\beta)=0,
\\
{\hskip -7.0 truemm}
E_2 & \equiv & \alpha (2 A_x + 4 A^2 -b C - 2 \sigma + 6 \nu)
- \alpha^{-1} (E_1 + 6 \alpha A)^2/6 
=0,
\\
{\hskip -7.0 truemm}
E_3 & \equiv & 
\hbox{p-mKdV}(R- \alpha \Log B)
-(3/2) \alpha^{-1} \sigma_x
\nonumber
\\
{\hskip -7.0 truemm}
& &
+(\sigma -4 A^2 -(1/3) \alpha^{-1} E_{1,x} -2 A_x) E_1
-2 A E_{1,x} -2 A E_2 -E_{2,x}
\\
{\hskip -7.0 truemm}
E_4 & \equiv & \hbox{expression vanishing with } E_1,E_2,E_3,E_5,
\\
{\hskip -7.0 truemm}
E_5 & \equiv & (3/4) \alpha \sigma \sigma_x + (1/4) \sigma^2 E_1=0,
\\
{\hskip -7.0 truemm}
X & \equiv & = S_t + C_{xxx} + 2 C_x S + C S_x = 0.
\end{eqnarray}
They depend on $(R,B)$ only through the combination $R - \alpha \Log B$.
Equation $j=4$ is a differential consequence of equations $j=1,2,3,5$,
because $4$ is a Fuchs index,
and the other equations have been written so as to display how they are
solved:
\begin{eqnarray}
& &
E_1:\
A=\alpha^{-1}((R - \alpha \Log B)_x - \beta),
\\
& &
E_5:\
\sigma=-2(\lambda(t)^2 - \nu),\ \lambda \hbox{ arbitrary function},
\\
& &
E_2:\
b C= 2 A_x - 2 A^2 + 4 \lambda(t)^2 + 2 \nu,
\\
& &
E_3:\
\hbox{p-mKdV}(R- \alpha \Log B)=0,
\\
& &
X:\
\lambda'(t)=0.
\end{eqnarray}
Thus, their general solution can be expressed in terms of a second solution
$W$ of the mKdV equation (\ref{eqmKdV}) 
and an arbitrary complex constant $\lambda$ \cite{Pickering1996}
\begin{eqnarray}
W & = & (R- \alpha \Log B)_x,\
A = (W - \beta) / \alpha,\
\nonumber
\\
b C & = & 2 W_{x} / \alpha -2 (W - \beta)^2 / \alpha^2 + 2 \nu +4 \lambda^2,\
\nonumber
\\
S & = &   2 W_{x} / \alpha -2 (W - \beta)^2 / \alpha^2 + 2 \nu -2 \lambda^2,\
\label{eqmKdVSCAB}
\end{eqnarray}
and the cross-derivative condition $X_1=0$ (Eq.~(\ref{eqOrder2GDX1})),
equivalent to the mKdV equation (\ref{eqmKdV}) for $W$,
proves that one has obtained a Darboux transformation and a Lax pair.

The SME, obtained by the elimination of $W$ between $S$ and $C$,
\index{singular manifold!equation}
\begin{eqnarray}
& &
b C - S - 6 \lambda^2 = 0,
\end{eqnarray}
is identical to that of the KdV equation (\ref{eqSMEKdV}).
\index{Korteweg-de Vries equation}

The auto-BT of mKdV
\index{B\"acklund transformation!auto--}
is obtained by the substitution
\begin{eqnarray}
& &
\Log (B Y) = \alpha^{-1} \int (w-W) \D x
\end{eqnarray}
in the two equations for the gradient of $ y=B Y$
\begin{eqnarray}
& &
{y_x \over y}
=
\lambda ({1 \over y} - y) - 2 {W - \beta \over \alpha},\
\\
& &
b {y_t \over y}
=
 {1 \over y} \left(
- 4 \lambda {W - \beta \over \alpha}
+ (2 {(W - \beta)^2 \over \alpha^2}
+ 2 {W_{x} \over \alpha}
- 4 \lambda^2) y
 \right)_x.
\end{eqnarray}

In the same manner as in the KdV truncation,
these two Riccati equations can also be interpreted as the hetero-BT
\index{B\"acklund transformation!hetero--}
between the mKdV equation and the PDE satisfied by the pseudopotential $y$,
called the Chen-Calogero-Degasperis-Fokas PDE.

% ==========================================================================
\subsubsection{The nonlinear Schr\"odinger equation}
\label{sectionNLS}
\indent

\index{nonlinear Schr\"odinger equation}
\index{AKNS system}

For the AKNS system of two second order equations in $(u,v)$
(whose reduction $\bar u=v$ is NLS, see (\ref{eqNLS11})),
no two-family truncation has yet been defined which strictly follows the 
method and provides the desired result.
It should be noted that the fourth order equation for $u$ resulting from the
elimination of $v$,
known as the Broer-Kaup equation or classical Boussinesq system,
admits a two-family truncation without any problem \cite{CMP1995}.
\index{Broer-Kaup equation}

The full result (DT, BT) can be found \cite{CMGalli95} for the AKNS system
by performing the one-family truncation \cite{Weiss1985Bq}
and then applying four involutions to the result of Weiss.

A second open problem for this PDE is that its bilinear BT is not yet known.
\index{B\"acklund transformation!bilinear}

% ==========================================================================
\subsection{Integrable equations with a third order Lax pair}
\label{sectionTwoFamThirdOrderIntegrable}
\indent

In principle,
there is no additional difficulty to extend the method to a scattering order
higher than two.
A good equation to process would be the modified Boussinesq equation 
\index{Boussinesq equation!modified}
\begin{eqnarray}
& &
E \equiv \left\lbrace
\begin{array}{ll}
\displaystyle{
-u_t + (v-(3/2) a^2 u^2)_x=0,\
}
\\
\displaystyle{
-v_t - 3 a^2 (u_{xx} - u v + a^2 u^3)_x=0,
}
\end{array}
\right.
\end{eqnarray}
which has two opposite families
\begin{eqnarray}
& &
u \sim (2/a) \chi^{-1},\
v \sim 6 \chi^{-2}
\end{eqnarray}
and a third order Lax pair like the Boussinesq equation.

The one-family assumption \cite{FP1999}
\begin{eqnarray}
{\hskip -5.0 truemm}
& &
u = U + (2/a) \partial_x   \Log \tau,\
v = V -6      \partial_x^2 \Log \tau,\
E(u,v)=
E(U,V)=0,\
\label{eqDTMBqOneFamily}
\end{eqnarray}
with the identity link $\tau=\psi$ and the choice 
of the scalar Lax pair (\ref{eqLaxScalar3xG})--(\ref{eqLaxScalar3tG}),
already leads to the solution % unique? No precision in \cite{FP1999}
\begin{eqnarray}
& &
f=-(3/2) a U,\
a=(V-3 a^2 U^2 - 3 a U_x)/4,\
b=\lambda,\
\nonumber
\\
& &
c=-3 a,\
d=-3 a^2 U,\
e=0.
\label{eqMBqOneFamilyResult}
\end{eqnarray}

Despite this success,
it would be more consistent with the two-family structure
to process this PDE with a two-family assumption,
removing in passing the restriction $E(U,V)=0$ in (\ref{eqDTMBqOneFamily}).
This could make the coefficients $(f,a,b)$ linear in $(U,V)$,
which is not the case in (\ref{eqMBqOneFamilyResult}).

Table \ref{table} summarizes,
for a sample of PDEs,
the currently best method to obtain its Lax pair, Darboux and B\"acklund
transformations from a truncation.

\tabcolsep=1.5truemm
\tabcolsep=0.5truemm

\begin{table}[h] % [p]
\caption[garbage]{
The relevant truncation for some $1+1$-dimensional PDEs.
The successive columns are~:
the usual name of the PDE
(a p means the potential equation),
its number of families
(a * indicates that only one family is relevant, see details in Ref),
the order of its Lax pair,
the truncation variable(s),
% (notation $Y_1=\psi_x / \psi, Y_2=\psi_{xx} / \psi$,
% except for Tzitz\'eica $Y_2=\psi_t / \psi$),
the link between $\tau$ and $\psi$,
the singularity orders of $u$ and $E(u)$,
the Fuchs indices (without the ever present $-1$),
the number of determining equations,
the reference to the place where the right method was first applied
(earlier references may be found in it).
The ``?'' in the AKNS system entry 
(the one whose NLS is a reduction)
means that the method has not yet been applied to it, see text.
\index{Liouville equation}
\index{Korteweg-de Vries equation}
\index{AKNS equation}
\index{Korteweg-de Vries equation!modified}
\index{sine-Gordon equation}
\index{Broer-Kaup equation}
\index{Sawada-Kotera equation}
\index{Kaup-Kupershmidt equation}
\index{Tzitz\'eica equation}
\index{nonlinear Schr\"odinger equation}
\index{AKNS system}
{}
}
\vspace{0.2truecm}
\begin{center}
\begin{tabular}{| l | l | l | l | l | l | l | l | l |}
\hline % \hline % ********************************************************
\hline % \hline % ********************************************************
Name
&
f
&
Lax
&
Trunc.~var.
&
$\tau$
&
$-p:-q$
&
indices
&
nb. det. eq.
&
Ref
\\ \hline % \hline % ********************************************************
Liouville % Name
&
$1$     % Nb of families
&
        % Scattering order
&
$\tau$  % Truncation variable
&
        % Link between $\tau$ and $\psi$
&
$0:2$   % -p:-q
&
$2$   % Fuchs indices
&
$3$     % Number of determining equations 3, 5
&
\cite{CMG1999} % Ref
\\ \hline % \hline % ********************************************************
KdV     % Name
&
$1$     % Nb of families
&
$2$     % Scattering order
&
$\chi$  % Truncation variable
&
$\psi$  % Link between $\tau$ and $\psi$
&
$2:5$   % -p:-q
&
$4,6$   % Fuchs indices
&
$2$     % Number of determining equations 3, 5
&
\cite{WTC} % Ref
\\ \hline % \hline % ********************************************************
AKNS eq.% Name
&
$1$     % Nb of families
&
$2$     % Scattering order
&
$\chi$  % Truncation variable
&
$\psi$  % Link between $\tau$ and $\psi$
&
$1:5$   % -p:-q
&
$4,6$   % Fuchs indices
&
$3$     % Number of determining equations 2, 3, 5
&
\cite{MusetteSainteAdele} % Ref
\\ \hline % \hline % ********************************************************
p-mKdV     % Name
&
$2$     % Nb of families
&
$2$     % Scattering order
&
$Y$     % Truncation variable
&
$\psi$  % Link between $\tau$ and $\psi$
&
$0:3$   % -p:-q
&
$0,4$   % Fuchs indices
&
$4$     % Number of determining equations 1, 2, 3, 5 % erratum to Gallipoli
&
\cite{Pickering1996} % Ref
\\ \hline % \hline % ********************************************************
sine-Gordon    % Name
&
$2$     % Nb of families
&
$2$     % Scattering order
&
$Y$     % Truncation variable
&
$\psi$  % Link between $\tau$ and $\psi$
&
$0:2$   % -p:-q
&
$2$     % Fuchs indices
&
$4$     % Number of determining equations 1,3
&
\cite{Pickering1996} % Ref
\\ \hline % \hline % ********************************************************
Broer-Kaup     % Name
&
$2$     % Nb of families
&
$2$     % Scattering order
&
$Y$     % Truncation variable
&
$\psi$  % Link between $\tau$ and $\psi$
&
$0:4$   % -p:-q
&
$0,3,4$ % Fuchs indices
&
$4$     % Number of determining equations 1, 2, 6, 7
&
\cite{Pickering1996} % Ref
\\ \hline % \hline % ********************************************************
pp-Boussinesq    % Name
&
$1$     % Nb of families
&
$3$     % Scattering order
&
$(\psi_{x}/ \psi, \psi_{xx}/ \psi)$  % Truncation variable
&
$\psi$  % Link between $\tau$ and $\psi$
&
$0:4$   % -p:-q
&
$0,1,6$ % Fuchs indices
&
$6$     % Number of determining equations
&
\cite{MCGallipoli1991} % Ref
\\ \hline % \hline % ********************************************************
p-SK    % Name
&
$1*$     % Nb of families
&
$3$     % Scattering order
&
$(\psi_{x}/ \psi, \psi_{xx}/ \psi)$  % Truncation variable
&
$\psi$  % Link between $\tau$ and $\psi$
&
$1:6$   % -p:-q
&
$1,2,3,10$   % Fuchs indices
&
$6$     % Number of determining equations
&
\cite{MC1998} % Ref
\\ \hline % \hline % ********************************************************
p-KK    % Name
&
$1*$     % Nb of families
&
$3$     % Scattering order
&
$(\psi_{x}/ \psi, \psi_{xx}/ \psi)$  % Truncation variable
&
G25($\psi$) % Link between $\tau$ and $\psi$
&
$1:6$   % -p:-q
&
$1,3,5,7$   % Fuchs indices
&
$14$     % Number of determining equations
&
\cite{CetraroMusette} % Ref
\\ \hline % \hline % ********************************************************
Tzitz\'eica % Name
&
$1*$     % Nb of families
&
$3$     % Scattering order
&
$(\psi_{x}/ \psi, \psi_{t}/ \psi)$  % Truncation variable
&
$\psi$  % Link between $\tau$ and $\psi$
&
$2:6$   % -p:-q
&
$2$   % Fuchs indices
&
$10$     % Number of determining equations
&
\cite{CMG1999} % Ref
\\ \hline % \hline % ********************************************************
AKNS system % Name
&
$4$     % Nb of families
&
$2$     % Scattering order
&
$?$  % Truncation variable
&
$ $  % Link between $\tau$ and $\psi$
&
$1:3,1:3$   % -p:-q
&
$0,3,4$   % Fuchs indices
&
$ $     % Number of determining equations
&
\cite{CMGalli95} % Ref
\\ \hline % \hline % ********************************************************
\end{tabular}
\end{center}
\label{table}
\end{table}

%\vfill \eject
% ==========================================================================
\subsection{Nonintegrable equations, second and third scattering order}
\label{sectionTwoFamAnyOrderNonintegrable}
\indent

A nonintegrable equation has no determined scattering order,
so this section cannot be split according to the scattering order.

% ==========================================================================
\subsubsection{The KPP equation}
\label{sectionKPP}
\indent

\index{KPP equation}
The KPP equation (\ref{eqKPP})
possesses the two opposite families 
(\ref{eqKPPFamilies})--(\ref{eqLaurentKPP})
and it fails the test at index $4$,
so there can only exist particular solutions.
Let us first review all the known solutions to this equation.

In addition to the notation (\ref{eqKPPNotation}),
it is convenient to introduce the symmetric constant
\begin{eqnarray}
& &
%s_1=e_1 + e_2 + e_3,\ 
%s_2=e_2 e_3 + e_3 e_1 + e_1 e_2,\ 
%s_3=e_1 e_2 e_3,\
%18 d^2 a_2=2(s_1^2 - 3 s_2) = (e_2-e_3)^2 + (e_3-e_1)^2 + (e_1-e_2)^2,\
a_1 % =(2 s_1^3 - 9 s_1 s_2 + 27 s_3)/(3 d)^3
   =(2e_1-e_2-e_3)(2e_2-e_3-e_1)(2e_3-e_1-e_1)/(3 d)^3
\end{eqnarray}
and the entire function
\begin{eqnarray}
& &
\Psi_3=\sum\limits_{n=1}^3 C_n e^{\displaystyle k_n (x + (3/b) k_n t)},\
 k_n = {3 e_n - s_1 \over 3 d},\
 C_n \hbox{ arbitrary},
\label{eqKPPPsi3}
\end{eqnarray}
i.e.~the general solution of the third order linear system
(\ref{eqLaxScalar3xG})--(\ref{eqLaxScalar3tG})
with constant coefficients \cite{CM1993}
\begin{eqnarray}
& &
(S) \equiv \left\lbrace
\begin{array}{ll}
\displaystyle{
\psi_{xxx} - 3 a_2 \psi_x - a_1 \psi=0,
}
\\
\displaystyle{
b \psi_t  - 3 \psi_{xx} =0.
}
\end{array}
\right.
\label{eqLaxScalar3KPP}
\end{eqnarray}

Let us also denote $(j,l,m)$ any permutation of $(1,2,3)$.
Three distinct solutions are presently known.

The first solution is trigonometric,
this is a \textit{collision of two fronts} \cite{KawaharaTanaka}
\begin{eqnarray}
& &
u = {s_1 \over 3} + d \partial_x \Log \Psi_3,\ C_1 C_2 C_3 \not=0
\label{eqKPP2f}
\end{eqnarray}
which depends on two arbitrary constants $C_1 / C_3, C_2 / C_3$.
For $C_j=0, C_l C_m \not=0$,
it degenerates into three heteroclinic 
(i.e.~with different limits at both infinities)
\textit{propagating fronts} 
which depend on one arbitrary constant $x_0$
\begin{eqnarray}
 u  & = & 
       {e_l + e_m \over 2}
    + d {k \over 2} \tanh {k \over 2}(x-ct-x_0),
\label{eqKPPFront}
\\
k^2 & = & (k_l - k_m)^2,\ 
c= -3  (k_l + k_m)/b.
\nonumber
\end{eqnarray} 

The second solution is elliptic \cite{CT1991},
\begin{eqnarray}
& & u = s_1/3 + d \psi_x \sqrt{\wp(\psi)},\
\psi=\Psi_3,\
g_3=0,\
g_2 \hbox{ arbitrary},\
a_1=0,
\label{eqKPPElliptic}
\end{eqnarray} 
it only exists under the constraint 
(codimension is one)
that one root $e_j$ be at the middle of the two others
and it depends on the four arbitrary constants $C_1,C_2,C_3,g_2$.
Its degeneracy $g_2=0$ (i.e.~$\wp(\psi)=\psi^{-2}$)
is the degeneracy $a_1=0$ of the collision of two fronts solution 
(\ref{eqKPP2f}).

The third and last solution is the stationary elliptic solution $u(x)$
\begin{eqnarray}
& &
u(x):\
-u'' + 2 d^{-2} (u-e_1) (u-e_2) (u-e_3)=0.
\label{eqKPPJacobi}
\end{eqnarray}
A trigonometric degeneracy bounded at infinity is 
made of the three homoclinic \textit{stationary pulses}
\begin{eqnarray}
{\hskip -4.0 truemm}
& &
 u= e_j +  {e_l - e_m \over \sqrt 2}
\sech i{e_l - e_m \over d \sqrt 2}(x-x_0),\ 
a_1=0,\
2 e_j - e_l - e_m=0,
\label{eqKPPPulse}
\end{eqnarray}
it has codimension one 
and it depends on the arbitrary constant $x_0$.

Let us now apply the various methods we have seen,
in order to retrieve these solutions, namely
\begin{enumerate}
\item
enforcement of one of the two no-log conditions (\ref{eqKPPQ4}),

\item
enforcement of the two no-log conditions (\ref{eqKPPQ4}),

\item
one-family truncation with a second order assumption,

\item
one-family truncation with a third order assumption,

\item
two-family truncation with a second order assumption,

\item
two-family truncation with a third order assumption.

\end{enumerate}

The single no-log condition (\ref{eqKPPQ4}) has two solutions.
The first one $C=0$ implies $u_t=0$ and thus defines the reduction $u(x)$,
i.e.~the elliptic equation (\ref{eqKPPJacobi}).
The second one is a first order nonlinear PDE for $C(x,t)$,
integrated by the method of characteristics as \cite{CT1989}
\index{reduction}
\begin{eqnarray}
& &
F(I_1,I_2)=0,\
c_n=3 k_n/b=(3 e_n-s_1)/(b d),\ 
\nonumber
\\
& &
I_1=e^x (C - c_1)^{p_1} (C - c_2)^{p_2} (C - c_3)^{p_3},\ 
\label{eqKPPOneNolog}
\\
& &
I_2=e^t (C - c_1)^{q_1} (C - c_2)^{q_2} (C - c_3)^{q_3},\ 
\nonumber
\end{eqnarray}
in which $p_n,q_n$ depend on $e_j$.
Unless some specific choice of the arbitrary function $F$ is made,
or $(x,t)$ are no more taken as the independent variables,
one cannot integrate further the system (\ref{eqChix})--(\ref{eqChit})
for $\chi$ and $S$.

The two no-log conditions (\ref{eqKPPQ4}) together,
apart the already encountered solution $C=0$,
provide the two relations
\begin{eqnarray}
& &
a_1=0,\
e_1=(e_2+e_3)/2,\
\nonumber
\\
& &
b^2 d^2 C^3 - (9/4) (e_2-e_3)^2 C - 3 b d^2 (C_t + C C_x)=0,
%\label{eqKPPTwoNolog}
\end{eqnarray}
whose solution is similarly
\begin{eqnarray}
& &
a_1=0,\
F(I_1,I_2)=0,\
\nonumber
\\
& &
I_1= \frac{b C - (3(e_2-e_3)/(2d))}{b C + (3(e_2-e_3)/(2d))}
     e^{\displaystyle{(e_2-e_3)x/d}},\
%                 e^{?(e_2-e_3)x/d}
\\
& &
I_2= \frac{C^2}{(b C)^2 - (3(e_2-e_3)/(2d))^2}
     e^{\displaystyle{(e_2-e_3)^2 t/(b d^2)}}.
%                 e^{?(e_2-e_3)^2 t/(b d^2)}
\nonumber
\end{eqnarray}

The one-family truncation (\ref{eqDTOneFamily}) with $\tau=\psi$ 
and the second order assumption (\ref{eqLaxScalar2x})--(\ref{eqLaxScalar2t})
generates three determining equations $E_j(S,C,U)=0,j=1,2,3$.
After solving the first one for $U$
\begin{eqnarray}
& & U=s_1/3 - b d C/6,
\end{eqnarray}
the two remaining equations,
with the ever present condition (\ref{eqCrossXT}), are \cite{ConteComo}
\begin{eqnarray}
E_2 & \equiv & - 6 a_2 - S + (b^2 / 6) C^2 - b C_x = 0
\\
E_3 & \equiv & 
a_2 b C - 2 a_1 + b S C/2 - b^3 C^3/108
\nonumber
\\
& &
+ S_x/2 - b^2 C_t/6 + 2 b C_{xx}/3 = 0.
\end{eqnarray}
The elimination of $S$ and $C_t$ yields a factorized equation
\begin{eqnarray}
& & - b^2 C_t  + b C_{xx} - 2 b^2 C C_x +(4/9) (b C)^3 - 12 a_2 b C - 12 a_1=0,
\\
& & [b C_{xx} - b^2 C C_x + b^3 C^3/9 - 3 a_2 b C - 3 a_1] C=0.
\end{eqnarray}
The subcase $C=0$, hence $S=-6 a_2, a_1=0$,
yields the degeneracy $a_1=0$ of the three fronts (\ref{eqKPPFront}).
In the other subcase,
the system for $C$ is linearizable 
into the third order system $(S)$ (\ref{eqLaxScalar3KPP}) 
in which both $\partial_x$ and $\partial_t$ change sign,
the generic solution $(S,C)$ of $(E_2,E_3)$ is therefore
\begin{eqnarray}
& & 
b C = - 3 \partial_x \Log \Psi_3(-x,-t),\
S = - 6 a_2 + (b C)^2 / 6 - b C_x,
\end{eqnarray}
and there only remains to integrate (\ref{eqChix})--(\ref{eqChit}) for $\chi$
or (\ref{eqS})--(\ref{eqC}) for $\varphi$.
Since the one-form $\D x - C \D t$ possesses an integrating factor
\cite{CT1989},
the PDE (\ref{eqC}) for $\varphi$ can be integrated by the method of 
characteristics,
\begin{eqnarray}
& & 
\varphi=\Phi(F),\
F=\frac{\Psi_x + k_2 \Psi} {\Psi_x + k_3 \Psi}
 \ \  \frac{e^{-k_2 x -k_2^2 t/b}}{e^{-k_3 x -k_3^2 t/b}}.
\end{eqnarray}
Note that the cyclic permutation of the roots $e_j$ is broken when going from
$(S,C)$ to $\varphi$.
With the classical identity on Schwarzians
\begin{eqnarray}
& & 
\lbrace \varphi;x \rbrace
\equiv
\lbrace \Phi;F \rbrace F_x^2 + \lbrace F;x \rbrace,
\end{eqnarray}
the third order ODE (\ref{eqS}) for $\varphi$ becomes
\begin{eqnarray}
& & 
\lbrace \Phi;F \rbrace=0,
\end{eqnarray}
which integrates as
\begin{eqnarray}
{\hskip -7.0 truemm}
& & 
\varphi=\Phi(F)= \frac{A_1 F + A_2}{A_3 F + A_4},\
A_j \hbox{ arbitrary constants},\
A_1 A_4 - A_2 A_3 \not=0.
\end{eqnarray}
The value of $\chi^{-1}$
\begin{eqnarray}
& & 
\chi^{-1}=\frac{F_x}{F - F_0} - \frac{F_{xx}}{2 F},\
F_0 \hbox{ arbitrary constant},\
\end{eqnarray}
is again invariant under a cyclic permutation of the roots $e_j$,
and the solution $u$ finally obtained is (\ref{eqKPP2f}).
% NDLR Establish a computer printout of the whole resolution.

The one-family truncation (\ref{eqDTOneFamily}) with $\tau=\psi$ 
and the third order assumption 
(\ref{eqLaxScalar3xG})--(\ref{eqLaxScalar3tG})
generates five determining equations (\ref{eqDetermining3}),
their straightforward resolution yields
\begin{eqnarray}
& & u = d \partial_x \Log \psi + s_1/3 + d U,\
\\
& &
\psi_{xxx} + 3 U \psi_{xx}
- 3 (a_2 - U^2 - U_x) \psi_x 
\nonumber 
\\ 
& &
\phantom{xxxxx}
-(a_1+3 a_2 U -U^3 - b U_t/2 -6 U U_x + U_{xx}/2) \psi=0,
\\
& &
b \psi_t  - 3 \psi_{xx} - 3 U \psi_x - 6 U^2 \psi=0,
\end{eqnarray}
in which $U$ is constrained by two relations.
But, since the coefficient $f$ can be set to zero without loss of generality,
the choice $U=0$ represents the general solution
(just like in \cite{CM1993} where constant values were assumed 
\textit{ab initio} for the coefficients in 
(\ref{eqLaxScalar3xG})--(\ref{eqLaxScalar3tG})),
and it represents again the collision of two fronts (\ref{eqKPP2f}).

The contrast of difficulty between the second order assumption (laborious)
and the third order assumption (immediate) is the signature that the good
scattering order of KPP is three,
despite the irrelevance of such a notion for nonintegrable equations.

The two-family truncation with a second order assumption
(\ref{eqTruncationTwoFamily1})--(\ref{eqTruncationTwoFamily4})
\cite{CM1992,CM1993,Pickering1993}
generates five determining equations.
Despite the factorized form of $E_5$,
we have not yet found their general solution.
The three particular solutions for which $(S,C,A)$ are constant
provide immediately the three pulses (\ref{eqKPPPulse}),
for they belong to the class of polynomials in $\tanh$ and $\sech$,
generated by negative and positive powers of $\chi$
according to the elementary identities (\ref{eqIdentitiestanhsech}).

The elliptic solution (\ref{eqKPPElliptic}) can also be written
\begin{eqnarray}
{\hskip -9.0 truemm}
& & u = s_1/3 + \partial_x \Log (\ns(\psi) - \cs(\psi)),\
\psi=\Psi_3,\
g_3=0,\
g_2 \hbox{ arb.},\
a_1=0,
\label{eqKPPEllipticWithJacobi}
\end{eqnarray} 
a relation in which the argument of the logarithm is the ratio of two
entire functions.
Therefore it could be possible to find it
by a suitable extension of a two-family truncation 
with a third order assumption.

This solution was first found by the following two-step procedure 
\cite{CT1991},
which, unfortunately for this nice method,
only works for a restricted class of PDEs
(those with $p=-1,u_0=c_0,u_1=c_1 C + c_2, c_j=\hbox{constant}$,
see (\ref{eqLaurentKPP})).
The first step is to define the truncation
\begin{eqnarray}
& &
u= d \partial_x \Log (\varphi - \varphi_0) + U,\
U=\hbox{constant},\
\nonumber
\\
& &
E(u)=\sum_{j=0}^{3}
 E_j(\varphi_{xx} / \varphi_x,\varphi_t / \varphi_x)
\left(\frac{\varphi - \varphi_0}{\varphi_x}\right)^{j-3},\
\forall j\ :\ E_j=0,
%\label{eqKPPTruncationCT1}
\end{eqnarray}
whose general solution is $U=s_1/3,\varphi- \varphi_0=\Psi_3$
(indeed, comparing with (\ref{eqLaurentKPP}),
 this assumption \textit{a priori} implies $b \varphi_t -3 \varphi_{xx}=0$).
The second step is not a truncation, but the change of function $u \mapsto f$
\begin{eqnarray}
& &
u= s_1/3 + \left(d \partial_x \Log \Psi_3 \right) f(\Psi_3),
\label{eqKPPTruncationCT2}
\end{eqnarray}
which transforms (\ref{eqKPP}) into 
\begin{eqnarray}
& &
U'' - 2 U^3 + 2 a_1 \Psi_{3,x}^{-3}=0,\
U(\psi)=f(\psi)/ \psi.
\end{eqnarray}
This is an ODE iff $a_1=0$, in which case its solution is
(\ref{eqKPPElliptic}).
Therefore, the assumption (\ref{eqKPPTruncationCT2}) has defined a reduction
of the PDE to an ODE. 
This subject will be further examined in Section \ref{sectionReduction}.
\index{reduction}

%\vfill \eject
% ==========================================================================
\subsubsection{The cubic complex Ginzburg-Landau equation}
\label{sectionCGL3}
\indent

The cubic complex Ginzburg-Landau equation (CGL3)
\index{complex Ginzburg-Landau equation}
\begin{eqnarray}
{\hskip -6.0 truemm}
& & E(u) \equiv 
i u_t + p u_{xx} + q \vert u \vert^2 u - i \gamma u = 0,
\ p q \ne 0,\ (u, p, q) \in {\mathcal C},\ \gamma \in {\mathcal R},
\label{eqCGL3}
\end{eqnarray}
with $p,q,\gamma$ constant,
is a generic PDE describing the propagation of the signal in an optical fiber
as well as 
superfluidity,
spatiotemporal intermittency,
pattern formation, etc.

One easily checks that $\mod{\GLA}$ generically behaves like a simple pole.
The dominant behaviour
\begin{eqnarray}
& & 
          \GLA  \sim           a _0 \chi^{-1+i \alpha},\
\overline{\GLA} \sim \overline{a}_0 \chi^{-1-i \alpha},\
\label{eqCGL3Leading}
\end{eqnarray}
in which $a_0$ is a complex constant, $\alpha$ a real constant,
is solution of the nonlinear algebraic system
\begin{eqnarray}
   & &  p  (-1+i \alpha)(-2+i \alpha) +  q  a_2 =0,\
\label{eqCGL3Leading1}
\\ & & \pc (-1-i \alpha)(-2-i \alpha) + \qc a_2 =0,\
\label{eqCGL3Leading2}
\end{eqnarray}
with $a_2=\mod{a_0}^2$.
This defines two families for $\mod{\GLA}^2$ (four for $\mod{\GLA}$)
\cite{CT1989}
\begin{eqnarray}
& & 
a_2={9 \mod{p}^2 \over 2 \mod{q}^2 d_i^2} \lbrack d_r + \Delta \rbrack,\
\alpha={3 \over 2 d_i} (d_r + \Delta),\
\label{eqAlpha}
\\
& &
{p \over q}=d_r - i d_i,\
\Delta^2=d_r^2 + (8/9) d_i^2.
%q=q_r + i q_i,\
\end{eqnarray}
To prevent these irrational expressions to mess up all subsequent computations
(Fuchs indices, no-log conditions, truncations),
the system (\ref{eqCGL3Leading1})--(\ref{eqCGL3Leading2})
can equivalently be solved as a \textit{linear} system
on ${\mathcal C}$
\cite{CM1993,CM2000b}
\begin{eqnarray}
& & a_2=- {p \over q} (1-i \alpha)(2-i \alpha),
\label{eqa2onC}
\\
& & \pc = K p (1-i \alpha)(2-i \alpha),\
    \qc = K q (1+i \alpha)(2+i \alpha),
\label{eqpcqconC}
\end{eqnarray}
in which $K$ is an irrelevant arbitrary nonzero complex constant.

The indicial equation is the determinant \cite{Cargese96Conte}
of the second order matrix
\begin{eqnarray}
{\hskip -3.0 truemm}
\bfP(j)
& = &
\pmatrix{
  (2 a_0 \overline{a_0}) q
&    a_0^2               q            
\cr
         \qc \overline{a_0}^2
& (2 a_0 \overline{a_0}) \qc
\cr
}
\nonumber
\\
{\hskip -3.0 truemm}
& &
+\diag( p  (j-1 + i \alpha)(j-2 + i \alpha),
       \pc (j-1 - i \alpha)(j-2 - i \alpha)),
\end{eqnarray}
and with the resolution (\ref{eqa2onC})--(\ref{eqpcqconC}) it evaluates to
\begin{eqnarray}
\det \bfP(j)
& = &
(j+1) j (j^2 - 7 j + 6 \alpha^2 + 12)=0.
\end{eqnarray}
For generic values of $(p,q)$, two of the four indices are irrational.

Let us consider, for simplification, the solitary wave reduction
\index{reduction}
\begin{eqnarray}
& &
\GLA(x,t)=U(\xi)
 e^{i \displaystyle{\left(\omega t + \varphi(\xi)\right)}},\
\xi=x-ct,\
\label{eq1CGL3red}
\end{eqnarray}
in which $(U,\varphi)$ are functions of 
the reduced independent variable $\xi$,
and let us restrict to the pure CGL3 case $\Im(p/q)\not=0$.
%For other reductions, see \cite{CM1993}.
The general solution of the fourth order system of ODEs for $(U,\varphi)$
\textit{a priori} depends on six arbitrary constants,
the four constants of integration plus the two reduction parameters 
$(c,\omega)$.
{}From these six constants,
one must subtract
\begin{enumerate}
\item
the irrelevant origin $\xi_0$ of $\xi$ (Fuchs index $-1$),
which represents the invariance under a space translation,

\item
the irrelevant origin $\varphi_0$ of the phase (Fuchs index $0$),
which represents the invariance under a phase shift,

\item
and
the number of irrational Fuchs indices, generically two.
Indeed, these irrational indices represent the chaotic nature of CGL3
(see the expansion (48) in \cite{CM1993})
and they cannot contribute to any analytic solution.
\end{enumerate}
Therefore only two relevant arbitrary constants are present in what can be 
called the \textit{general analytic solution}
            \index{general analytic solution}
of the reduction (\ref{eq1CGL3red}).

Presently, 
one only knows four particular solutions of the reduction $\xi=x-ct$ 
with a zero codimension (no constraint on $(p,q,\gamma)$).
These are 
\begin{enumerate}
\item
a pulse or solitary wave \cite{PS1977}
\begin{eqnarray}
& &
 \GLA=- i a_0 k \sech k x 
 e^{\displaystyle{i[\expon \Log \cosh k x + K_1 t]}},\
K_2 k^2 - \gamma =0,\
\label{eqCGL3Pulse}
\end{eqnarray}

\item
a front or shock \cite{NB1984}
\begin{eqnarray} 
\GLA
& = &
a_0 {k \over 2} \left[\tanh {k \over 2} \xi \pm 1 \right]
e^{\displaystyle i[\expon \Log \cosh {k \over 2} \xi + K_3 c \xi - K_4 c^2 t]},
\end{eqnarray} 

\item
a source or propagating hole \cite{BN1985}
\begin{eqnarray} 
{\hskip -10.0 truemm}
\GLA
& = &
a_0 \left[{k \over 2} \tanh {k \over 2} \xi + (K_1 + i K_2) c \right]
\nonumber
\\
{\hskip -10.0 truemm}
& &
\times 
e^{\displaystyle
i[\expon \Log \cosh {k \over 2} \xi + K_3 c \xi - (K_4 k^2 + K_5 c^2) t]},\
K_6 k^2 + K_7 c^2 = \gamma,\
\label{eqCGLHole}
\end{eqnarray} 

\item
an unbounded solution \cite{CM1993}
\begin{eqnarray}
\mod{\GLA}^2 & = & a_2 (\tan^2 {k \over 2} \xi + K^2),\ c=0.
\end{eqnarray}

\end{enumerate}
In the above expressions, all parameters $(a_2,\expon,k,c,K_l)$ are real
and only depend on $(p,q,\gamma),$
except in (\ref{eqCGLHole}) where the velocity $c$ is arbitrary.

These four particular solutions 
are four different degeneracies \cite{CM1993}
of the yet unknown general analytic solution.

In experiments or computer simulations,
one has observed 
\cite{BN1985,Lega1992,Chate1994,vanHecke}
other regular patterns which should correspond to 
other degeneracies of the general analytic solution.
One of them \cite{vanHecke} is a homoclinic hole solution,
complementing the heteroclinic hole (\ref{eqCGLHole}).
Another one, of the highest interest in fiber optics,
is a propagating pulse,
extrapolating (\ref{eqCGL3Pulse}) to $c\not=0$
and reducing in the NLS limit ($p,q$ real, $\gamma=0$)
to a ``bright soliton'' of arbitrary velocity.
\index{nonlinear Schr\"odinger equation}

Let us now address the question of retrieving these four solutions
(and ideally of finding the unknown one ar at least other degeneracies)
by some truncation.
For a truncation to be successful,
the truncated variables should be free of any multivaluedness in
their dominant behaviour.
This is not the case of the natural physical variables
$(\GLA,\overline{\GLA})$ or $(\Re \GLA, \Im \GLA)$,
which are \textit{always} locally multivalued as seen from 
(\ref{eqCGL3Leading}).
A more detailed study \cite{CM1993} uncovers the best representation for
this purpose,
namely a \textit{complex modulus} $Z$ and a real argument $\Theta$
uniquely defined by
\begin{eqnarray}
& &           \GLA  =           Z  e^{ i \Theta},\
    \overline{\GLA} = \overline{Z} e^{-i \Theta},
\end{eqnarray}
and the above four exact solutions are written in this notation.
For each family,
if one excludes the contribution of the irrational Fuchs indices,
the three fields $(Z,\overline{Z},\grad \Theta)$ are locally singlevalued
and they behave like simple poles.
The physical variables $(\mod{\GLA}^2, \grad \arg \GLA)$ also have this
nice property of being locally singlevalued 
(they respectively behave like a double pole and a simple pole),
but they are not as elementary as $(Z,\overline{Z},\grad \Theta)$.

The one-family truncation of the third order ODE satisfied by $\mod{\GLA}^2$ 
(after elimination of $\varphi$),
with a constant coefficient second order assumption,
evidently captures all four solutions,
since $\mod{\GLA}^2$ is a degree-two polynomial in $\tanh \kappa \xi$.
Such a truncation generates cumbersome computations and provides no
additional solution.

The one-family truncation of $(Z,\overline{Z},\grad \Theta)$ 
with the same constant coefficient second order assumption
is defined as \cite{CM1993,CM1995}
\begin{eqnarray}
& &
\left\lbrace
\begin{array}{ll}
\displaystyle{
Z = a_0 (\chi^{-1} + X + i Y),\
}
\\
\displaystyle{
\overline{Z} = \overline{a}_0 (\chi^{-1} + X - i Y),\
}
\\
\displaystyle{
\Theta = \omega t + \alpha \Log \psi + K \xi,\
}
\\
\displaystyle{
(\Log \psi)'=\chi^{-1},\
}
\displaystyle{
\chi' = 1 - (k^2/4) \chi^2,\
}
\\
E e^{-i \Theta} = \sum_{j=0}^{3} E_j \chi^{j-3},
\end{array}
\right.
%\label{eq1CGL3HoleTrunc}
\end{eqnarray}
in which $\chi$ and $\psi$ are functions of $\xi=x-ct$,
$(\omega,X,Y,K,k^2)$ are real constants.
One has to solve the four complex (eight real) equations $E_j=0$
in the
eight real unknowns $(a_2,\alpha,\omega,X,Y,K,c,k^2)$,
the two complex parameters $(p,q)$,
and the real parameter $\gamma$.
If there exists a solution, the elementary building block functions 
evaluate to
\begin{eqnarray}
& & 
\chi={k \over 2} \tanh {k \xi \over 2},\
\psi=\cosh {k \xi \over 2}.
\end{eqnarray}
The good methodology \cite{CM1993,CM2000b} is again to select,
among the eleven complex variables considered as equivalent,
four variables which make the system a \textit{linear} one of Cramer type.
The system $(E_0,E_1,E_2)$ is of Cramer type in $(a_2,K,\omega)$,
and after its resolution
the last equation $E_3$ is independent of $(p,q,\gamma,c)$ and factorizes
into a product of \textit{linear} factors
\begin{eqnarray}
& & E_3 \equiv  [k^2 - 4 (X+iY)^2] (\alpha Y - 2 X)=0.
\end{eqnarray}
Finally, this one-family truncation recovers all four solutions except the
pulse (\ref{eqCGL3Pulse}).

The two-family truncation of $(Z,\overline{Z},\grad \Theta)$ 
with the same constant coefficient second order assumption
retrieves the pulse solution (\ref{eqCGL3Pulse}), but finds nothing new.

Similar truncations for two coupled CGL3 equations can be found in
\cite{CM2000b,CM2000c}.

% ==========================================================================
\subsubsection{The nonintegrable Kundu-Eckhaus equation}
\label{sectionKunduEckhaus}
\indent

The PDE for the complex field $U(x,t)$ \cite{Kundu1984,CE1987}
\index{Kundu-Eckhaus equation}
\begin{eqnarray}
{\hskip -3.0 truemm}
& & 
 i U_t + \alpha U_{xx} 
+({\beta^2 \over \alpha}{\mod{U}}^4 +2 b e^{i \gamma}({\mod{U}}^2)_x)U=0,\
(\alpha, \beta, b, \gamma) \in {\mathcal R},\ 
%\theta = \arg U,\
%u_x = \mod{U}^2
\end{eqnarray}
with $\alpha \beta \cos \gamma \not=0$,
is linearizable when $b^2=\beta^2$ into the Schr\"odinger equation
\begin{eqnarray}
& & 
 i V_t + \alpha V_{xx} = 0,\
%\arg U = \arg V,\
   U=\sqrt{{\alpha \over 2 \beta \cos \gamma}}
          {     V    \over \sqrt{\int \mod{V}^2 \D x}}.
\end{eqnarray}
This suggests considering the PDE for $u=\int \mod{V}^2 \D x$
\begin{eqnarray}
& & 
 {\alpha \over 2} (u_{xxxx} u_x^2 + u_{xx}^3 - 2 u_x u_{xx} u_{xxx})
+ 2 {\beta^2 - (b \sin \gamma)^2 \over \alpha} u_x^4 u_{xx}
\nonumber 
\\
& &
+ 2 (b \cos \gamma) u_x^3 u_{xxx}
+{1 \over 2 \alpha} (u_{tt} u_x^2 + u_{xx} u_t^2 - 2 u_t u_x u_{xt})
=0.
\end{eqnarray}
When $b^2\not=\beta^2$, this PDE fails the test \cite{CC1987} because, 
for each of the two families for $u$, one index is generically irrational.
However, its one-family truncation with a second order assumption
(i.e.~the usual WTC truncation) is a very rich exercise \cite{CMNEEDS93}
which yields quite unusual solutions,
among them an elliptic one involving the ODE of class III of Chazy
\cite{ChazyThese}.

%\vfill \eject

% ==========================================================================
\section{Singular manifold method \textit{versus} reduction methods}
\label{sectionReduction}
\indent

In order to find exact solutions of PDEs,
there exist two main classes of methods.
The first class, which has been detailed in these lectures,
is based on the structure of the movable singularities
and it can be called, in short, the singular manifold method.
\index{singular manifold!method}

The second class,
presented in another course of this school \cite{CetraroWinternitz},
basically relies on group theory
and consists in finding the \textit{reductions} 
to a PDE in a lesser number of independent variables,
and at the end to an ODE.
These reductions are obtained
either by looking for the infinitesimal symmetries of the PDE
(space translation, etc)
and by integrating them,
or by a direct search not involving any group theory.
The main methods in this second class are known as
(see references in \cite{CetraroWinternitz})
the \textit{classical method} (point symmetries),
the \textit{nonclassical method} (conditional point symmetries),
the \textit{direct method} (direct search).
\index{classical!method}
\index{nonclassical!method}
\index{direct!method}
\index{reduction}

The question of the comparison of these four methods by their results
is an active research subject 
\cite{EstevezGordoa1995,EstevezGordoa1998,GB2000},
and its current state is given in \cite{Cargese96CW,CetraroWinternitz}.

Let us take as an example a second order nonintegrable PDE,
this is enough to give an idea of the comparison.
The KPP equation (\ref{eqKPP}) has been studied in detail
with the singular manifold method, Section \ref{sectionKPP}.
It has also been investigated with the three other methods,
and the results are the following.
\index{KPP equation}

In the classical and nonclassical methods, let us denote
\begin{eqnarray}
& & 
\tau(x,t,u) u_t + \xi(x,t,u) u_x - \eta(x,t,u)=0
\end{eqnarray}
the PDE for $u(x,t)$ which, 
after computation of the symmetries $(\tau,\xi,\eta)$,
defines the constraint on $u$ susceptible to yield a reduction if the
constraint can be integrated.
\index{reduction}

\underbar{Classical method}. 
It yields only two reductions $u(x,t) \mapsto U(z)$ \cite{CT1989,CT1991},
one noncharacteristic (i.e.~conserving the differential order two)
\index{characteristic}
\begin{eqnarray}
{\hskip -9.0 truemm}
& & 
z=x-ct,\
u=U,\
-U'' - b c U' + 2 d^{-2} (U-e_1) (U-e_2) (U-e_3)=0,
\label{eqKPPReducWave}
\end{eqnarray}
one characteristic (i.e.~lowering the differential order two)
\begin{eqnarray}
& & 
z=t,\
u=U,\
b U' + 2 d^{-2} (U-e_1) (U-e_2) (U-e_3)=0.
\end{eqnarray}
This is in fact a unique reduction $z=\lambda x + \mu t$,
but the splitting according to the characteristic nature is relevant for
the Painlev\'e property.
None of these two ODEs has the Painlev\'e property,
unless $c=(3 e_j - s_1)/(b d)$ in (\ref{eqKPPReducWave}).

\underbar{Nonclassical method}.
It yields three sets of values for $(\tau,\xi,\eta)$ \cite{NC1992},
% Cargese96CW page 637
two with $\tau \not=0$ and one with $\tau=0,\xi \not=0$.

The first one \cite{CT1991} has codimension one 
% To be verified. NC1992 seems to disagree with Cargese96CW
% CT1991 (5.4), their particular solution (5.12) is the general solution
% NC1992 (29), (34) depends on t
% Clarkson and Mansfield 1993 Physica D (4.7)--(4.9)
% Cargese96CW page 638 Table 10.4 does not depend on t. ERROR?
\begin{eqnarray}
{\hskip -11.0 truemm}
& & 
a_1=0,\
e_1=(e_2+e_3)/2,\
% \tau=1,\
% \xi=- 3 b^{-1} (\Log \cosh \frac{e_2-e_3}{2 d} x)_x
% \eta=- (u-e_1) \xi_x,\
b u_t - 3 d^{-2} \left((\Log \psi)_x (u-e_1)\right)_x=0,\
\psi=\Psi_3,
\end{eqnarray}
its integration defines the noncharacteristic reduction to an elliptic
equation
\begin{eqnarray}
& & 
a_1=0,\
z     =  \Psi_3,\ 
u=e_1 + d z_x U(z),\
U''-2 U^3 =0,
\end{eqnarray}
and one finds the solution (\ref{eqKPPElliptic}).

The second one \cite{NC1992} has codimension zero
% NC1992 (39)
\begin{eqnarray}
& & 
b u_t + d^{-1}(u - s_1/3)u_x + 3 d^{-2}(u-e_1)(u-e_2)(u-e_3)=0,\ 
\label{eqKPPNonClassicalSol2}
\end{eqnarray}
and, remarkably, this first order PDE, which fails the test,
identifies to the no-log condition (\ref{eqKPPQ4}) with $3 u-s_1=b d C$.
Its integration (\ref{eqKPPOneNolog}) cannot define a reduction unless some
choice of the arbitrary function is made.
Nevertheless, the common solution to the PDEs (\ref{eqKPP}) and 
(\ref{eqKPPNonClassicalSol2}) is (\ref{eqKPP2f}).

The third one \cite{NC1992} is
% NC1992 (49)
% Nothing on $\tau=0$ in Cargese96CW
\begin{eqnarray}
& & 
u_x - \eta=0,\
\label{eqKPPNonClassicalSol3}
\end{eqnarray}
in which $\eta$ satisfies the second order PDE 
\begin{eqnarray}
& & 
\eta_{xx} + 2 \eta \eta_{xu} + \eta^2 \eta_{uu}
+ 2 d^{-2} (u-e_1)(u-e_2)(u-e_3) \eta_u
\nonumber
\\
& &
- 2 d^{-2} (3 u^2 - 2 s_1 u + s_1^2/3 - 3 d^2 a_2) \eta
+ b \eta_t
=0.
\label{eqKPPNonClassicalSol3Eta}
\end{eqnarray}
Integrating (\ref{eqKPPNonClassicalSol3Eta}) is equivalent to integrating the
original PDE (\ref{eqKPP}),
since the transformation (\ref{eqKPPNonClassicalSol3}) simply exchanges them,
so one is stranded.
The only way out is to put some additional constraints on $\eta$.
% The constraints 4.2.1, 4.2.2, 4.2.3 in Clarkson and Mansfield 1993
% generate nothing new. 4.2.1 is very restrictive.
The consistent way to do that 
(\cite{Cargese96CW} page 634)
is to eliminate $u_x$ and its derivatives (in this case $u_{xx}$ only)
between (\ref{eqKPPNonClassicalSol3}) and (\ref{eqKPP}),
which results in a nonlinear first order ODE for the function 
$t \mapsto u(x,t)$ (i.e.~with $x$ as a parameter)
\begin{eqnarray}
& & 
b u_t - (\eta_{x} + \eta \eta_{u}) + 2 d^{-2} (u-e_1)(u-e_2)(u-e_3)=0.
\label{eqKPPNonClassicalSol3ODEut}
\end{eqnarray}
Requiring the invariance of this ODE under the infinitesimal transformation
$(\tau=0,\xi=1,\eta)$ of the classical method creates constraints on $\eta$,
an exercise which is left to the reader.
%[NDLR Computation to be done. Just follow page 634]

\underbar{Direct method}.
% NC1992 (49)
% Cargese96CW page 640 lines 4--11, no details
The search for a reduction $u(x,t) \mapsto U(z)$ in the class
\begin{eqnarray}
& &
u(x,t)=\lambda(x,t) U(z(x,t)) + \mu(x,t),
\end{eqnarray}
apart from the characteristic reduction $z=t,u=U$,
yields the noncharacteristic reduction
\cite{NC1992}
\begin{eqnarray}
& &
U'' - 2 U^3 + g_1(z) U' + g_2(z) U + g_3(z)=0,
\label{eqKPPDirectODE}
\\
& &
u = d z_x U + s_1/3,\
z_x \not=0,
\end{eqnarray}
provided $z,g_1,g_2,g_3$ satisfy the system
\begin{eqnarray}
& &
\left\lbrace
\begin{array}{ll}
\displaystyle{
b z_{xt} - z_{xxx} - 6 a_2 z_{x} + z_x^3 g_2=0,
}
\\
\displaystyle{
b z_t - 3 z_{xx} + z_x^2 g_1=0,
}
\\
\displaystyle{
2 a_1 - z_x^3 g_3=0.
}
\end{array}
\right.
\label{eqKPPDirectM}
\end{eqnarray}
The constraint $z_x \not=0$ splits the discussion into $a_1=0$ and 
$a_1 \not=0$.
The case $a_1$ arbitrary defines the reduction \cite{NC1992}
\begin{eqnarray}
& &
z=x-ct,\
u=d U + s_1/3,\
U'' + c U' - 2 (U^3 - 3 a_2 U - a_1)=0,
\end{eqnarray}
identical to (\ref{eqKPPReducWave}).
In the case $a_1=0$, hence $g_3=0$,
the system is solved for $(z_{xxx},z_t)$,
and the condition $(z_{xxx})_t=(z_t)_{xxx}$ reads 
\begin{eqnarray}
& &
\left(18 \left(\frac{z_{xx}}{z_x^2}\right)^2
+18       \frac{z_{xx}}{z_x^2}
+2 g_1'(z) - g_1(z) \partial_z + 3 \partial_z^2\right)
Q(z)=0,\
\\
& &
Q(z)=9 g_2 - 2 g_1^2 - 3 g_1'.
\end{eqnarray}

For $Q(z) \not=0$, the condition integrates as
\begin{eqnarray}
& &
z=G(x f_1(t)+f_2(t)),
\end{eqnarray}
in which $G$ is an arbitrary function,
and $(f_1,f_2)$ are further constrained by $f_1'=0, f_2'=0$.
The result is $a_1=0,z=G(x)$, 
and the ODE (\ref{eqKPPDirectODE}) transforms to an elliptic equation
under
\begin{eqnarray}
& &
U(z)=\frac{f(G^{-1}(z))}{G'(z)},\
f'' - 2 f^3 + 6 a_2 f=0,
\end{eqnarray}
in which $G^{-1}$ denotes the inverse function of $G$.
This solution is not distinct from the stationary elliptic reduction
(\ref{eqKPPJacobi}).

For $Q(z)=0$, 
if one defines the function $G$ by
\begin{eqnarray}
& &
g_1(z)=-3 (\Log(G'(z)))',
\end{eqnarray}
the system (\ref{eqKPPDirectM}) is equivalent to
\begin{eqnarray}
& &
\left\lbrace
\begin{array}{ll}
\displaystyle{
g_2=(2/9) g_1^2+(1/3) g_1',
}
\\
\displaystyle{
(\partial_x^3 - 3 a_2 \partial_x) G(z(x,t))=0,\
}
\\
\displaystyle{
(b \partial_t - 3 \partial_x^2) G(z(x,t))=0,\
}
\\
\displaystyle{
f''(Z) - 2 f(Z)^3=0,\
U(z)=G'(z) f(Z),\
Z=G(z),
}
\end{array}
\right.
\end{eqnarray}
and this proves that the particular solution $g_1=g_2=0$ considered in
\cite{NC1992} is the general solution,
equivalent to the reduction $z=\Psi_3$ in (\ref{eqKPPElliptic}).

% ==========================================================================

\section{Truncation of the unknown, not of the equation}
\label{sectionCJP}
\indent

When applied for instance to the second Painlev\'e equation (P2)
\index{Painlev\'e!equation (P2)}
\begin{eqnarray} 
\hbox{(P2)} &  &
E(u) \equiv u'' - 2 u^3 - x u - \alpha =0,
\label{eqP2}
\end{eqnarray}
the one-family singular manifold method
in the case of a second order scattering problem,
i.e.~the one originally performed by WTC,
see Sections 
\ref{sectionOneFamSecondOrderIntegrable}
and 
\ref{sectionOneFamSecondOrderNonintegrable},
presents the following drawback.
The (P2) ODE has two families $u \sim \varepsilon \chi^{-1}, \varepsilon^2=1$.
With the definition of $\grad \chi$
\begin{eqnarray} 
& &
\chi'=1 + (S/2) \chi^2,
\label{eqChiPrime}
\end{eqnarray}
the one-family truncated expansion for $u$ is found to be
\begin{eqnarray} 
& &
u= \varepsilon \chi^{-1},\
E(u) = \varepsilon (S-x) \chi^{-1} + (- \alpha - \varepsilon S'/2) \chi^{0},
\label{eqWTCTrunc}
\\
& &
E_2 \equiv \varepsilon (S-x)=0,
\\
& &
E_3 \equiv - \alpha - \varepsilon S'/2 =0,
\end{eqnarray}
and its general solution is
\begin{eqnarray} 
& &
S=x,\
2 \alpha + \varepsilon=0,\
u=\varepsilon (\Log \Ai(x))',\
\Ai'' + (x/2) \Ai=0.
\end{eqnarray}
One therefore finds only a one-parameter particular solution 
in terms of the Airy function,
at the price of one constraint on the parameter $\alpha$.
This is unsatisfactory because the method fails to find the highest
information on (P2) (highest in the context of these lectures),
namely its \textit{Schlesinger transformation}.
\index{birational transformation}
Such a transformation is by definition a \textit{birational transformation}
between two different copies of (P2), denoted $u(x,\alpha)$ and $U(X,A)$,
and it reads \cite{Luka1971}
\begin{eqnarray} 
& &
x=X,\
u+U
={-2 A    -1 \over 2(U'+U^2) + X}
={2 \alpha-1 \over 2(u'-u^2) - x},\
\alpha=A+1.
\label{eqP2STLuka}
\end{eqnarray}

A method to remedy this drawback is the following \cite{CJP}.
We rephrase it in the homographically invariant formalism,
which simplifies the exposition.
Firstly,
rather than splitting $E(u)$, defined in (\ref{eqWTCTrunc}),
into one equation per power of $\chi$,
one retains the single information $E(u)=0$,
and one eliminates $u$ and $\chi$ between the three equations
(\ref{eqChiPrime}) 
and
(\ref{eqWTCTrunc})
to obtain the second order ODE for $S(x)$
\begin{eqnarray} 
& &
2 (S-x) S'' -S'^2 + 2 S' 
+ 2 S^3 - 4 x S^2 + 2 x^2 S + 4 \alpha (\alpha+\varepsilon)=0.
\end{eqnarray}
This ODE for $S(x)$, which is birationally equivalent to (P2) under
the transformation
\begin{eqnarray} 
& &
{\hskip -5.0 truemm}
u=\varepsilon \chi^{-1},\
S=-2 (\chi^{-1})' -2 \chi^{-2},\
\chi^{-1}={S'+ 2 \varepsilon \alpha \over 2(S-x)},
\label{eqODES}
\end{eqnarray}
bears the number $34$ in the classification of Painlev\'e and Gambier 
\cite{GambierThese}.

Secondly,
despite the fact that one already knows the general solution $S(x)$
in terms of the (P2) function $u(x,\alpha)$,
one takes advantage of the two-family structure of (P2)
(the sign $\varepsilon$ is $\pm 1$)
to perform an involution
by representing $S(x)$ with another (P2) function $U(X,A)$ as
\begin{eqnarray} 
{\hskip -10.0 truemm}
& &
S=-2 V' - 2 V^2,\
U=\varepsilon_2 V,\
\varepsilon_2^2=1,\
U'' - 2 U^3 - X U - A=0,\
X=x.
\label{eqSnew}
\end{eqnarray}
The elimination of $S$ between (\ref{eqODES}) and (\ref{eqSnew})
provides a relation between $(\varepsilon \alpha,\varepsilon_2 A)$ only
\begin{eqnarray} 
& &
(A+\varepsilon_2/2)^2 = (\alpha + \varepsilon/2)^2.
\end{eqnarray}
The solution $A=-\varepsilon_2 (\varepsilon \alpha +1)$
 is the Schlesinger transformation.

\index{Painlev\'e!equation (P2)}
\index{Painlev\'e!equation (P4)}

An equivalent presentation can be found in Ref.~\cite{GJP1999a}.
In the latter,
one first computes the two coefficients $u_0,u_1$ of the Laurent expansion
\begin{eqnarray} 
& &
u=u_0 \chi^{-1} + u_1,
\end{eqnarray}
then the Schlesinger transformation is readily obtained by
(more precisely, the computation of \cite{GJP1999a} reduces to)
the elimination of the three variables $u,Z,U''$ 
between the four equations 
($u_0,u_1,u,U,Z,s$ are functions of $X=x$)
\begin{eqnarray} 
& &
u=u_0 Z^{-1} +U,\
\label{eqGJPDT}
\\
& &
Z'=1 + 2 \frac{U - u_1}{u_0} Z + \frac{s}{2} Z^2,\
\label{eqGJPRiccati}
\\
& &
\hbox{(Pn})(u,x,\alpha,\beta,\gamma,\delta)=0,\
\label{eqGJPnu}
\\
& &
\hbox{(Pn)}(U,X,A,B,C,D)=0.
\label{eqGJPPnU}
\end{eqnarray}
Equation (\ref{eqGJPDT}) is an assumption for a Darboux transformation,
and (\ref{eqGJPRiccati}) defines a Riccati equation for the expansion
variable $Z$ which depends on a free function $s$.
The elimination is differential for $u$ and $Z$, algebraic for $U''$,
and it results in
\begin{eqnarray} 
& &
F(U',U;s,s',s'',\alpha,\beta,\gamma,\delta,A,B,C,D)=0.
\end{eqnarray}
The algebraic independence of $(U',U)$,
consequence of the irreducibility of (Pn),
requires the identical vanishing of $F$ as a polynomial of the
two variables $(U',U)$,
and this provides two solutions:
the identity ($u=U,Z^{-1}=0$)
and, at least for (P2) and (P4), the Schlesinger transformation.
The result for (P2) is
\begin{eqnarray} 
\hbox{(P2)}
& &
\varepsilon Z^{-1}=u-U
=\frac{\varepsilon (A - \alpha)}{2 U' + \varepsilon (2 U^2 +x)},\
\alpha+A+\varepsilon=0,\
s=0,
\label{eqP2STu}
\end{eqnarray}
and the inverse transformation
\begin{eqnarray} 
\hbox{(P2)}
& &
u-U=\frac{\varepsilon (A - \alpha)}{2 u' + \varepsilon (2 u^2 +x)}
\end{eqnarray}
follows from the elimination of $U'$ between (\ref{eqP2STu}) and
\begin{eqnarray} 
& &
(U-u)' + \varepsilon (U^2 - u^2)=0,
\end{eqnarray}
itself obtained by the elimination of $Z$ between 
(\ref{eqGJPDT}) and (\ref{eqGJPRiccati}).
This Schlesin\-ger transformation is identical, thanks to the parity invariance
of (P2), to (\ref{eqP2STLuka}).

The result for (P4) is
\begin{eqnarray} 
{\hskip -10.0 truemm}
& &
\hbox{(P4)}\
u''-u'^2/(2 u) -(3/2) u^3 - 4 x u^2 - 2 x^2 u + 2 \alpha u - \beta/u=0,
\\
{\hskip -10.0 truemm}
& &
\varepsilon Z^{-1}=u-U
=\frac{4 \varepsilon (\alpha - A) U} 
  {3 U' + \varepsilon (3 U^2 + 6 x U  - 2 A - 4 \alpha) + 6},\
\nonumber
\\
{\hskip -10.0 truemm}
& &
\phantom{\varepsilon Z^{-1}=u-U}
=\frac{4 \varepsilon (\alpha - A) u} 
  {3 u' + \varepsilon (3 u^2 + 6 x u  - 2 \alpha - 4 A) + 6},\
\\
{\hskip -10.0 truemm}
& &
(U-u)' + \varepsilon (U^2 - u^2 + 2 x (U-u) + 2 (\alpha - A)/3)=0,
\\
{\hskip -10.0 truemm}
& &
9 \beta + 2 (\alpha + 2 A -3 \varepsilon)^2=0,\
9 B     + 2 (A + 2 \alpha -3 \varepsilon)^2=0,\
\\
{\hskip -10.0 truemm}
& &
s=4 (A - \alpha)/3.
\end{eqnarray}

% ==========================================================================
\section{Conclusion, open problems}
\indent

The singular manifold method, which is based on the singularity structure,
is quite powerful to provide exact solutions or other analytic results.
There still exist many challenging problems,
in particular in nonlinear optics and spatiotemporal intermittency 
\cite{BN1985,vanHecke},
in which the equations, although nonintegrable, possess some regular
``patterns'' which could well be described by exact particular solutions.
The difficulty to find them \cite{CM1993} comes from the good guess which
must be made for the functions $\psi$,
which do not necessarily satisfy a linear system any more.
Methods from group theory usually provide complementary results,
although they also fail in the two just quoted examples.

% ==========================================================================
\section*{Acknowledgments}
\indent

The author thanks the CIME Fundazione and l'Universit\`a di Palermo
for invitation.

\vfill \eject

\printindex

% ***************************************************************** References

%                                 SAMPLE FILE OF SPRINGER MODIFIED BY CONTE
\clearpage
\addtocmark[2]{Subject Index}% additional numbered TOC entry
\markboth{Subject Index}{Subject Index}%
\renewcommand{\indexname}{Subject Index}%
\threecolindex  % starts the next index in three column mode
% \printindex
\end{document}